\documentclass[12pt]{iopart}
\usepackage{graphicx}
\usepackage{amssymb}
\usepackage{psfrag}
\usepackage{epstopdf}
\usepackage{color}
\def\ga{\mathrel{\mathpalette\fun >}}
\def\fun#1#2{\lower3.6pt\vbox{\baselineskip0pt\lineskip.9pt
  \ialign{$\mathsurround=0pt#1\hfil##\hfil$\crcr#2\crcr\sim\crcr}}}

 \definecolor{Black}{named}{Black}
 \definecolor{Blue}{named}{Blue}
 \definecolor{Red}{named}{Red}
 
 \newcommand{\red}[1]{\color{Black} #1\color{Black}}

 \newcommand{\J}[1]{\color{black} #1 \color{black}}

\def\nab{{\hbox{\boldmath$\nabla$}}}

\begin{document}

\title[Galactic Signatures of Decaying Dark Matter]{Galactic Signatures of Decaying Dark Matter}
\author{Le Zhang$^1$, Javier Redondo$^2$ and G\"unter Sigl$^1$}
\address{$^1$ II. Institut f\"ur theoretische Physik, Universit\"at Hamburg,
Luruper Chaussee 149, D-22761 Hamburg, Germany}
\address{$^2$ Deutsches Elektronen-Synchrotron, Notkestra\ss e 85,
22607 Hamburg, Germany}
 
\begin{abstract}
If dark matter decays into electrons and positrons, it can affect Galactic radio emissions and the local cosmic ray fluxes. We propose a new, more general analysis of constraints on dark matter. The constraints can be obtained for any decaying dark matter model by convolving the specific dark matter decay spectrum with a response function. We derive this response function from full-sky radio surveys at 408 MHz, 1.42 GHz and 23 GHz, as well as from the positron flux recently reported by PAMELA. 
We discuss the influence of astrophysical uncertainties on the response function, such as from propagation and from the profiles of the dark matter and the Galactic magnetic field. As an application, we find that some widely used dark matter decay scenarios can be ruled out under modest assumptions.     \end{abstract}


\maketitle

%
%

\section{Introduction}

The existence of non-baryonic dark matter is supported by many astronomical and cosmological observations such as rotation curves of galaxies, gravitational lensing, cluster dynamics, large scale structure surveys and the
cosmic microwave background (CMB) anisotropies~\cite{Bertone:2004pz}. 
However, the nature of dark matter remains elusive after decades of research 
- we only know that dark matter does not participate in the electromagnetic or strong interactions 
and that it has behaved as a non-relativistic fluid during the formation of the large scale structure of the universe.

Many extensions of the Standard Model of particle physics generally predict new dynamics between the electroweak and the Planck scales together with a number of new particles, sometimes with the required properties to be dark matter. 
If dark matter is composed of these particles, it can decay on cosmological time-scales or, in the case that a conservation law forbids the decay, it can annihilate in pairs. 
Dark matter decays or annihilations in our galaxy will produce electromagnetic radiation, anti-matter
and cosmic rays which we can be observed at Earth and serve as the first hints of dark matter non-gravitational interactions,
provided their fluxes can be disentangled from astrophysical backgrounds.

Such indirect detection of dark matter would not only provide us with a tangible glimpse of the
existence of particle physics beyond the Standard Model, but would
also allow a deeper understanding of structure formation. It may serve as a window
into the early universe, probing times much closer to the Big Bang than primordial nucleosynthesis. 
In addition, indirect signatures of dark matter can reveal its large scale distribution
and may thus serve as a diagnostic for astrophysics. Finally, decaying or annihilating dark matter
can even have interesting consequences for stellar evolution.

In this paper we develop a general formalism to derive information and constraints on decaying dark matter models. 
We focus on two observables, namely the cosmic positron flux on Earth and the synchrotron radiation from electrons and positrons in the galactic magnetic fields for both of which exist detailed observational data. 
The positron flux has recently been carefully measured by the PAMELA
collaboration~\cite{Adriani:2008zq,Adriani:2008zr}. For the synchrotron radiation, our analysis is based
on three full-sky maps at 408 MHz~\cite{map408}, 1.420 GHz~\cite{map142} and 22 GHz~\cite{map23} the latter from WMAP.
 
Usually, constraints are derived in the literature for specific dark matter scenarios with given decay or annihilation spectra and branching ratios into final state products. For a given dark matter scenario
one proceeds to compute the diffusion of electrons/positrons through the galaxy and the resulting
synchrotron maps. One can then compare with observations and derive constraints as in~\cite{Bertone:2001jv,Bertone:2002je,Ishiwata:2008qy,Finkbeiner:2004us,Hooper:2008zg}.
This method can be tedious if one is to use detailed propagation models, specifically when including re-acceleration because a numerical treatment is then mandatory, or if one has to repeat many simulations to fit data or scan a parameter space.

However, since the propagation equation is linear in the electron density, each injected electron energy evolves independently. Therefore, with a finite number of numerical simulations at different injected energies we can construct a numerical \emph{response  function} of signal to background. 
These response functions do depend on astrophysical parameters such as the cosmic ray propagation model and the dark matter halo profile, but not on the microscopic decay scenario. 
Constraints can then be simply obtained by requiring that the convolution of the response functions with a given dark matter decay spectrum be smaller than unity. 
In this paper, we have developed our own numerical propagation code to build such response functions and
provide constraints on decaying dark matter.
As a service to the model-builder we also provide analytic fits to our numerical response functions.

It should be noted that relatively large uncertainties from the propagation of cosmic rays and the dark matter
halo density profile could make dark matter constraints relatively uncertain~\cite{Donato:2008jk,Delahaye:2008ua}.
We therefore systematically investigate the model dependence of our response functions by taking into account different choices for propagation models and dark matter density profiles as well as for the Galactic magnetic field.     

Recently, the PAMELA~\cite{Adriani:2008zq,Adriani:2008zr} and ATIC~\cite{:2008zzr}
results show anomalously large fluxes of high energy positrons and electrons and indicate
the presence of previously unaccounted nearby sources. 
The results from the PAMELA anti-matter search show a hard change of slope in the positron to electron ratio in the range 10-100 GeV while at the same time the observed anti-proton flux nicely satisfies the expectations from being a secondary product of interactions of cosmic rays with the intergalactic medium. 
The ATIC collaboration reported an interesting bump in the energy range from 300 to 800 GeV and a sharp cutoff near $\sim$ 500 GeV in the flux of electrons plus positrons. 
However, more recent results from the FERMI/LAT experiment indicate at most a considerably smaller excess~\cite{Abdo:2009zk} than the suggested by ATIC.
In order to explain these features, much discussion of the possible sources has ensued in the literature. 
Some conventional astrophysics sources such as a nearby pulsar~\cite{Hooper:2008kg,Yuksel:2008rf,Malyshev:2009tw,Profumo:2008ms,Kobayashi:2003kp} can explain the
positron ``excess'' rather naturally. However, given the impact a discovery of dark matter would have, much
discussion has been on dark matter annihilation or decay as the origin of the PAMELA and ATIC
``anomalies'', see for instance~\cite{Cholis:2008vb,Bergstrom:2008gr,Cirelli:2008pk,Cholis:2008hb,ArkaniHamed:2008qn,Yin:2008bs,Hamaguchi:2008rv,Chen:2008fx,Zhang:2008tb} and~\cite{Ibarra:2008jk} for a model independent discussion. 
Appealing as they are, these scenarios also face a number of potential challenges. 
For instance, most dark matter models generate a softer spectrum in electrons and positrons than the PAMELA and ATIC data imply~\cite{Cholis:2008qq}. In the dark matter annihilation case, boost factors of $10^2-10^3$ are required to normalize the observed signals~\cite{Cholis:2008hb}, but at the same time are in serious conflict with other observations such as radio emissions~\cite{Bertone:2008xr}, $\gamma-$rays and anti-protons~\cite{Cirelli:2008pk,Donato:2008jk}. 
We believe that our study can help the model builder in the quest for explaining the PAMELA positron fluxes,
or in providing complementary bounds from the synchrotron radio maps.

The rest of this paper is organized as follows. In Sect.~\ref{sec:setup}, we provide general estimates of
positron and radio signals induced by dark matter, and describe our method of response functions. In Sect.~\ref{sec:res-radio} and Sect.~\ref{sec:res-positron} we present the response functions resulting from the observed radio emissions, and
from the flux of electrons and positrons, respectively. In Sect.~\ref{sec:constraint} we apply our method
to some concrete dark matter decay scenarios. A summary and discussion are given in Sect.~\ref{sec:conclusion}.
In Appendix~A we review the astrophysical parameters including propagation models, halo profile and magnetic fields and their influence on the distribution of electrons and positrons. In Appendix~B we provide analytical fits
to our response functions that can easily be applied to any dark matter decay model the reader may wish
to probe. We will use natural units in which $c=1$ throughout.

%
%

\section{Propagation of electron/positrons and Response Functions}
\label{sec:setup}

In this section we introduce response functions which describe the effects of propagation on the electron and positron fluxes and on the resulting synchrotron fluxes observable at Earth in terms of the injected electron or positron energy. 
To start, we give some simple estimates of these fluxes before embarking on a more detailed calculation of the response functions.

\subsection{Estimate of the electron-positron flux}
The energy loss time of electrons and positrons in a radiation field of energy
density $u_\gamma$ is
\begin{eqnarray}\label{eq:loss}
  t_{\rm loss}(E)=-\frac{E}{dE/dt}&\simeq&6.5\times10^{15}\,\left(\frac{\rm GeV}{E}\right)
  \left(\frac{u_\gamma}{{\rm eV}\,{\rm cm}^{-3}}\right)^{-1}\,{\rm s}\nonumber\\
  &\simeq&10^{16}\,\left(\frac{\rm GeV}{E}\right)\,{\rm s}\,,
\end{eqnarray}
where the latter expression is often assumed throughout the Galaxy~\cite{Baltz:1998xv}.
The confinement time due to diffusive propagation in the Galaxy is similar to
the confinement time of hadronic cosmic rays which at GeV energies can be estimated
from secondary beryllium isotopes in the Galactic cosmic ray flux~\cite{connell},
\begin{equation}\label{eq:conf}
  t_{\rm conf}(E)\simeq3\times10^7\,{\rm y}\simeq10^{15}\,{\rm s}\,.
\end{equation}
This is consistent with the diffusion time $t_{\rm diff}(E)\simeq h^2/K(E)$ in a
galactic disk of height $2h\sim4\,$kpc with the diffusion constant~\cite{Baltz:1998xv}
\begin{equation}\label{eq:diff_coeff}
  K(E)\simeq3\times10^{27}\left(\frac{E}{\rm GeV}\right)^{0.6}\,{\rm cm}^2\,{\rm s}^{-1}\,,
\end{equation}
which yields
\begin{equation}\label{eq:diff}
  t_{\rm diff}(E)\simeq3\times10^{15}\left(\frac{h}{2\,{\rm kpc}}\right)^2
  \left(\frac{E}{10\,{\rm GeV}}\right)^{-0.6}\,{\rm s}\,.
\end{equation}
The effective lifetime of electrons and positrons is thus
$\tau_e(E)\simeq{\rm min}\left[t_{\rm loss}(E),t_{\rm diff}(E)\right]$.
At $E\simeq10\,$GeV this is, therefore, $\tau_e(10\,{\rm GeV})\simeq10^{15}\,$s.

The differential flux of electrons and positrons $j_e(E)$ per energy interval
from dark matter of mass $m_X$ and
lifetime $\tau_X$ which produces on average $Y_e(E)$ electrons and positrons per decay
and has a local density $\rho_X$ can thus be estimated as
\begin{eqnarray}\label{eq:flux-model}
  \fl E^2j_e(E)&\simeq&E\,\frac{c_0}{4\pi}\frac{\rho_X}{m_X}\frac{Y_e(E)}{\tau_X}\tau_e(E)\\
  \fl &\simeq&
  7\times10^{-3}\,\left(\frac{\rho_X}{0.3\,{\rm GeV}\,{\rm cm}^{-3}}\right)\,
  \left(\frac{Y_e(E)E}{m_X}\right)\,
  \left(\frac{\tau_e(E)}{10^{15}\,{\rm s}}\right)\,
  \left(\frac{10^{26}\,{\rm s}}{\tau_X}\right)\,
  {\rm GeV}\,{\rm cm}^{-2}\,{\rm s}^{-1}\,{\rm sr}^{-1}\,.\nonumber
\end{eqnarray}
Here, $Y_e(E)(E/m_X)\le1$ depends on the particle physics of the decays and could
be of order unity.

The observed flux of electrons and positrons at $E\simeq10\,$GeV is~\cite{Baltz:1998xv,Moskalenko:1997gh,Abdo:2009zk}
\begin{equation}\label{eq:flux-obs}
  E^2j^{\rm obs}_e(E)\simeq2\times10^{-3}\,
  {\rm GeV}\,{\rm cm}^{-2}\,{\rm s}^{-1}\,{\rm sr}^{-1}\,. 
\end{equation}
The gravitino dark matter model discussed in Ref.~\cite{Ibarra:2008qg} was constructed
such that the gravitino decays could explain the EGRET excess, leading to
$m_X\simeq150\,$GeV, $\tau_X\simeq10^{26}\,$s. Comparing Eq.~(\ref{eq:flux-model})
with Eq.~(\ref{eq:flux-obs}) shows that the
electron-positron flux produced by the decays can be comparable to or even exceed
the locally observed electron-positron flux. A more detailed numerical simulation
is, therefore, called for.

Indirect effects such as radio emission in the Galactic magnetic field, can give complementary
constraints since they are sensitive not only to the local electron-positron flux but
also to the electron-positron flux induced by dark matter decay in remote parts of
the Galaxy which is not directly measurable. We, therefore, now turn to radio signatures.

\subsection{Estimate of the radio flux}
The power per frequency interval emitted by an electron or positron of energy $E$ in a magnetic field $B$,
averaged over magnetic field directions, is given by~\cite{Ghisellini:1988}
\begin{equation}\label{eq:P_radio}
P(\nu,E) = \frac{2\sqrt{3}e^3B}{m_e} x^2 \left\{ K_{4/3}(x)K_{1/3}(x)-\frac{3}{5} (K_{4/3}^2(x)-K_{1/3}^2(x))\right\}
\end{equation}
where $e$ and $m_e$ are the electron charge and mass, respectively, 
and we have abbreviated $x=\nu/(2\nu_c)$ where the critical frequency is
\begin{equation}\label{eq:nu_crit}
  \nu_c(E)=\frac{3}{4\pi}\frac{eB}{m_e}\left(\frac{E}{m_e}\right)^2=
  966\,\left(\frac{B}{6\mu{\rm G}}\right)\,\left(\frac{E}{100\,{\rm GeV}}\right)^2\,{\rm GHz}\, .
\end{equation}
For $\nu\ga10\,$MHz self-absorption is negligible and the emitted radio
intensity $J(\nu)$ in units of power per frequency interval along a given line of sight
is then given by
\begin{equation}\label{eq:Jnu}
  J(\nu)=\int ds\int dE j_e(E)P(\nu,E)\,,
\end{equation}
where $s$ is the distance along the line of sight. 
Inserting Eq.~(\ref{eq:P_radio}) with the approximation $2x^2\{... \}\sim  \delta(2x-1.5)$
simplifies Eq.~(\ref{eq:Jnu}) to
\begin{equation}\label{eq:Jnu2}
  \nu J(\nu)\simeq\frac{3e^{7/2}}{4(\pi\cdot0.29)^{1/2}}\frac{\nu^{1/2}}{m_e^{5/2}}
  \int ds B(s)^{3/2}\left.\left[E^2j_e(E)\right]\right|_{E_c(\nu)}\,,
\end{equation}
where the critical energy
\begin{equation}\label{eq:E_crit}
  E_c(\nu)=\left(\frac{4 \pi}{3\cdot0.29}\frac{m_e^3}{e}\frac{\nu}{B}\right)^{1/2}=
  5.9\left(\frac{\nu}{1\,{\rm GHz}}\right)^{1/2}\left(\frac{B}{6\,\mu{\rm G}}\right)^{-1/2}
  \,{\rm GeV}
\end{equation}
is the inversion of Eq.~(\ref{eq:nu_crit}). Assuming the magnetic field approximately
constant out to a distance $d$, for example for about 10 kpc towards the galactic anti-center,
Eq.~(\ref{eq:Jnu2}) can be quantified as
\begin{equation}\label{eq:Jnu3}
  \nu J(\nu)\simeq2.6\times10^{-4}\,\left(\frac{\nu}{\rm GHz}\right)^{1/2}\,
  \left(\frac{d}{10\,{\rm kpc}}\right)\,\left(\frac{B}{6\,\mu{\rm G}}\right)^{3/2}\,
  \left.\left[E^2j_e(E)\right]\right|_{E_c(\nu)}\,.
\end{equation}
Inserting now the estimate Eq.~(\ref{eq:flux-model}), we obtain
\begin{eqnarray}\label{eq:Jnu4}
  \nu J(\nu)&\simeq&2.9\times10^{-9}\,\left(\frac{\nu}{\rm GHz}\right)^{1/2}\,
  \left(\frac{d}{10\,{\rm kpc}}\right)\,\left(\frac{B}{6\,\mu{\rm G}}\right)^{3/2}\,\\
  &&\times\left(\frac{Y_e(E)E}{m_X}\right)\,
  \left(\frac{\tau_e(E)}{10^{15}\,{\rm s}}\right)\,
  \left(\frac{10^{26}\,{\rm s}}{\tau_X}\right)\,
  {\rm erg}\,{\rm cm}^{-2}\,{\rm s}^{-1}\,{\rm sr}^{-1}\,.\nonumber
\end{eqnarray}
This is comparable to or higher than the measured high Galactic latitude radio flux,
which is of order $10^{-9}\,{\rm erg}\,{\rm cm}^{-2}\,{\rm s}^{-1}\,{\rm sr}^{-1}$
at GHz frequencies. We now turn to more detailed numerical calculations of the
radio signatures and local flux of positrons and electrons. 

\subsection{Response Function and Constraints}
\label{sec:response}

The propagation of positrons and/or electrons in the Galactic magnetic field is usually described by a diffusion model. 
Under this approximation, the diffusion-loss equation for the relevant particle density per unit of momentum interval $n({\bf r},p,t)$ 
can be written in the form
\begin{equation}
\label{eq:transport_pre}
\frac{\partial n}{\partial t}-{\cal D}n= Q({\bf r},p)
\end{equation}
where the differential operator ${\cal D}$ is 
\begin{equation}
\label{eq:transport}
\fl {\cal D}n = \nab\cdot\left(D_{xx}\nab n-{\bf V_c}n\right)+\frac{\partial}{\partial p}\left(p^2D_{pp}\frac{\partial} {\partial p}\frac{n}{p^2}\right) - \frac{\partial}{\partial p}\left[\dot{p}\,n -\frac{p}{3}(\nab\cdot{\bf V_c}n)\right]\,.
\end{equation}

Here, $D_{xx}$ is the spatial diffusion coefficient, ${\bf V_c}$ is the convection velocity, re-acceleration is described as the diffusion in momentum space and is determined by the coefficient $D_{pp}$, and
$\dot{p}\equiv dp/dt$ is the momentum loss rate. Since we are interested in relativistic 
electrons and positrons, we will use energy $E$ and momentum $p$ indistinctly and
write $n({\bf r},E)$ in the following. Henceforth, we use $n_+$ and $n_-$ for the positron and electron density, respectively, and $n_e=n_++n_-$.
In the case of CP conserving decays one has $n_+=n_-=n_e/2$.

The source term for positrons/electrons due to decaying dark matter particles with mass $m_{\chi}$ and lifetime $\tau$ is given by
\begin{equation}\label{eq:source}
Q_\pm({\bf r},E_0) = \frac{\rho_X({\bf r})}{m_X\tau_X} \frac{dN_\pm}{dE_0} 
\end{equation}
where $m_X $ is the dark matter particle mass and $\tau_X$ its lifetime, $\rho_X({\bf r})$ is the dark matter density profile in our Galaxy, and $dN_\pm/dE_0$ is the spectrum of positrons/electrons per dark matter particle decay.

Consider the stationary solutions $n_\pm^{E_0}({\bf r},E)$ to the  propagation equation for monochromatic injection of positrons or electrons at $E_0$, i.e. the Green's function satisfying
\begin{equation} \label{eq:transport_mono}
-{\cal D}\, n_\pm^{E_0}({\bf r},E) = \frac{\rho_X({\bf r})}{m_X\tau_X} \delta(E-E_0) \ . 
\end{equation}
The solution of Eq.~(\ref{eq:transport}) for an arbitrary spectrum $dN_\pm/dE_0$ can then be written as
\begin{equation}
n_\pm({\bf r},E) = \int dE_0\, n_\pm^{E_0}({\bf r},E) \frac{dN_\pm}{dE_0}  \ . 
\end{equation}

The synchrotron flux arriving to the earth from a direction $\Omega$ characterized by galactic coordinates $\Omega=(\phi,\theta)$ has a contribution for each monochromatic injection $E_0$ given by Eq.~(\ref{eq:Jnu})
\begin{equation}
J^{E_0}(\Omega,\nu) = \frac{1}{4 \pi}\int ds \int dE \, n_e^{E_0}({\bf r},E) P(\nu, E) \ . 
\end{equation}

For an arbitrary injection spectrum $dN_e/dE_0$ the synchrotron flux at frequency $\nu$ is then
obtained by
\begin{equation}
J(\Omega,\nu) = \int\, d E_0 J^{E_0}(\Omega,\nu) \frac{d N_e}{dE_0}\,.
\end{equation}
It is then convenient to introduce the response functions for positrons 
$F_p(E;E_0)$ and for synchrotron emission $F_r(\Omega,\nu;E_0)$ as the ratio of the numerically computed 
$n^{E_0}_+(E)$ and $J^{E_0}(\phi,\theta,\nu)$, respectively, and the observed fluxes as
\begin{eqnarray}\label{eq:response}
F_p(E;E_0) &=& \frac{n_+^{E_0}({\bf r}_{\rm earth},E)}{n_+^{\rm obs}(E)} 
\left(\frac{\tau_X}{10^{26}\,{\rm s}}\right) \left(\frac{m_X}{100\,{\rm GeV}}\right)\nonumber\\
F_r(\Omega,\nu;E_0) &=& \frac{J^{E_0}(\Omega,\nu)}{J^{\rm obs}(\Omega,\nu)}
\left(\frac{\tau_X}{10^{26}\,{\rm s}}\right) \left(\frac{m_X}{100\,{\rm GeV}}\right) 
\end{eqnarray}
These functions depend neither on $\tau_X$ nor on $m_X$ and
constraints on a given dark matter decay model can then be easily cast in the form
\begin{eqnarray}\label{eq:constraint}
\int_{m_e}^{m_X} d E_0\, F_p(E;E_0)\frac{dN_+}{dE_0}&\leq&
\left(\frac{\tau_X}{10^{26}\,{\rm s}}\right)\left(\frac{m_X}{100\,{\rm GeV}}\right)\ ,\nonumber\\
\int_{m_e}^{m_X} d E_0\, F_r(\Omega,\nu;E_0)\frac{dN_e}{dE_0}&\leq&
\left(\frac{\tau_X}{10^{26}\,{\rm s}}\right)\left(\frac{m_X}{100\,{\rm GeV}}\right)		\ . 	
\end{eqnarray}

The desired response functions can be computed numerically by using the methods of~\cite{Strong:1998pw,Moskalenko:1997gh}. 
In order to do so, we have developed our own numerical code. 
Details on our code and computations are described in Appendix~A. 

Let us once more stress that our response functions do not depend on the specific decay spectrum,
but still depend on the characteristics of the propagation model and
the dark matter distribution. In this paper we use different halo models, see Appendix~A, always normalized such that
$\rho({\bf r}_{\rm earth})= 0.3$ GeV cm$^{-3}$. For other normalizations $\rho({\bf r}_{\rm earth})$, our
response functions have to be multiplied by $\rho({\bf r}_{\rm earth})/0.3$ GeV cm$^{-3}$.

Finally, apart from the injection energy $E_0$, the synchrotron response function $F_r(\Omega,\nu;E_0)$
depends on the observed frequency $\nu$ and
the direction in the sky $\Omega$, whereas the positron response function $F_r(E;E_0)$ depends on the
observed positron energy $E$.
In the latter case, we use the PAMELA data to get our constraints. It consists of seven different energies so we can construct seven different response functions, see Section~5. 
The synchrotron case is more complicated since in principle there are infinite directions to look at, and the optimal direction will depend on the injected spectrum $dN_e/dE_0$ and the observed frequency. We discuss this case in the next section.

%
%

\section{Response Functions for Radio Signals}\label{sec:res-radio}

\begin{figure}[tp]
\centering  
\includegraphics[width=7.3cm]{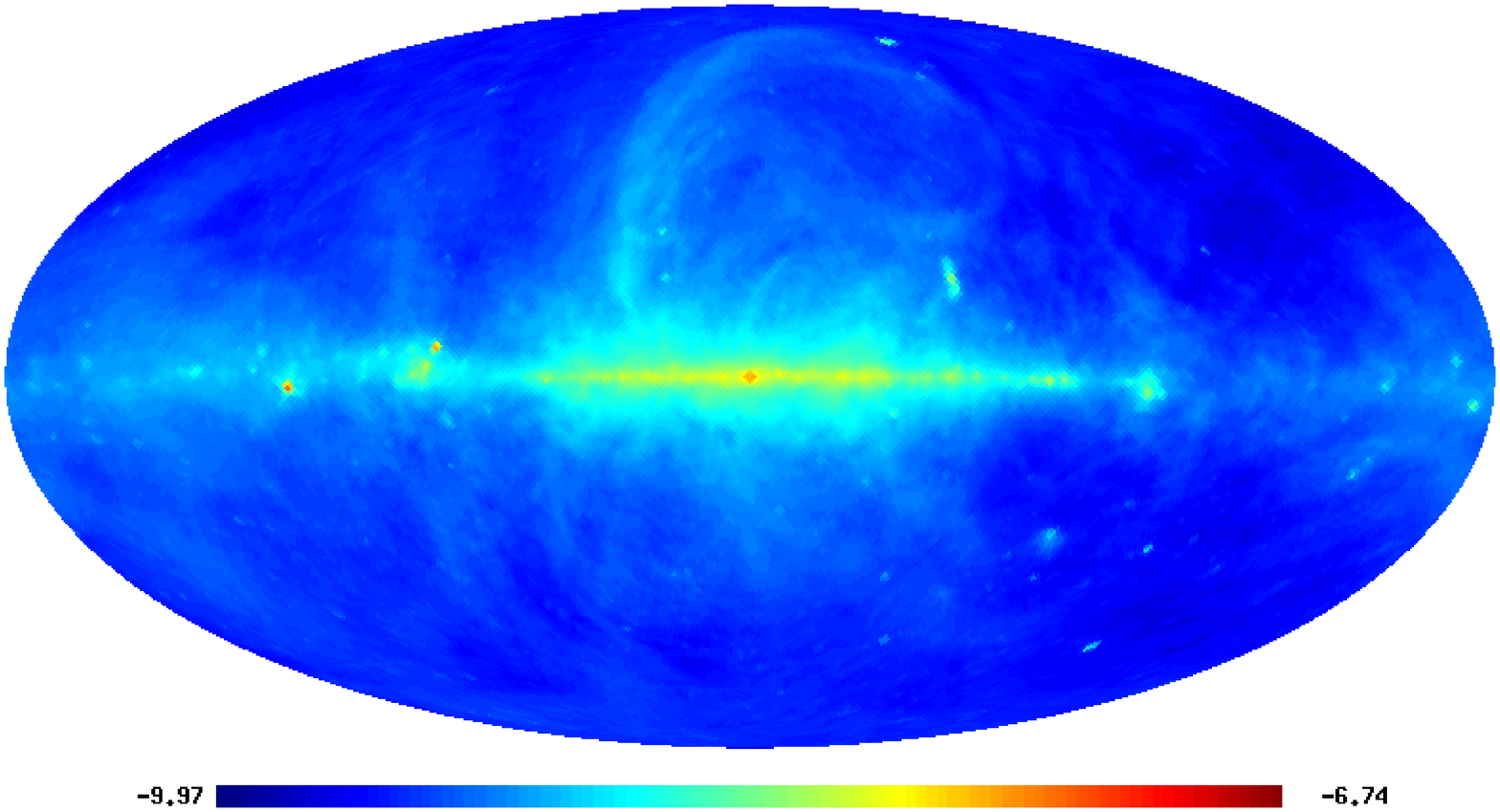}
\includegraphics[width=7.3cm]{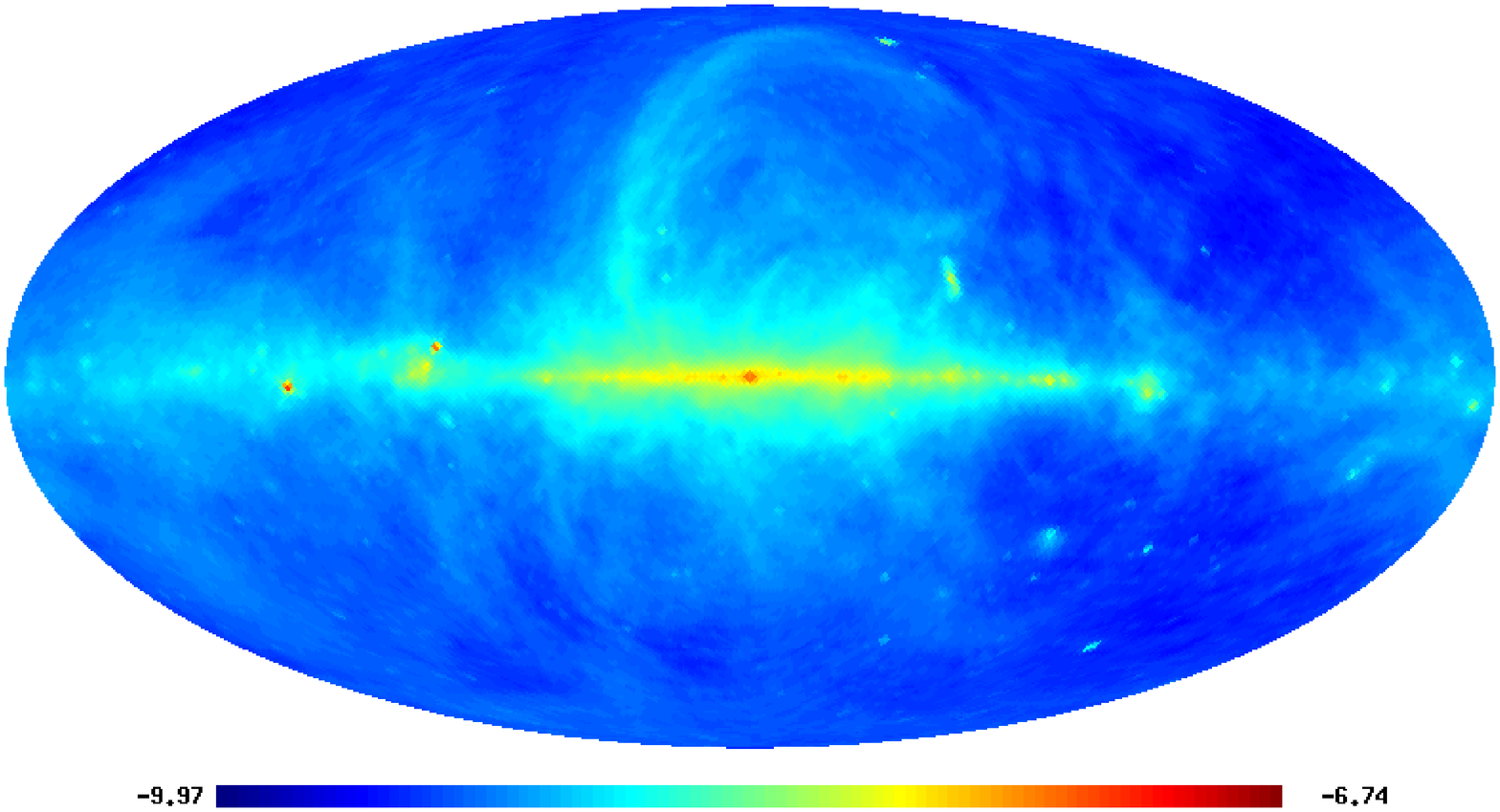}
\includegraphics[width=7.5cm]{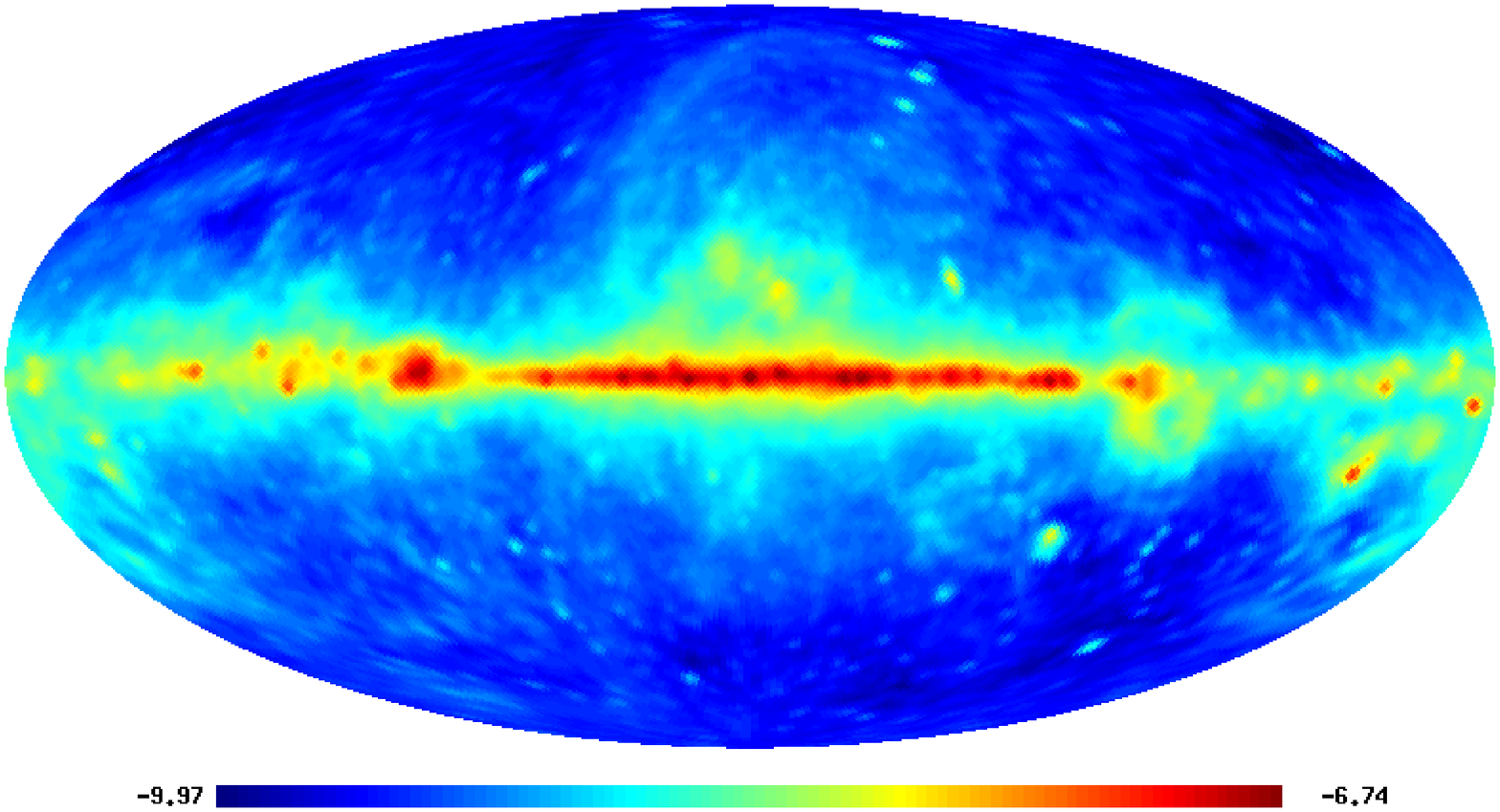}
\caption{Maps of the radio sky at frequencies 408 MHz, 1.42 GHz, and 23 GHz, from the top left and moving clockwise~\cite{map408,map142,map23}. The
color scaling is the logarithm to the base 10 of the flux in erg/s/$\rm{cm}^2$/sr.}
\label{fig:ob-map}
\end{figure}

In this section we compute the radio emission induced by dark matter decay and establish the response function by comparison with radio observations. 
As can be seen from Eq.~(\ref{eq:nu_crit}), the radio frequencies relevant for our study are between 0.1 and a few 100 GHz. 
Although the cosmic microwave background (CMB) would dominate the radio sky at frequencies above $\simeq1$ GHz, this signal can be removed from the foreground based on the sensitive multi-frequency survey performed by the WMAP satellite. 
In Fig.~\ref{fig:ob-map}, we show the full sky surveys at the frequencies 408 MHz~\cite{map408}, 1.42 GHz~\cite{map142}, and 23 GHz~\cite{map23}. 
We do not use the higher frequency channels (33 GHz, 41 GHz, 61 GHz, 94 GHz) observed by WMAP, as they are considerably more noisy and less robust to foreground subtraction than the lower frequency bands. 
In addition, we smoothed all maps to angular resolution of $1^\circ$.

\subsection{Observed Radio Sky}
Based on various global multi-frequency model fits, recently a public code has been made available~\cite{deOliveiraCosta:2008pb} which allows to fit most of the radio survey observation in the range 10 MHz-100 GHz to accuracy around $1\%-10\%$ depending on frequency and sky region. 
To obtain the strongest possible constraints on radio emission due to dark matter decay, one would perform a pixel-by-pixel scan over
the whole sky and over all frequencies between 0.1 MHz to 100 GHz. 
Most of the observed radio maps, however, have only partial sky coverage. 
Although these uncovered regions can be interpolated by global model fits, we want to rely on direct observations so we only use the 408 MHz, 1.42 GHz and 23 GHz full sky survey maps for our study.

Recently, an excess of microwave emission in the inner $20^\circ$ around the center of our Galaxy has been revealed in WMAP bands between 22 and 93 GHz, after a subtraction of the free-free, dust and standard synchrotron emissions. 
This excess, dubbed as ``WMAP haze''~\cite{Finkbeiner:2003im,Finkbeiner:2004je}, is distributed with approximate radial symmetry and is uncorrelated with the known foregrounds\footnote{The origin of the haze is currently a hot topic of debate. 
In principle it could be explained by conventionally astrophysical sources such as pulsars~\cite{Kaplinghat:2009ix} but an explanation in terms of dark matter annihilation has also been 
suggested~\cite{Finkbeiner:2004us,Cumberbatch:2009ji,Hooper:2007kb}. }.  
The use of this haze map with subtracted ``known'' foregrounds could further strengthen our constraints on decaying dark matter easily by an order of magnitude as has been shown already in the case of dark matter annihilation in~\cite{Borriello:2008gy}. \red{We have, in fact, also performed this analysis and found the same order of magnitude improvement in the constraints than Ref.~\cite{Borriello:2008gy}. In addition,
the 22 GHz emission is likely dominated by spinning dust, as well as synchrotron and
thermal dust emission. The synchrotron emission can be constrained from its polarized 
emission~\cite{miville-deschenes}. However, it should be kept in mind that the astrophysical backgrounds themselves depend to some extent on the not very well known galactic magnetic field and cosmic ray propagation parameters. Therefore, in this paper we want to be conservative and stick to the real observations so we do not present response functions based on the haze maps or other background subtractions. }
The subtraction of foregrounds will be improved by forthcoming radio data from Planck and at low frequencies from LOFAR and, in a more distant future, from SKA.

\subsection{Radio emission from dark matter electrons}

\begin{figure}[tp]
\centering
\includegraphics[height=2.5 cm]{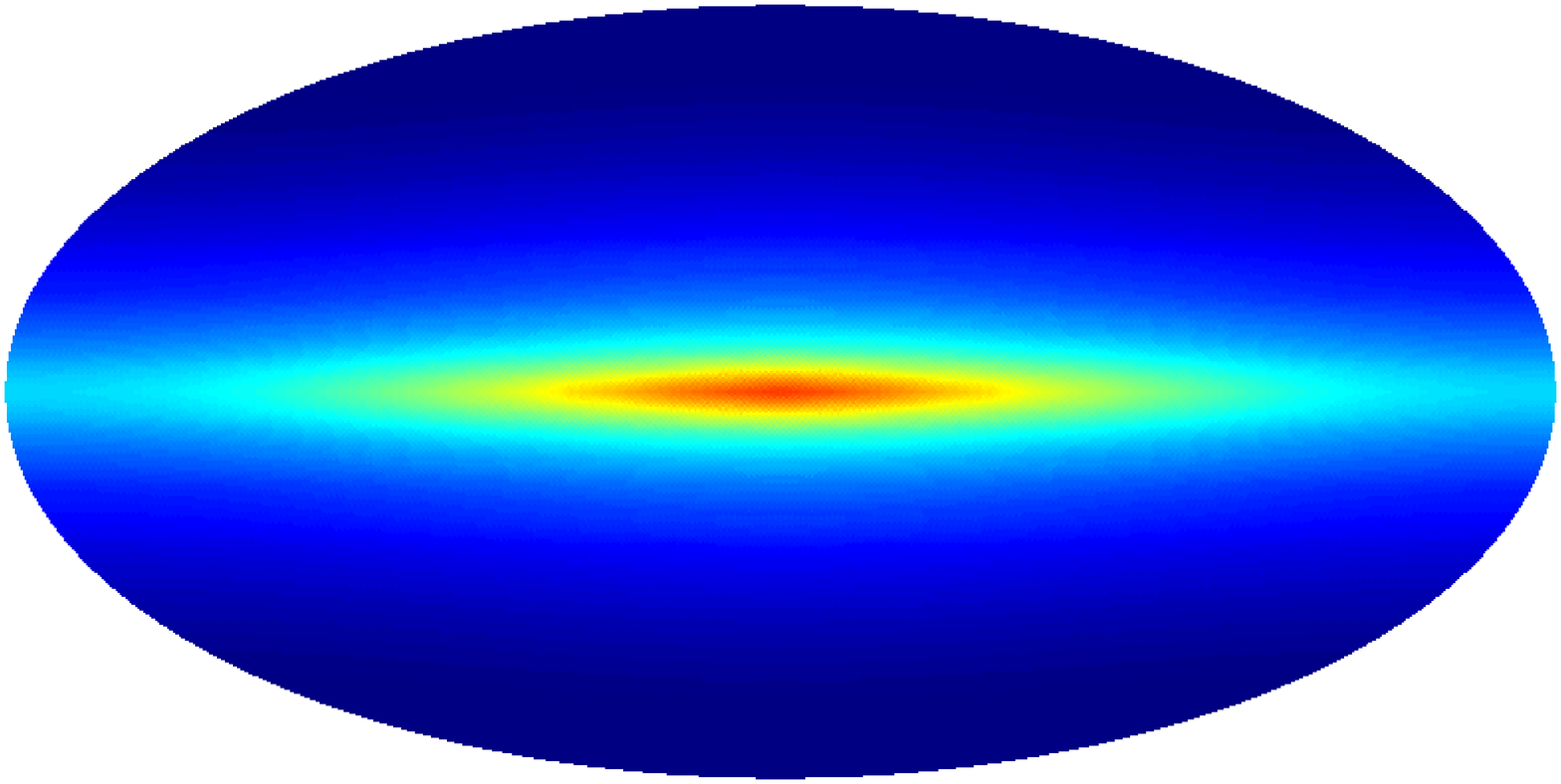}
\includegraphics[height=2.5 cm]{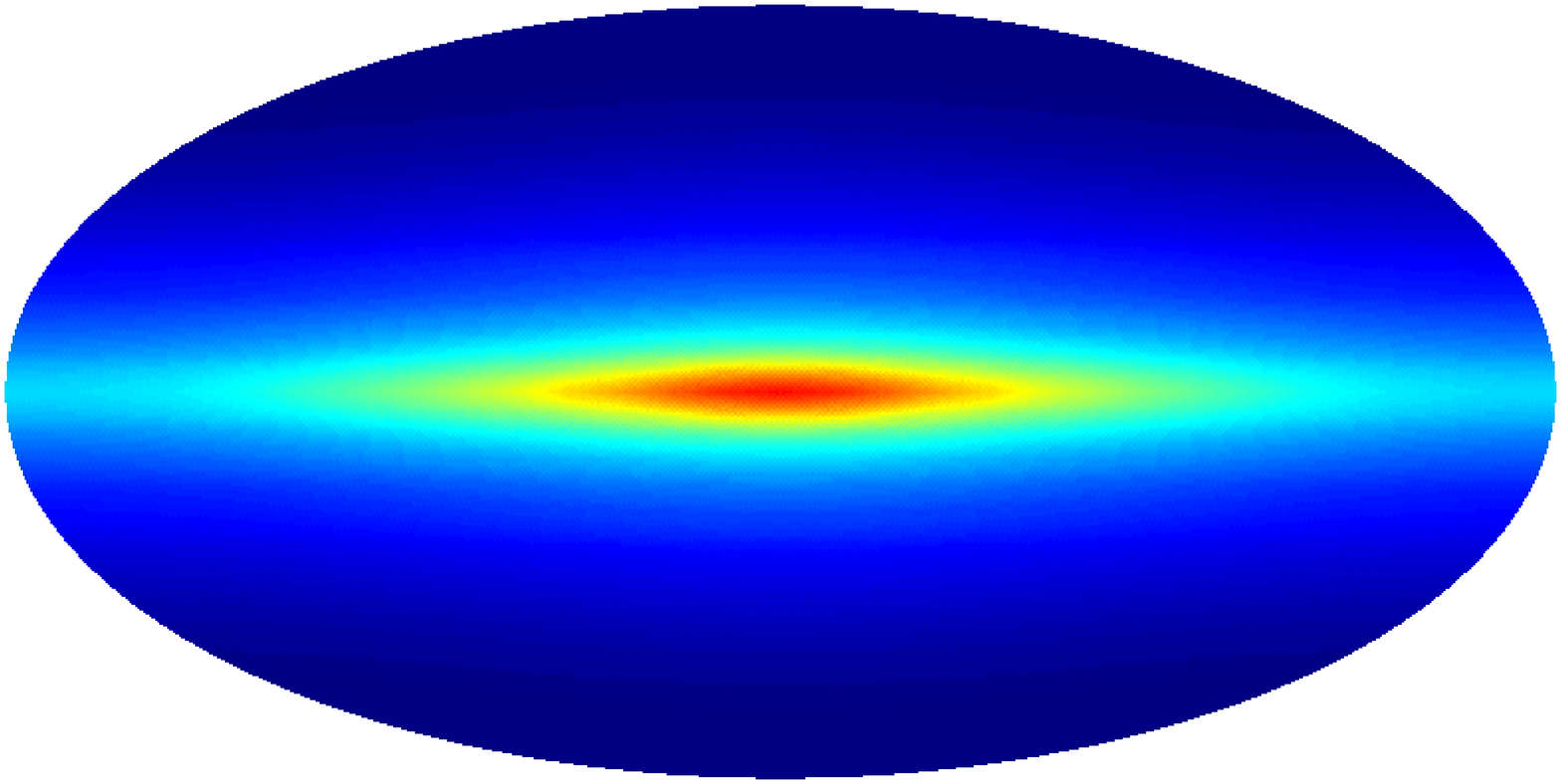}
\includegraphics[height=2.5 cm]{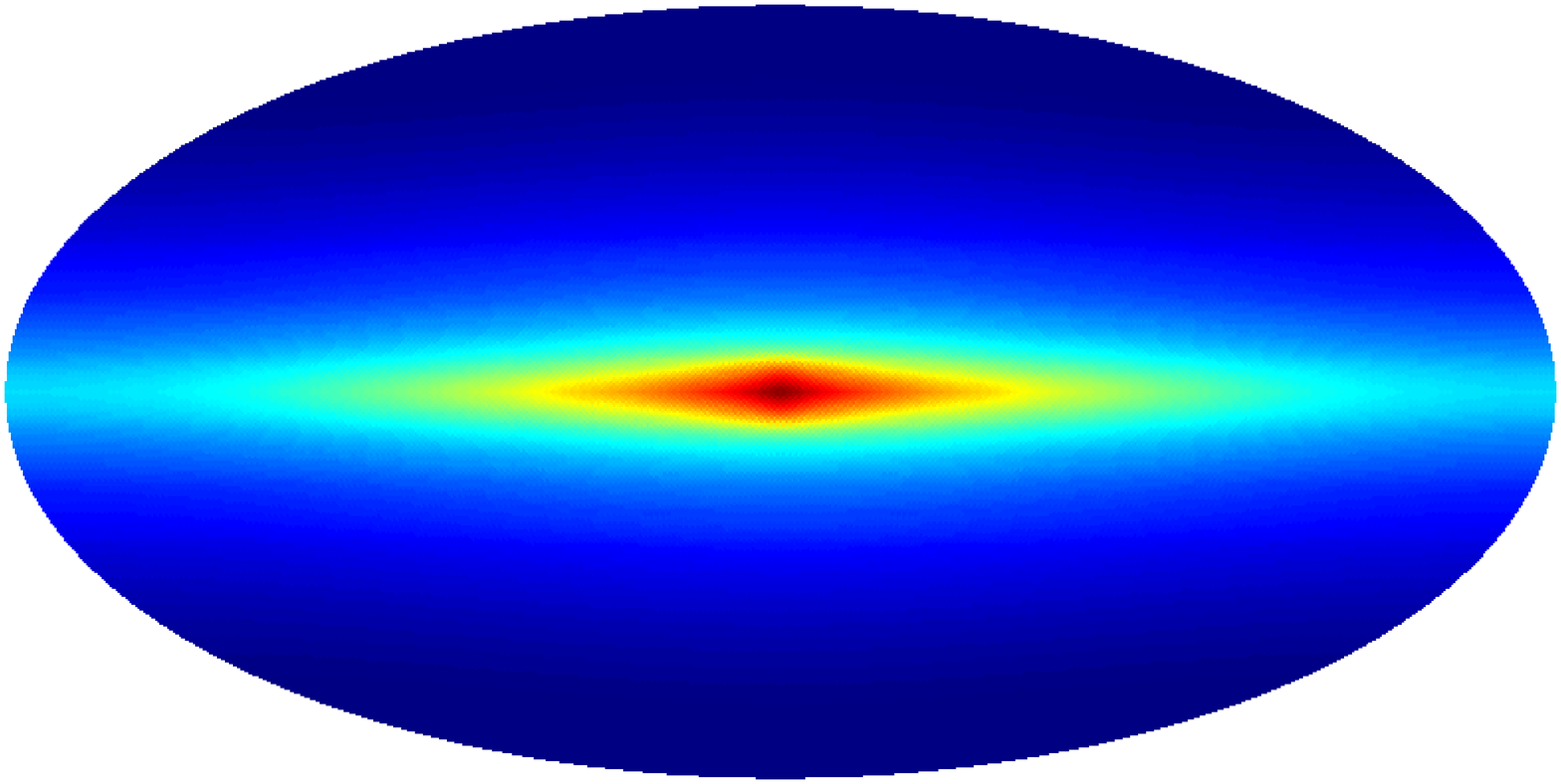}

\includegraphics[height=2.5 cm]{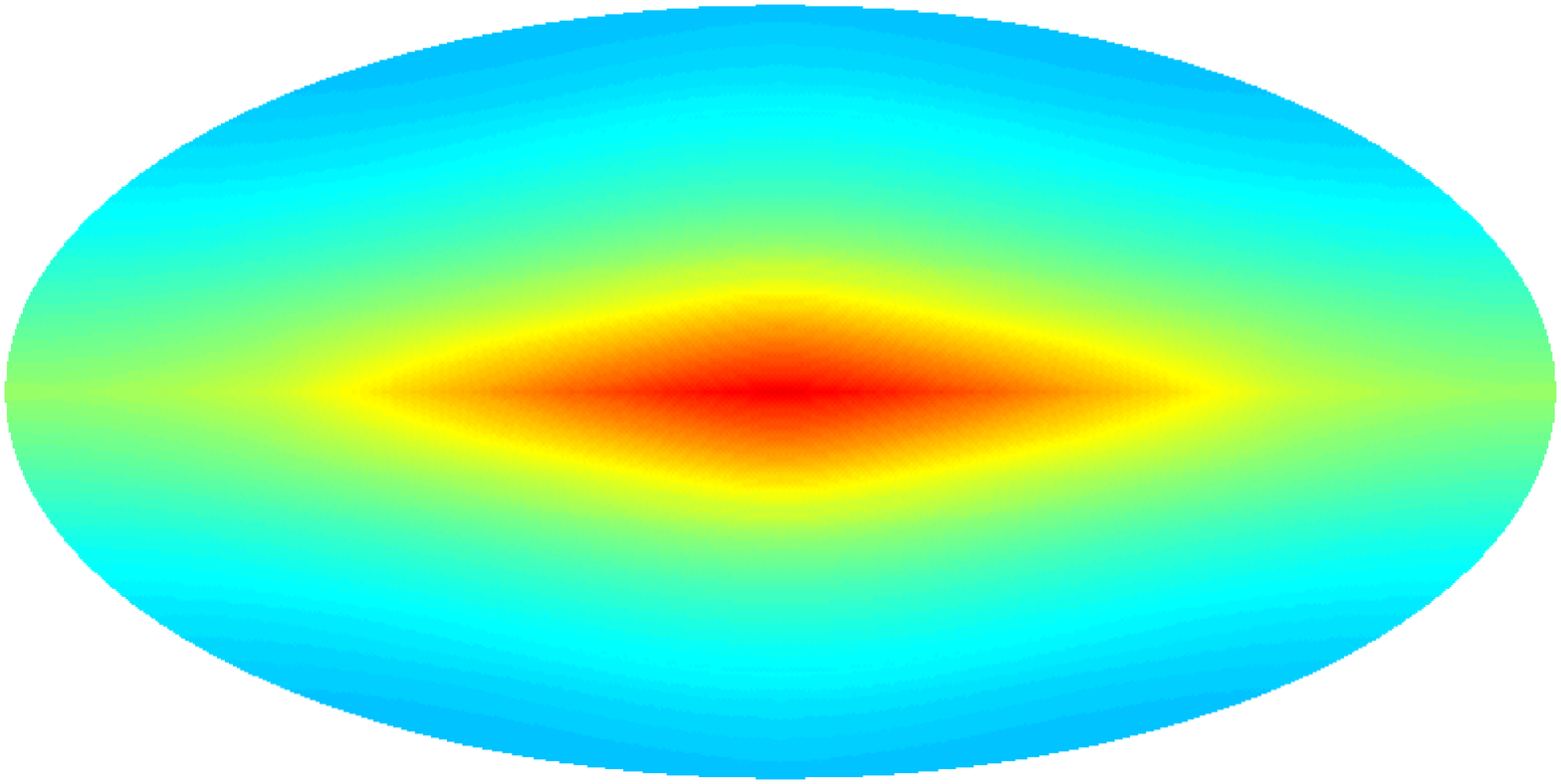}
\includegraphics[height=2.5 cm]{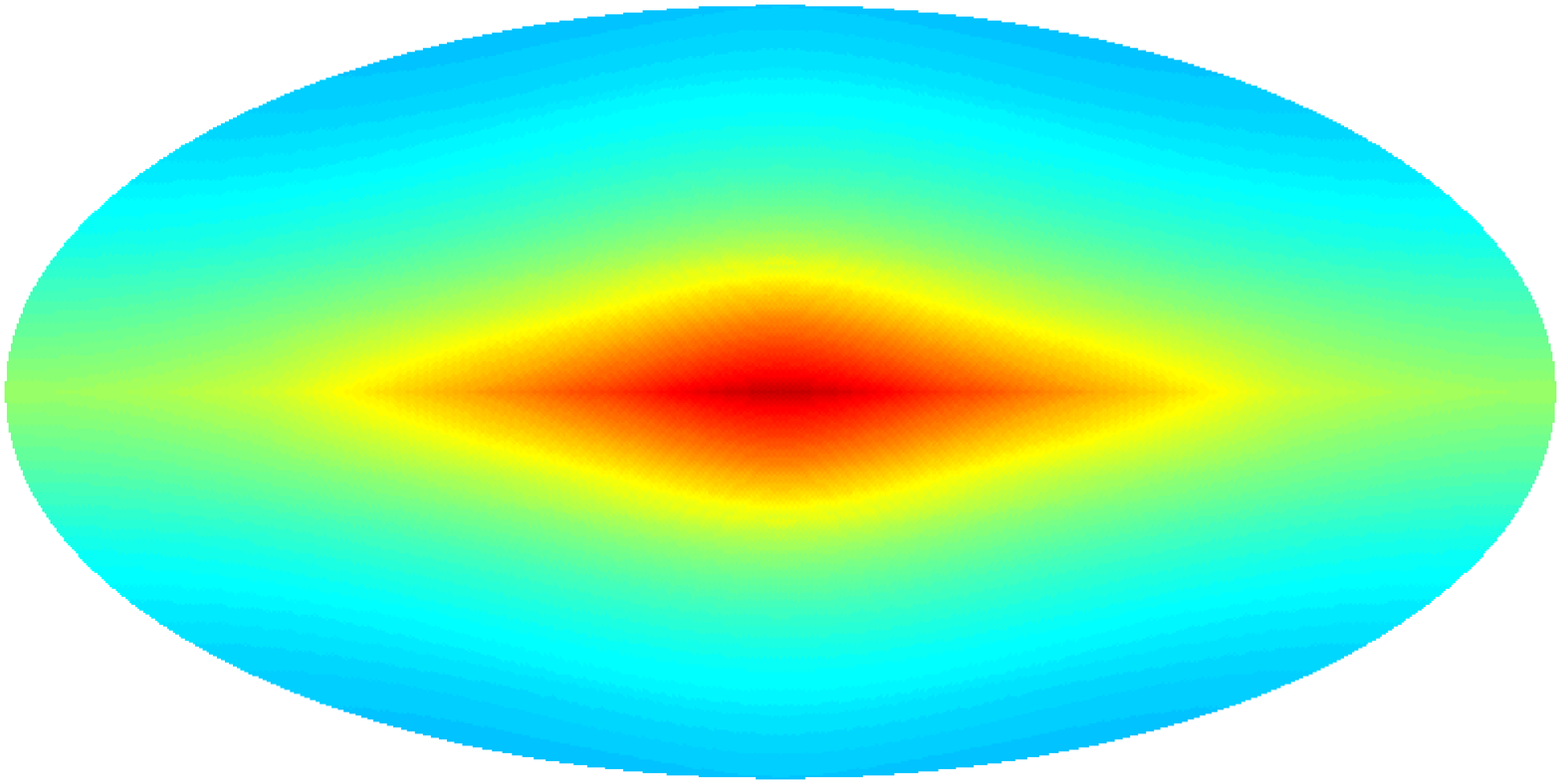}
\includegraphics[height=2.5 cm]{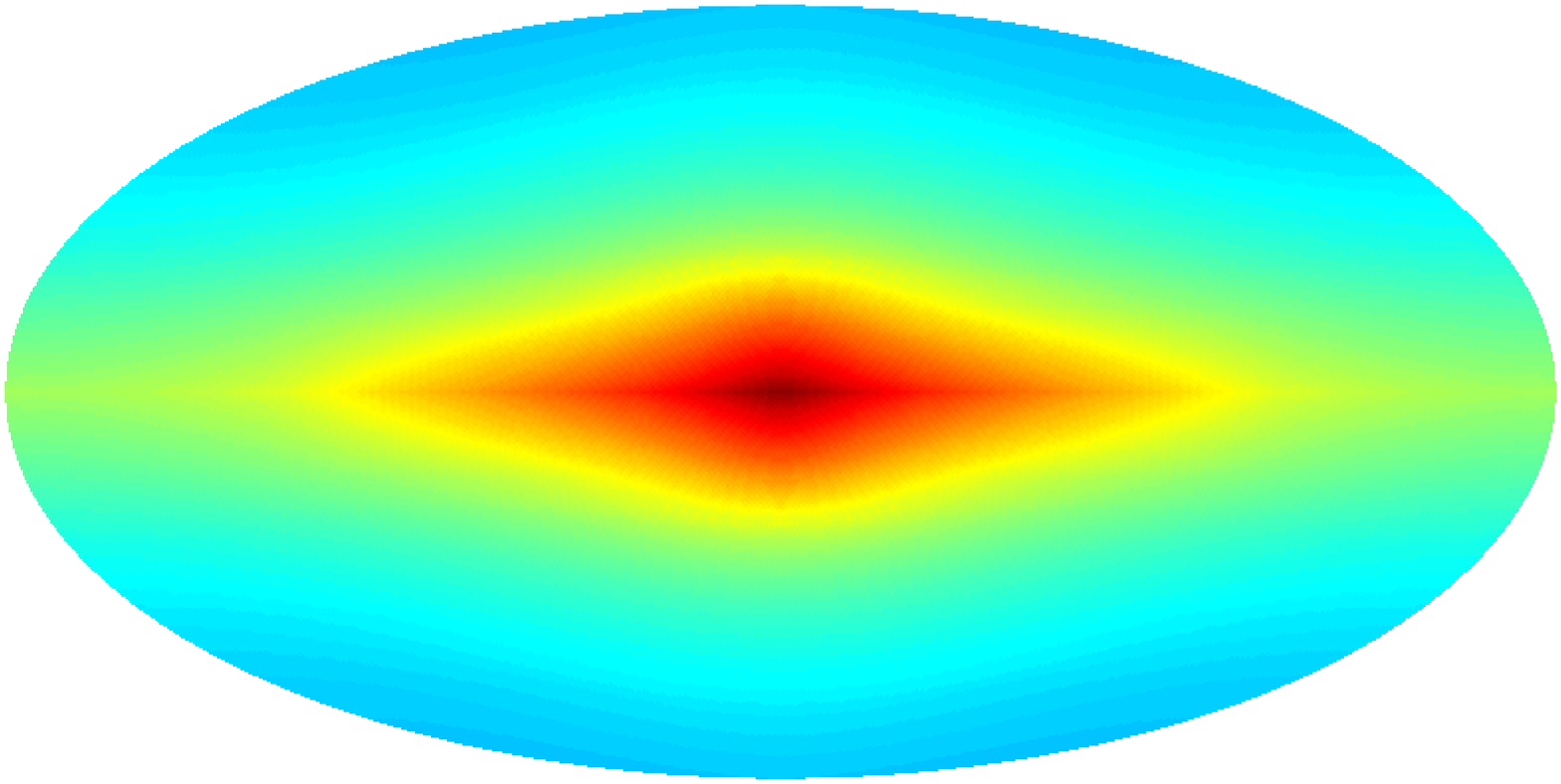}

\includegraphics[height=2.5 cm]{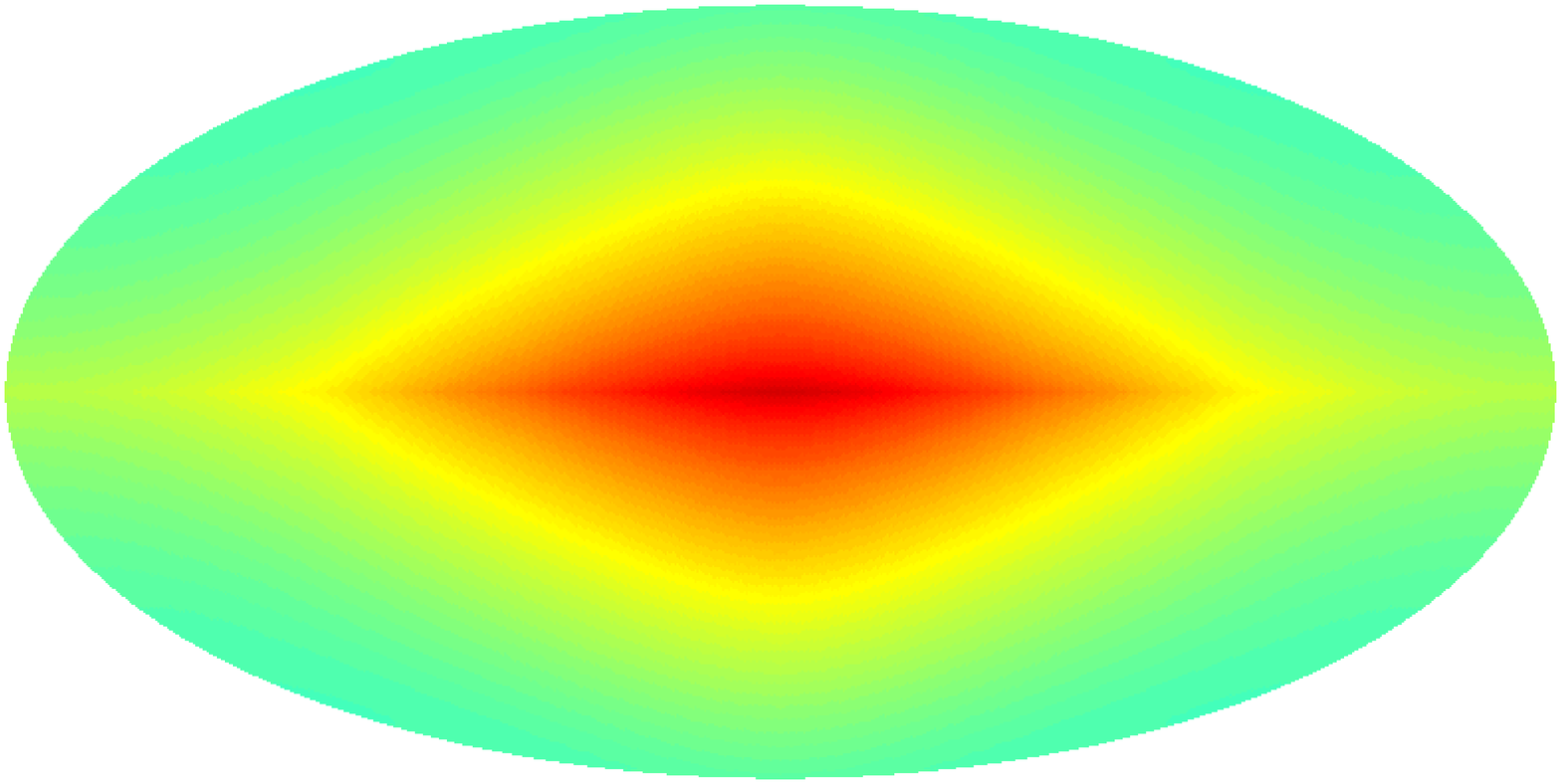}
\includegraphics[height=2.5 cm]{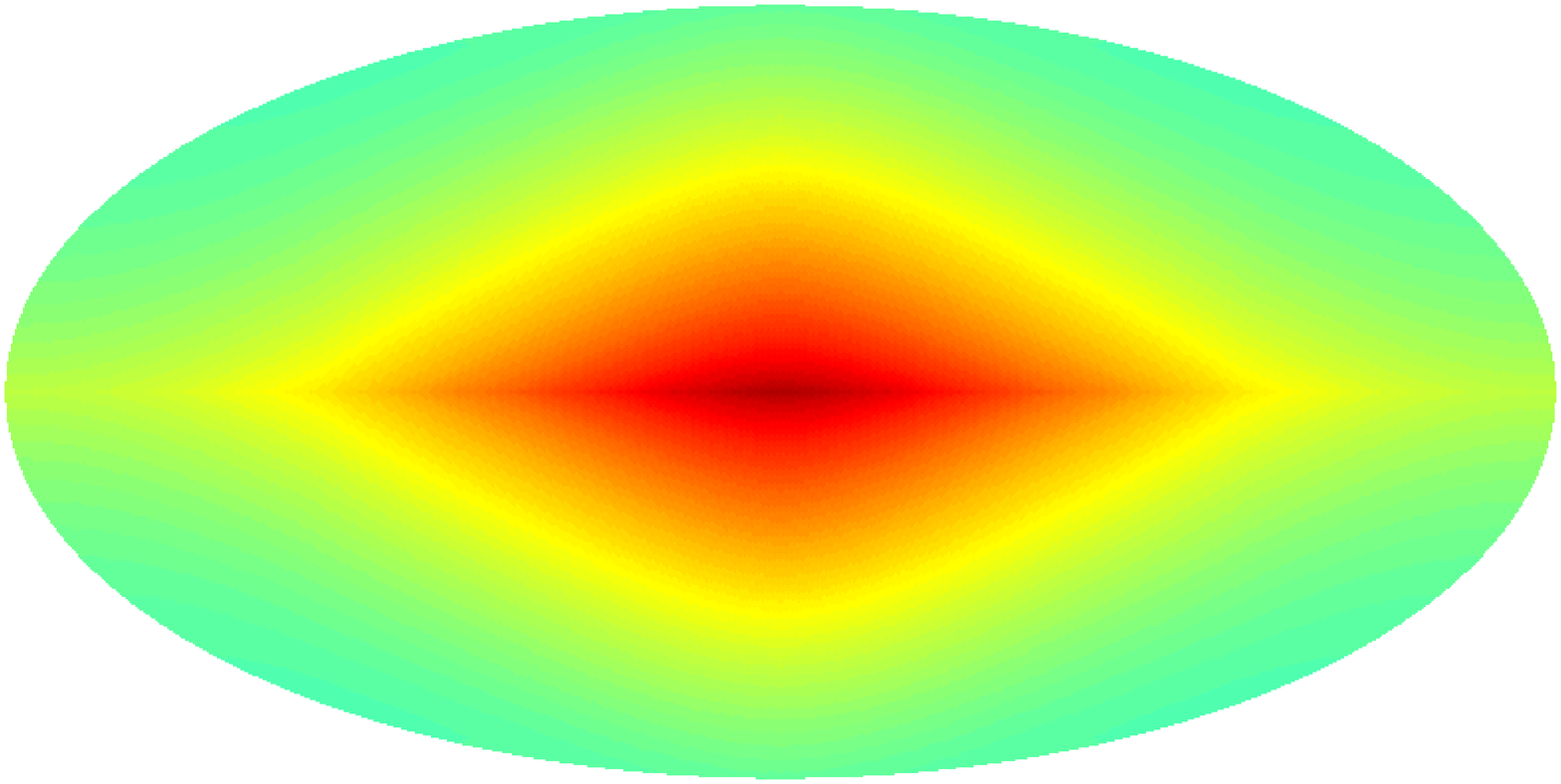}
\includegraphics[height=2.5 cm]{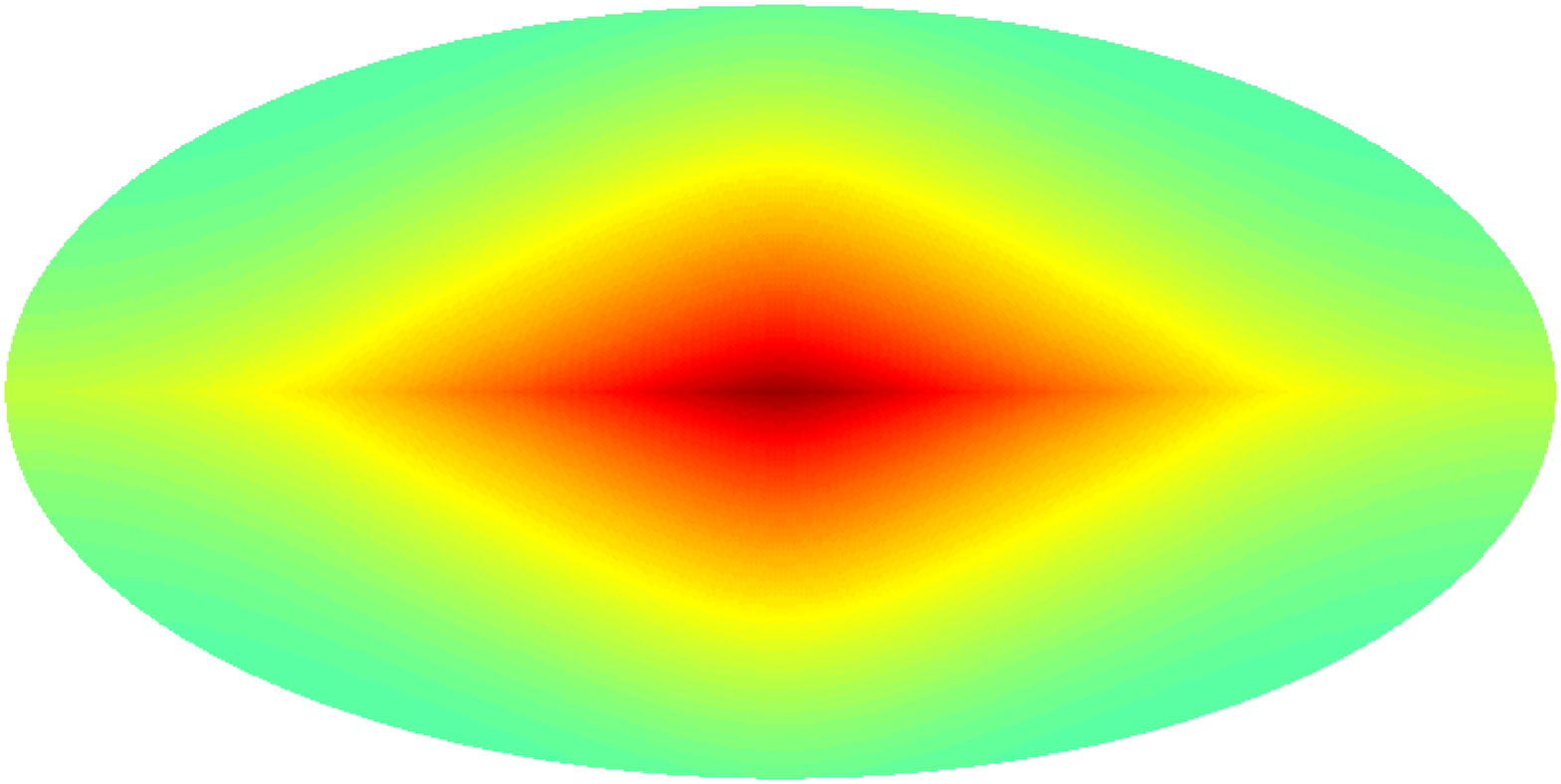}

\includegraphics[height=2.5 cm]{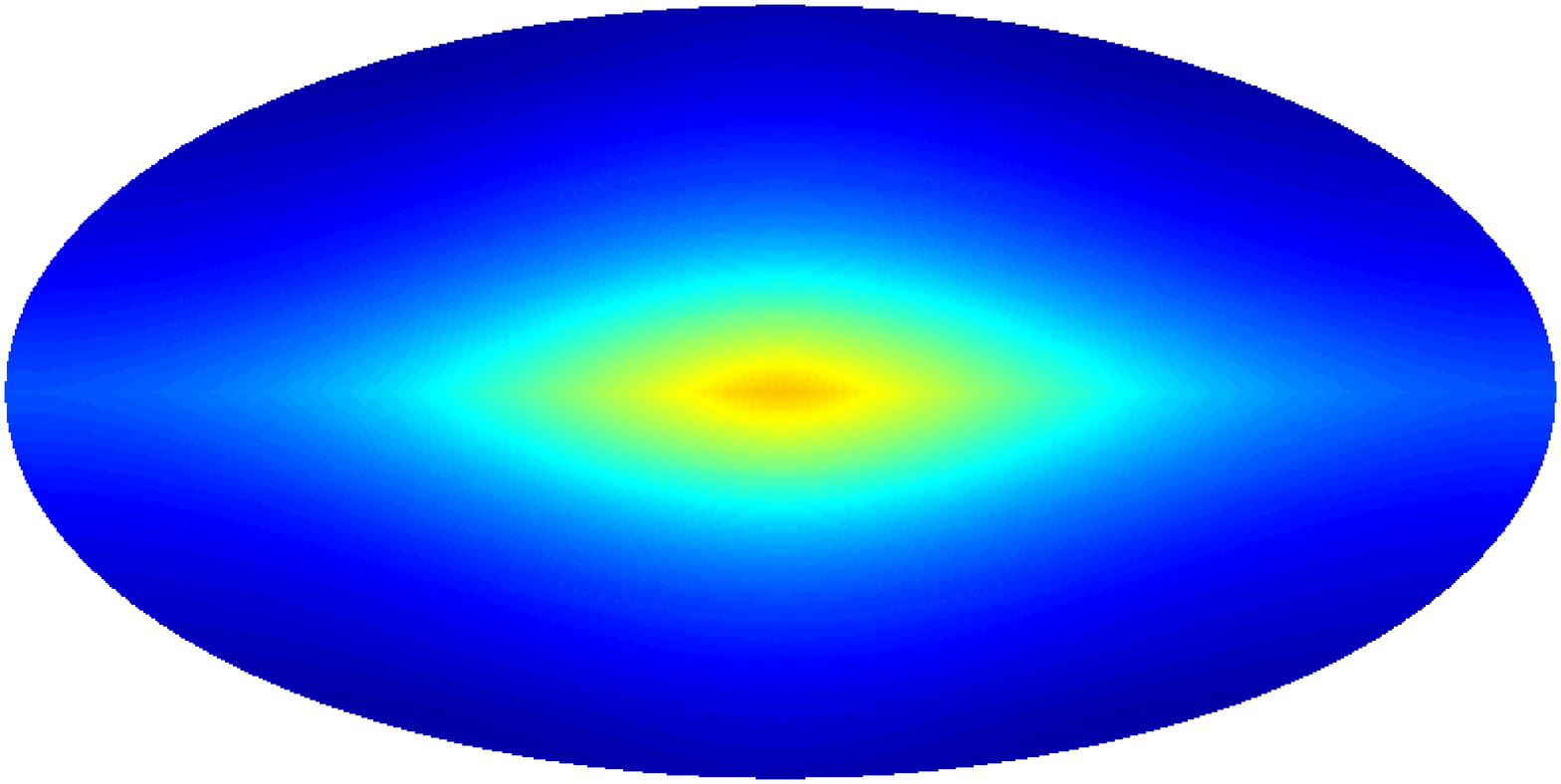}
\includegraphics[height=2.5 cm]{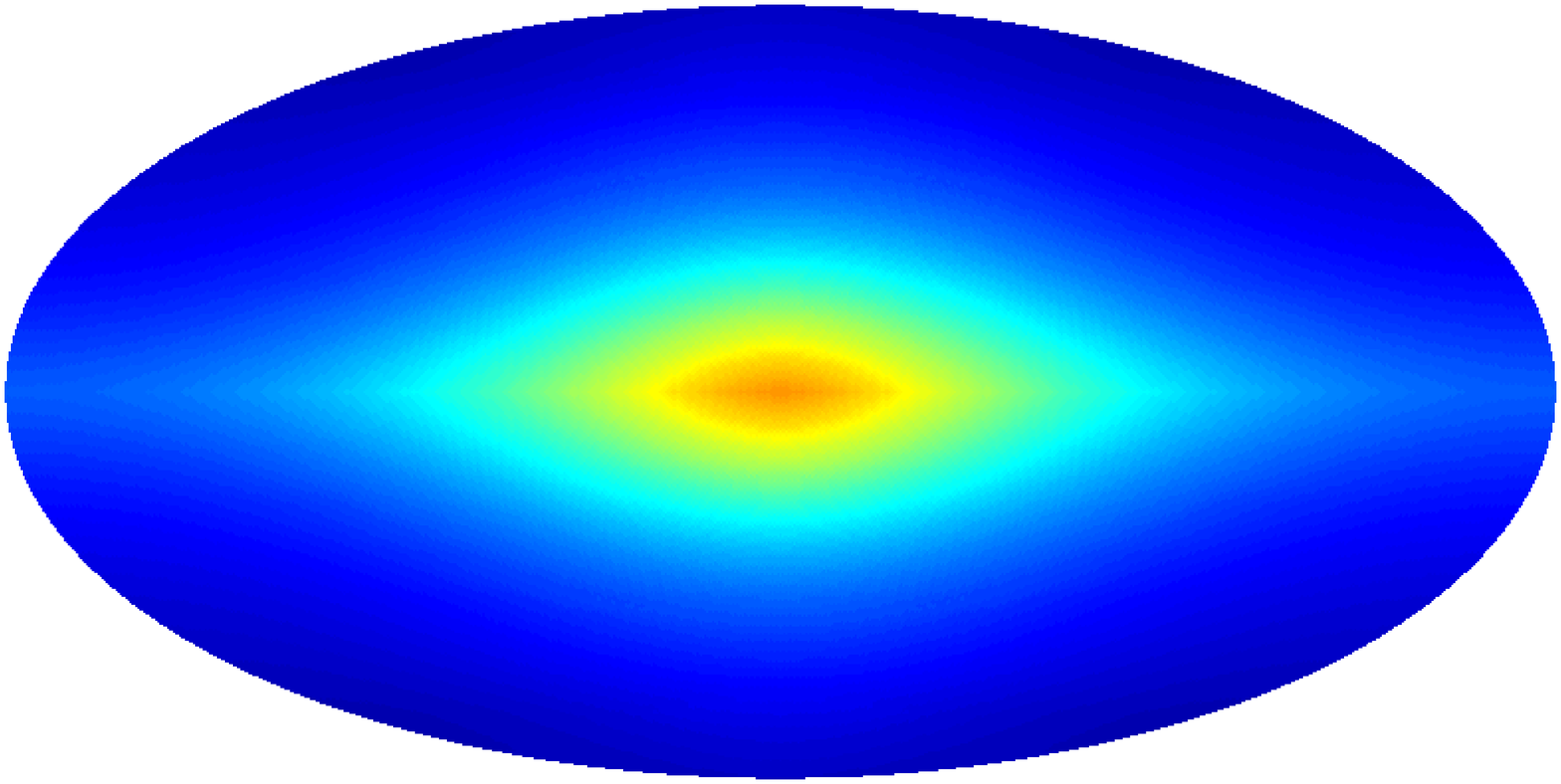}
\includegraphics[height=2.5 cm]{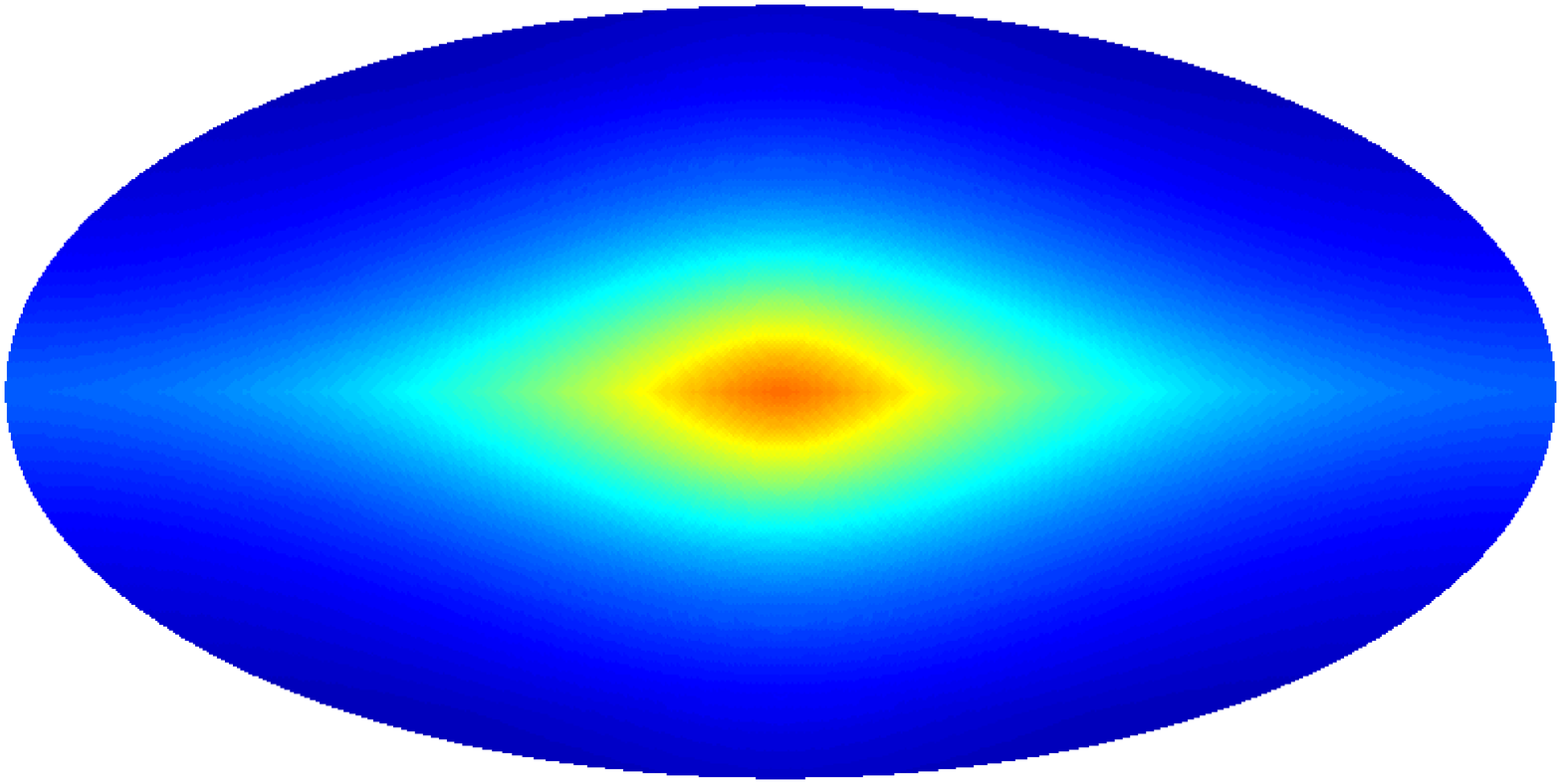}

\includegraphics[height=2.5 cm]{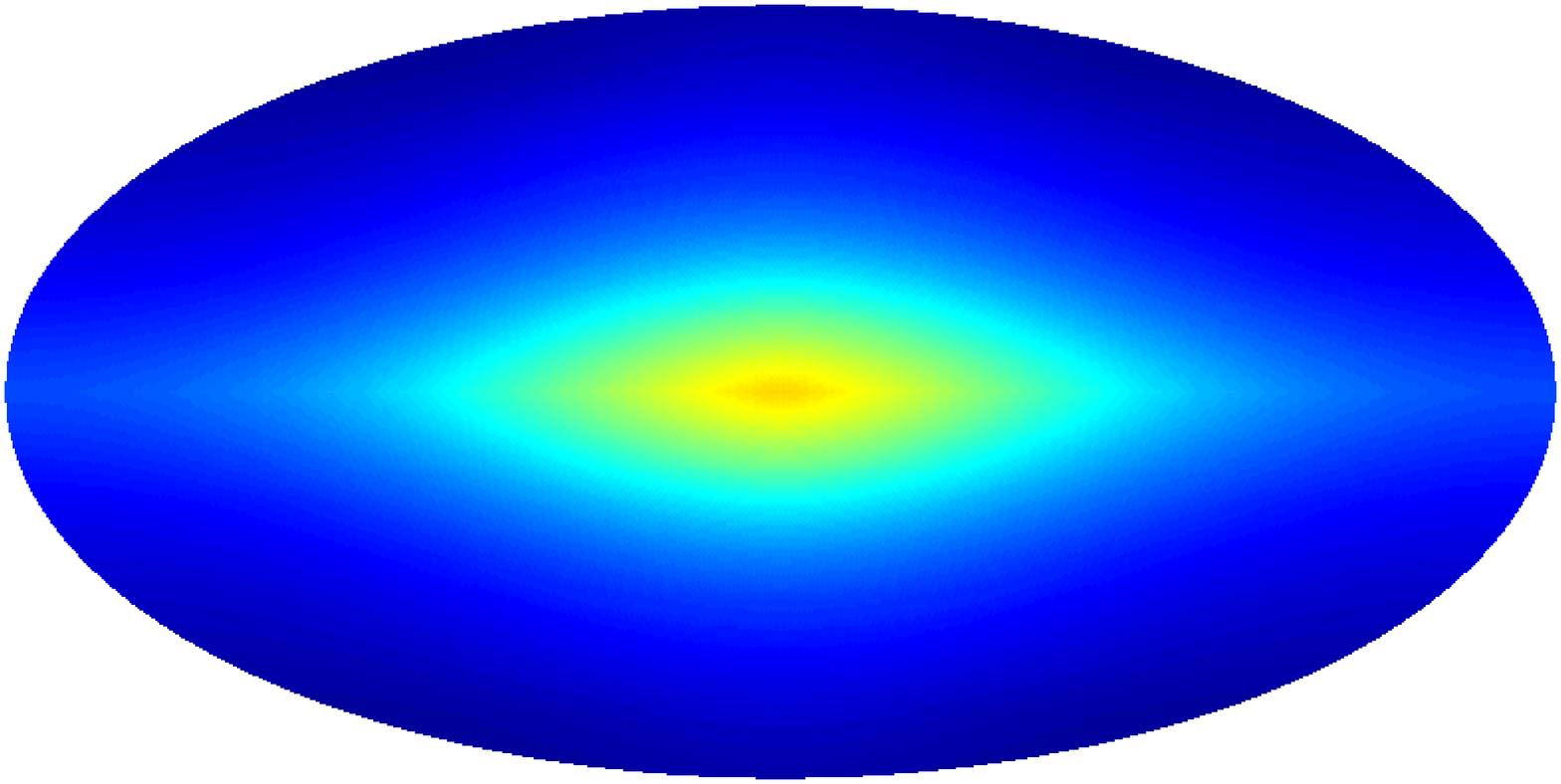}
\includegraphics[height=2.5 cm]{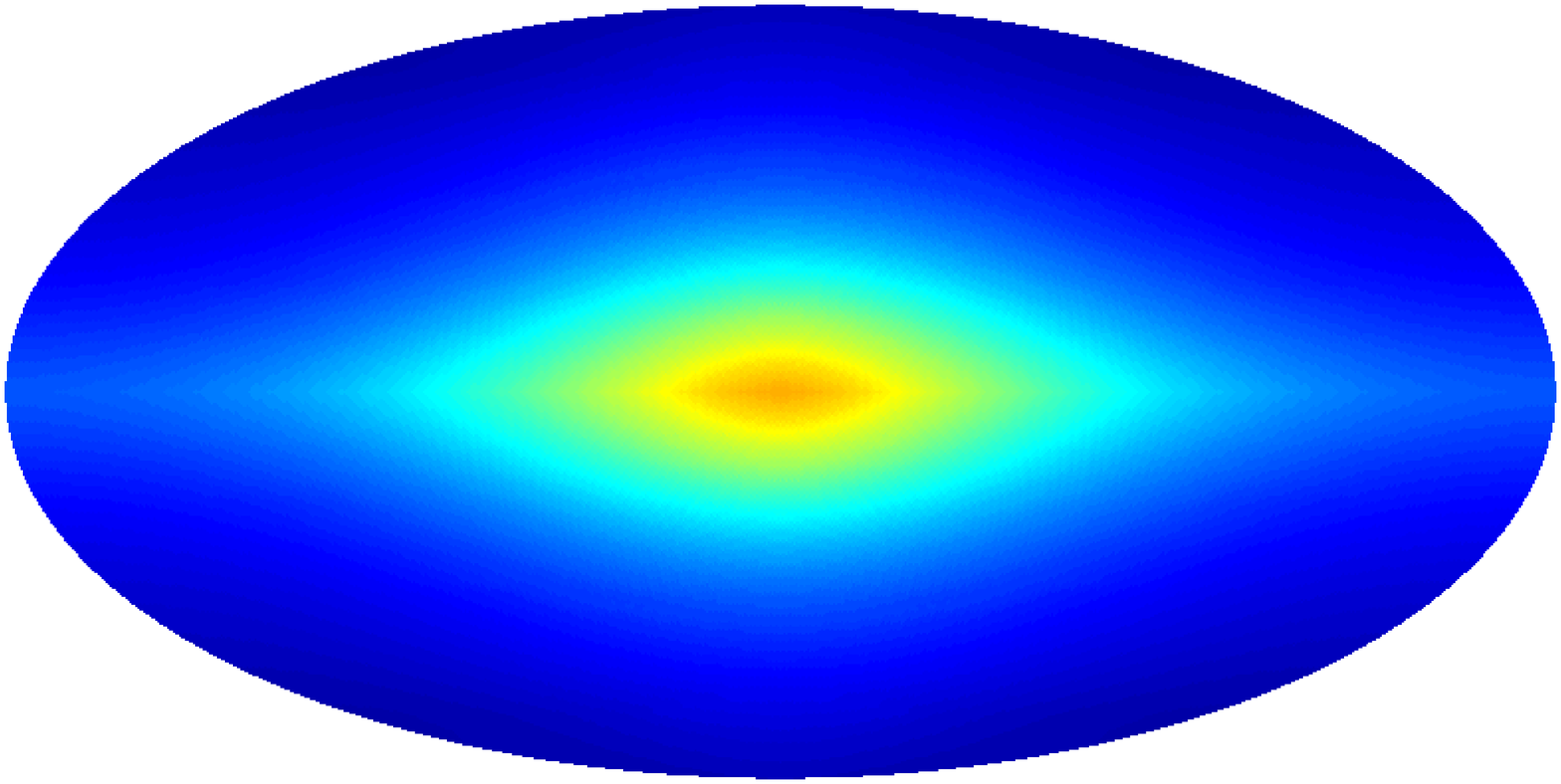}
\includegraphics[height=2.5 cm]{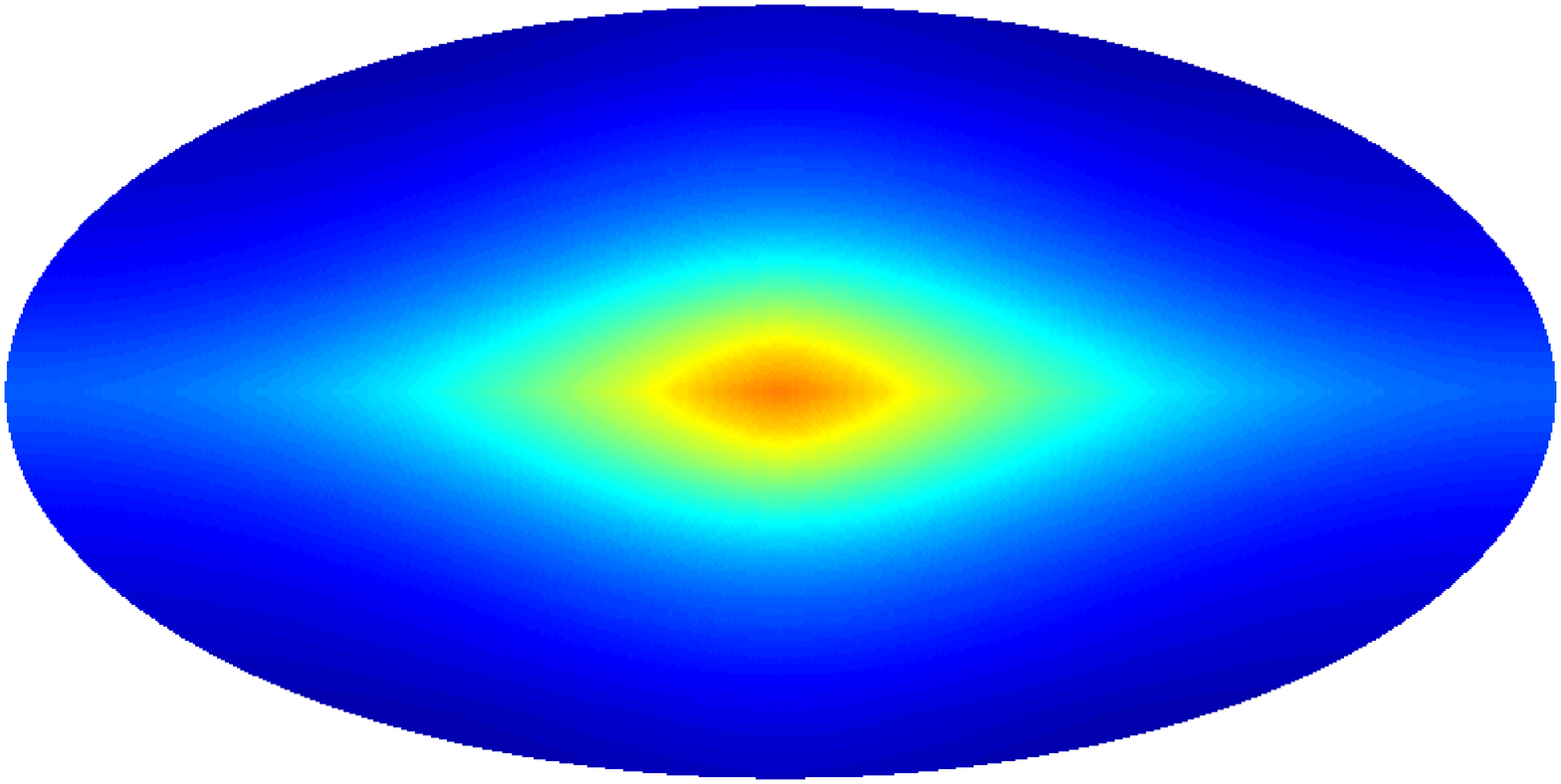}
\includegraphics[height=0.4 cm]{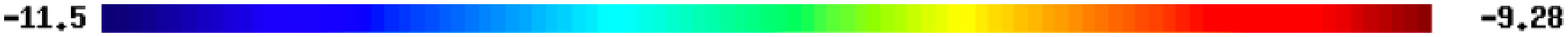}
\caption{Model dependence of radio signatures at 1.42 GHz induced by decays of dark matter particles
with $m_X=100\,$GeV, $\tau_X= 10^{26}\,$s, for an injection spectrum $dN_e/dE = \delta(E-m_X)$. Results
for the five different diffusion models of Tab.~\ref{tab:prop_model} (from top to bottom: MIN, MED, MAX, DC and DR) and for the three dark matter halo profiles of Tab.~\ref{tab:halo} (from left to right: Kramers, Isothermal and NFW) are shown. 
The color scaling corresponds to the logarithm to the base 10 of the flux in erg/s/$\rm{cm}^2$/sr. 
Note that the color scale corresponds to the same flux range in all panels for convenient comparison.}
\label{fig:1.42G_models}
\end{figure}

In Fig.~\ref{fig:1.42G_models} we show the radio emission from dark matter decay for the five propagation models of Tab.~\ref{tab:prop_model} and for the three halo density profiles of Tab.~\ref{tab:halo}. 
For the sake of illustration, we adopted a dark matter of 100 GeV, a lifetime of $10^{26}\,$s, and we use a decay spectrum $dN_e/dE = \delta(E-m_X)$ such that the total energy goes into one electron.

From our numerical calculation illustrated in Fig.~\ref{fig:1.42G_models}, it is clear that the largest uncertainty of synchrotron radiation comes from the propagation models. The average radio flux can differ by a factor of ten. For the MIN model, since the height of the diffusion zone is smallest, most of the radio emissions occurs at low latitudes. In other propagation models the radio emission is more extended because of the larger diffusion coefficient and the larger scale height of the diffusion zone which leads to more dark matter decays
contributing.

Compared to the diffusion models MIN, MED and MAX, the DC and DR models always produce smaller signals over the whole diffusion zone. This is mostly due to the larger diffusion coefficient which allows the electrons to escape
more easily from the diffusion zone corresponding to fewer confined electrons. 
Meanwhile, the power of re-acceleration, described by $D_{pp}$ in Eq.~(\ref{eq:transport}), is also weaker since 
it is inversely proportional to the diffusion constant ($D_{pp}\propto D^{-1}_{xx}$), see Eq.~(\ref{eq:Dpp}). 
This implies that re-acceleration plays an important role in propagation models.

We also study the variation of the synchrotron emission due to different halo profiles. In general, decaying
dark matter with the NFW profile produces the largest average diffuse radio signals due to the relatively steeper slope of the density distribution. For the other two profiles the emissions are comparable with each other. Since the decay rate is only proportional to the density, uncertainties from the halo profiles do not alter the
resulting dark matter constraints significantly. We do not take into account any possible
small-scale structure of dark matter halos since due to the linear scaling of injection rates with dark
matter density it has much smaller influence on fluxes than in annihilation scenarios.

\subsection{Response Functions} 
\label{sec:con-r}

\begin{figure}[htp]
\centering
\includegraphics[height=2.5cm]{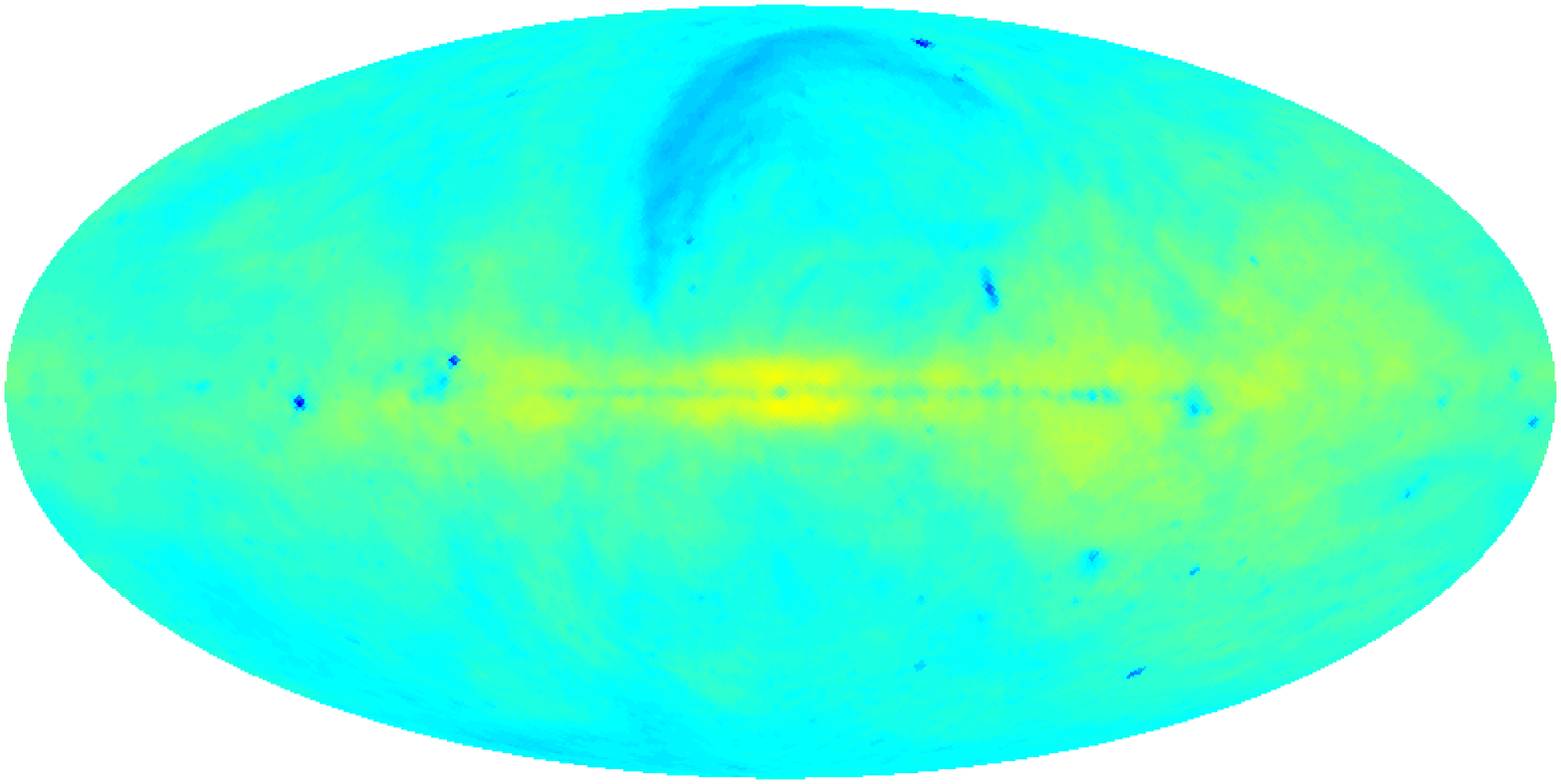}
\includegraphics[height=2.5cm]{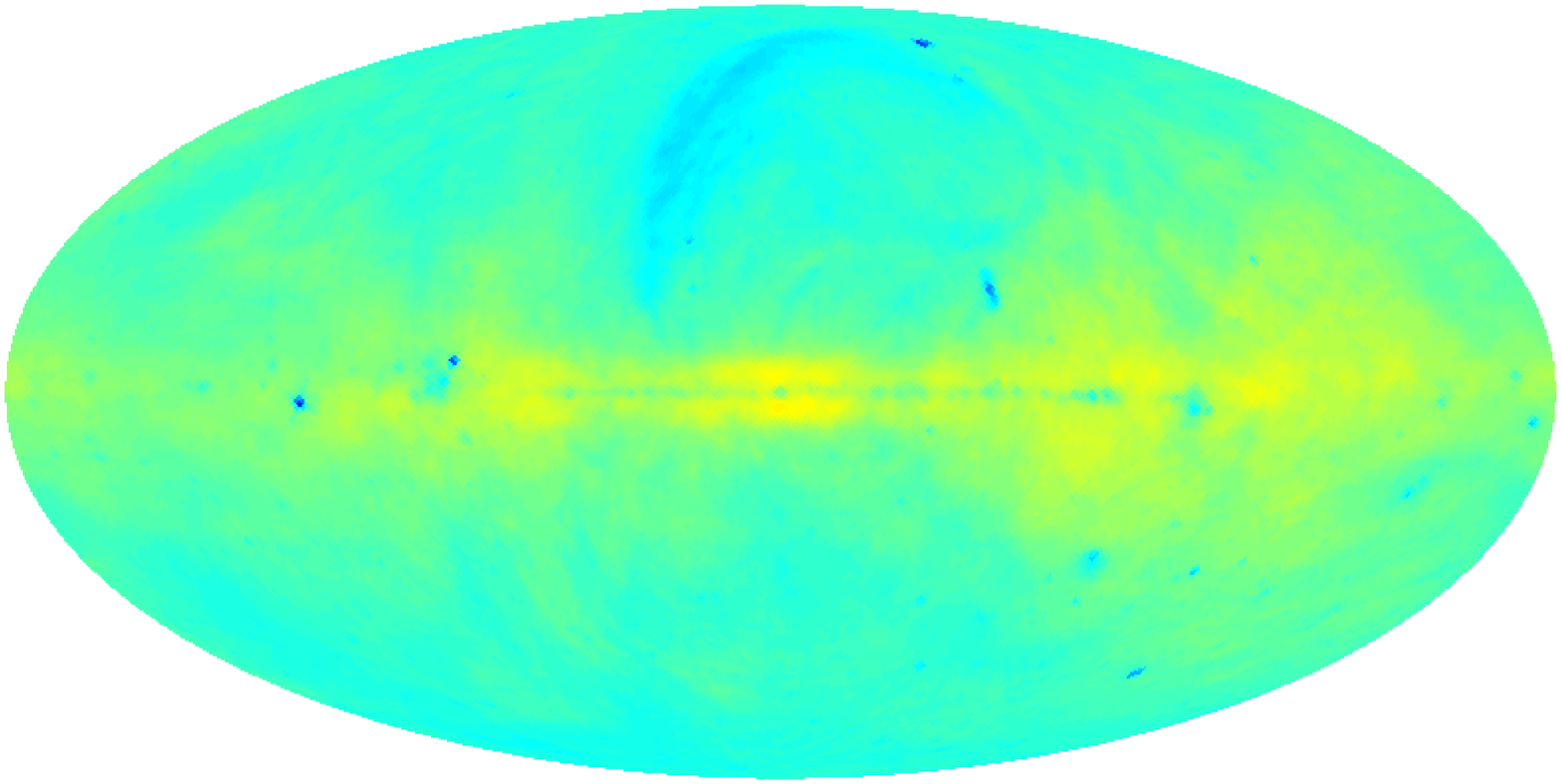}
\includegraphics[height=2.5cm]{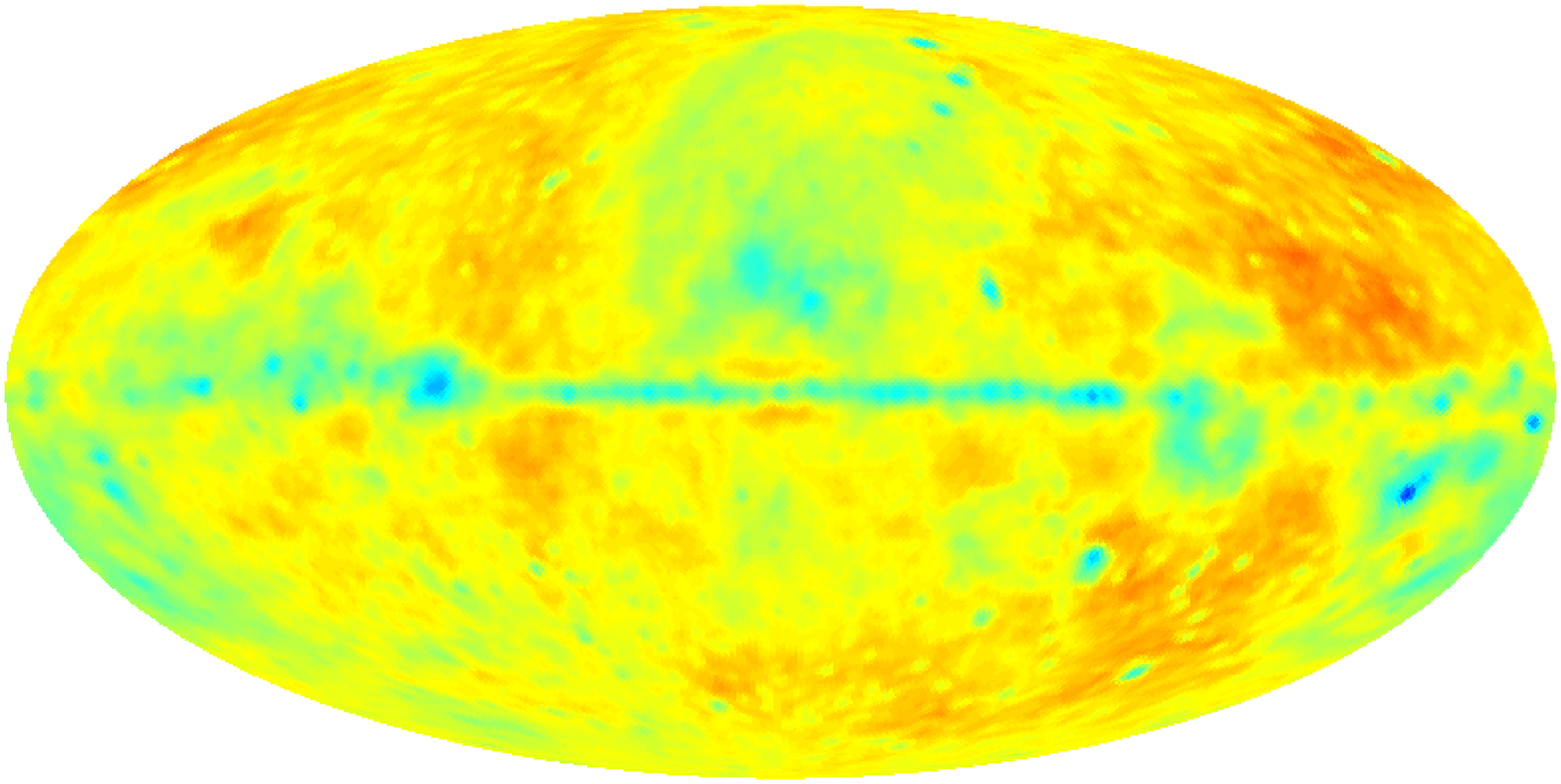}

\includegraphics[height=2.5 cm]{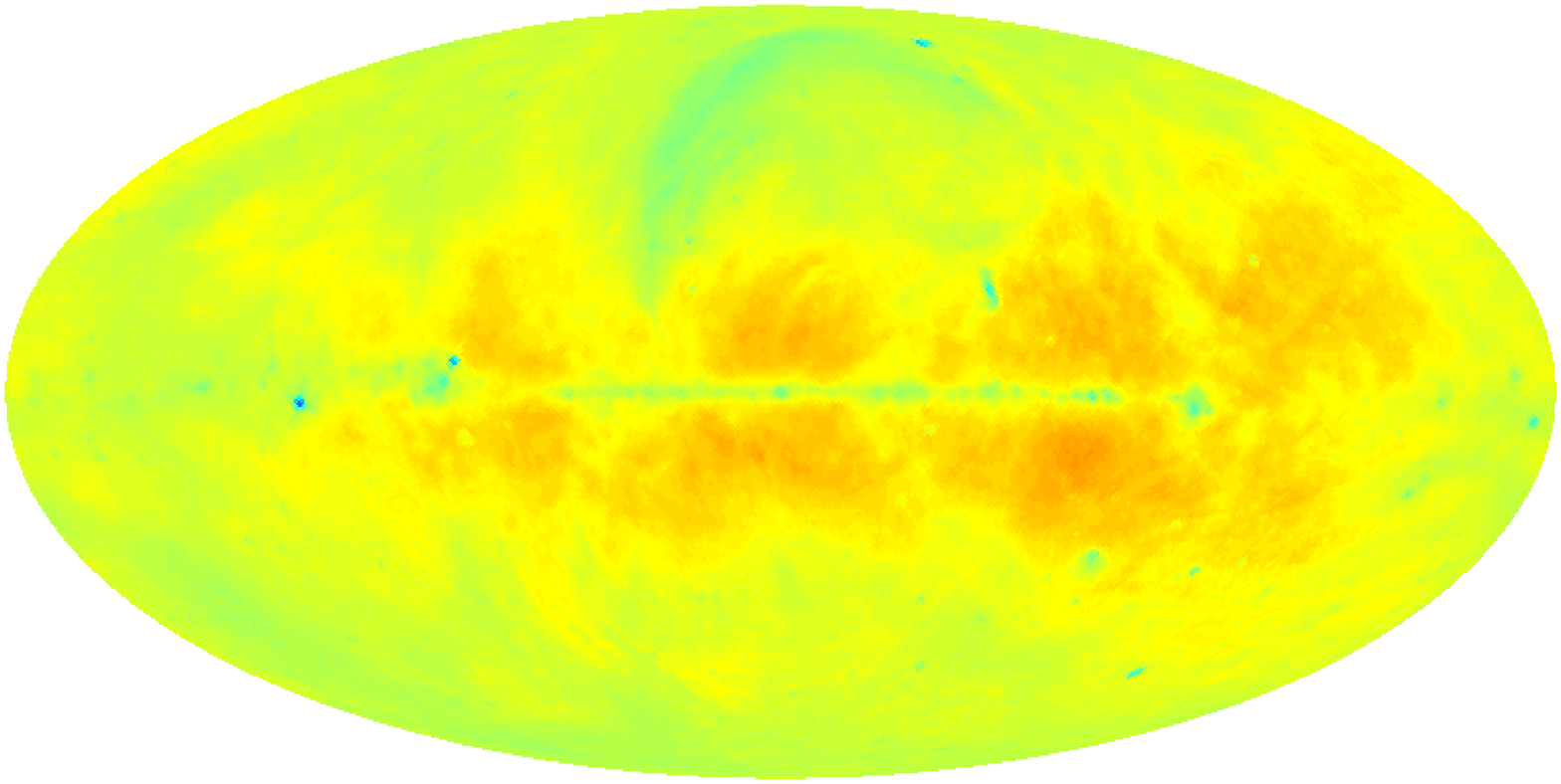}
\includegraphics[height=2.5 cm]{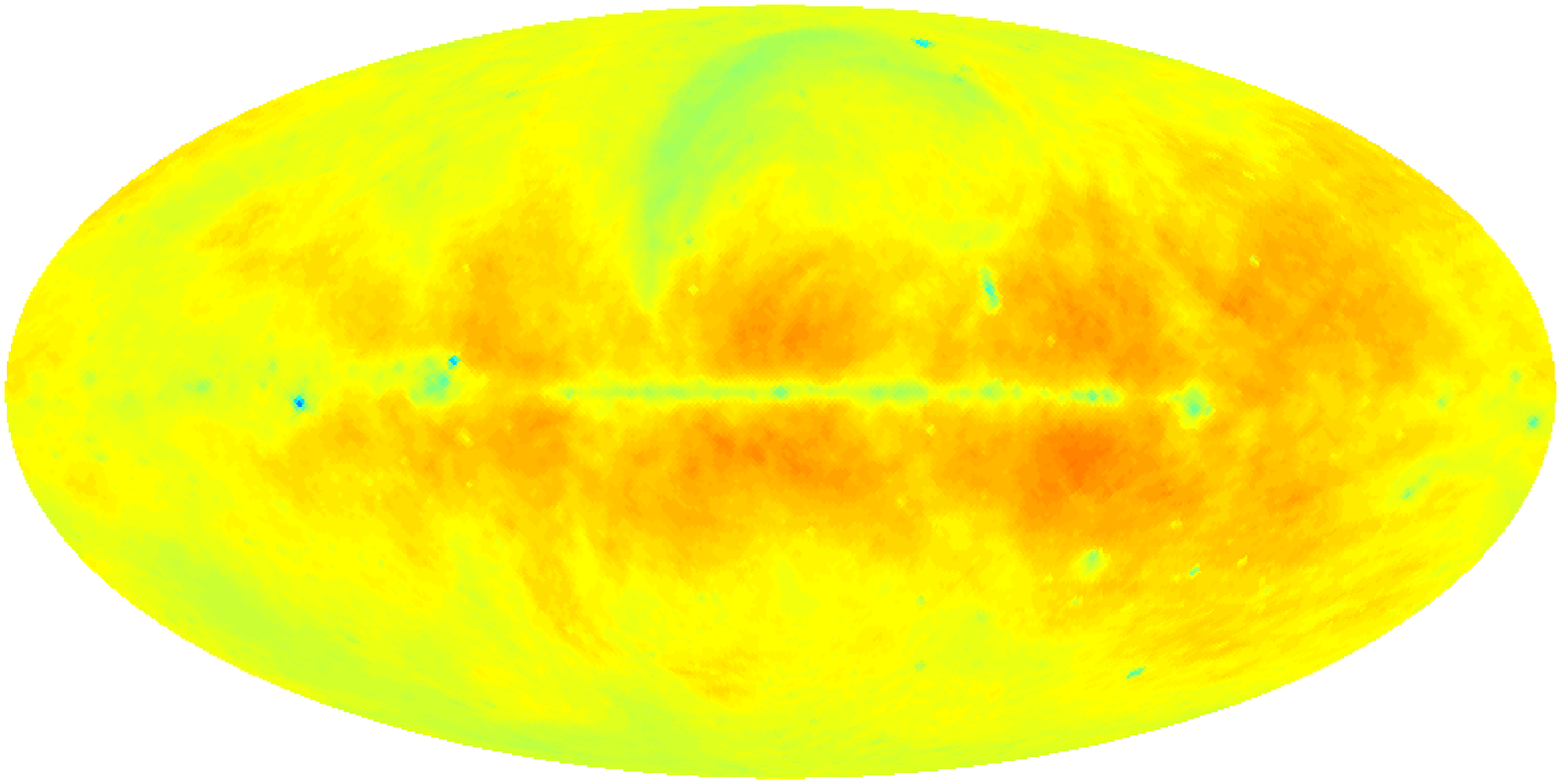}
\includegraphics[height=2.5 cm]{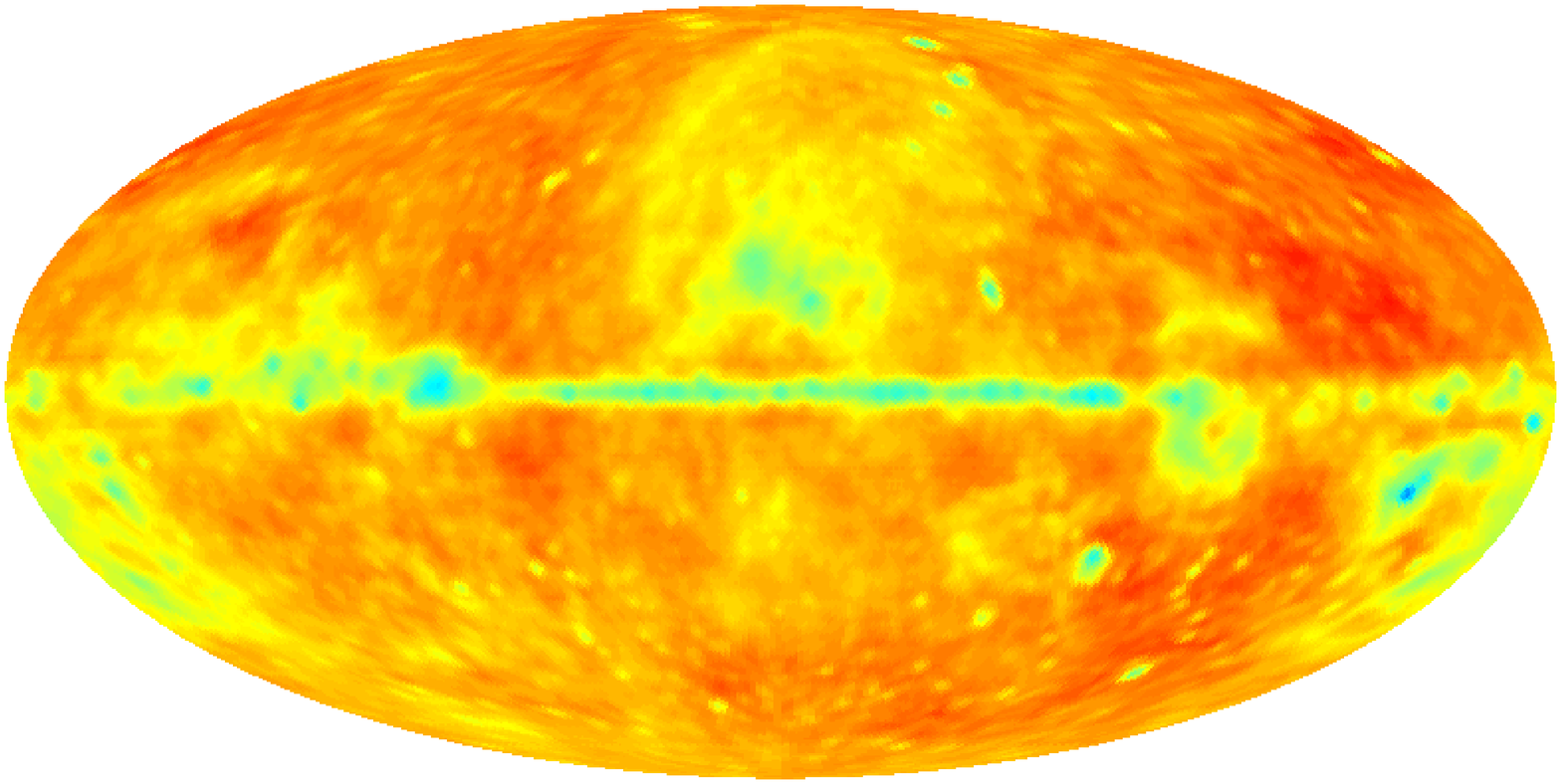}

\includegraphics[height=2.5 cm]{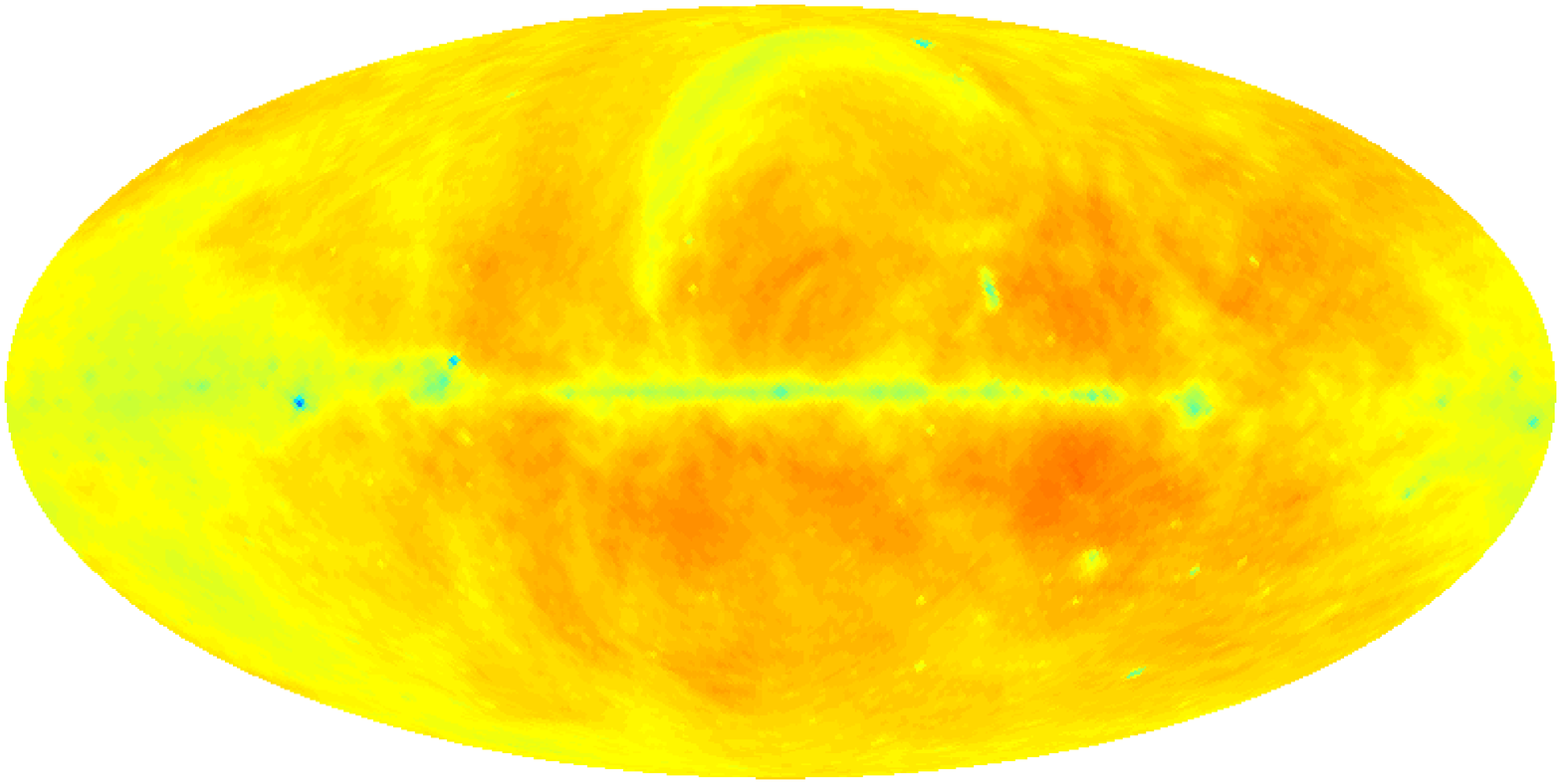}
\includegraphics[height=2.5 cm]{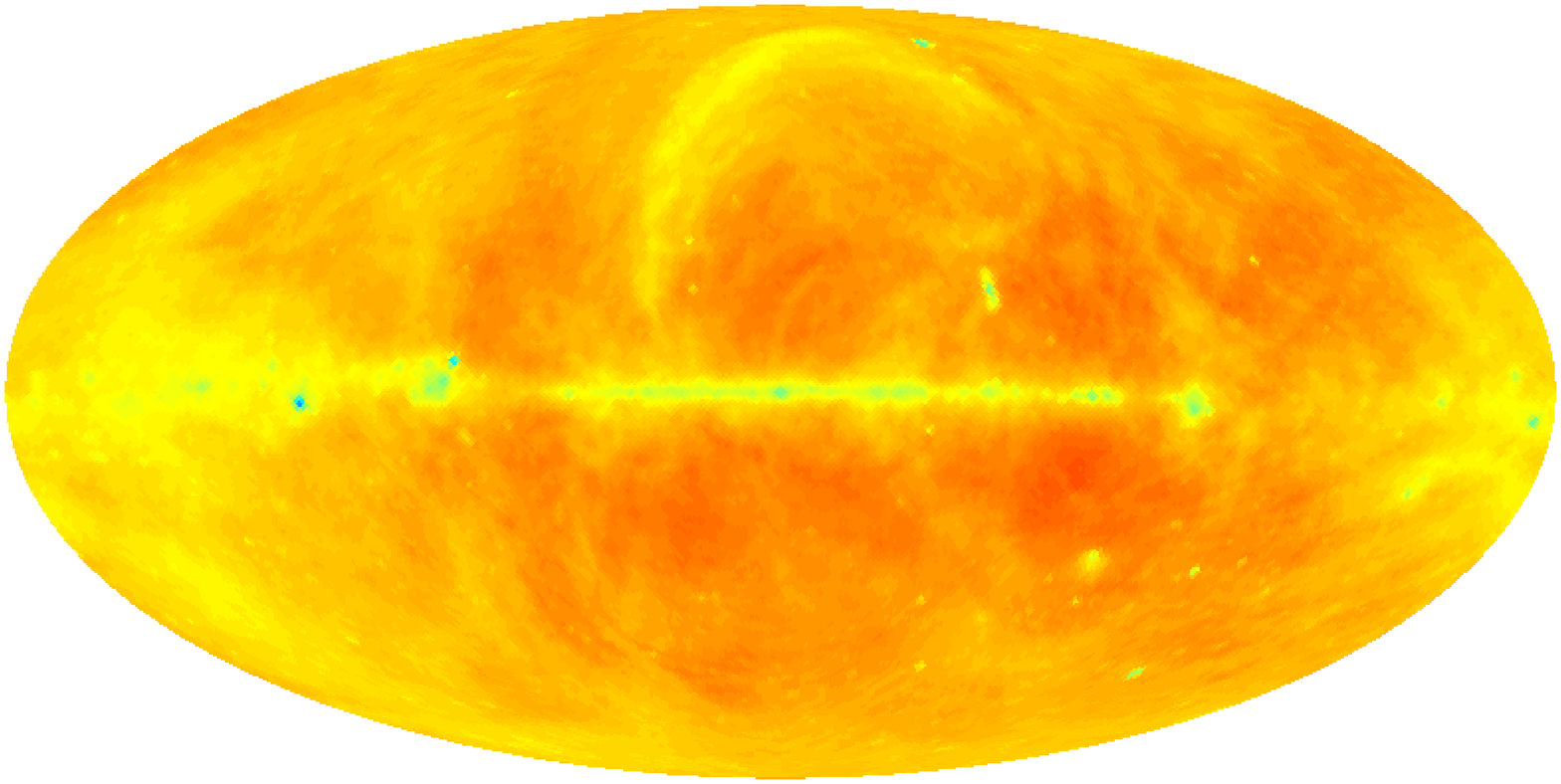}
\includegraphics[height=2.5 cm]{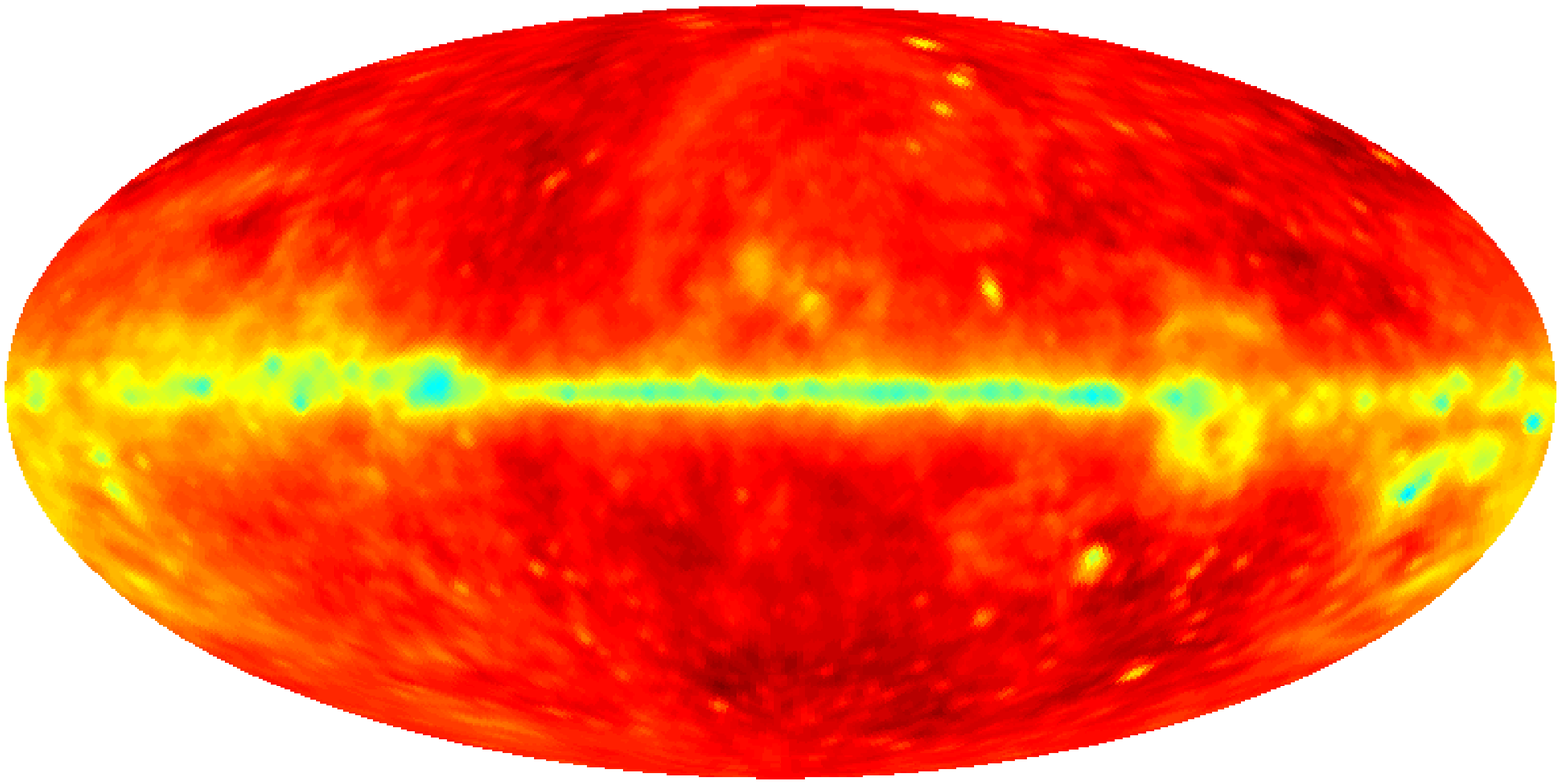}

\includegraphics[height=2.5 cm]{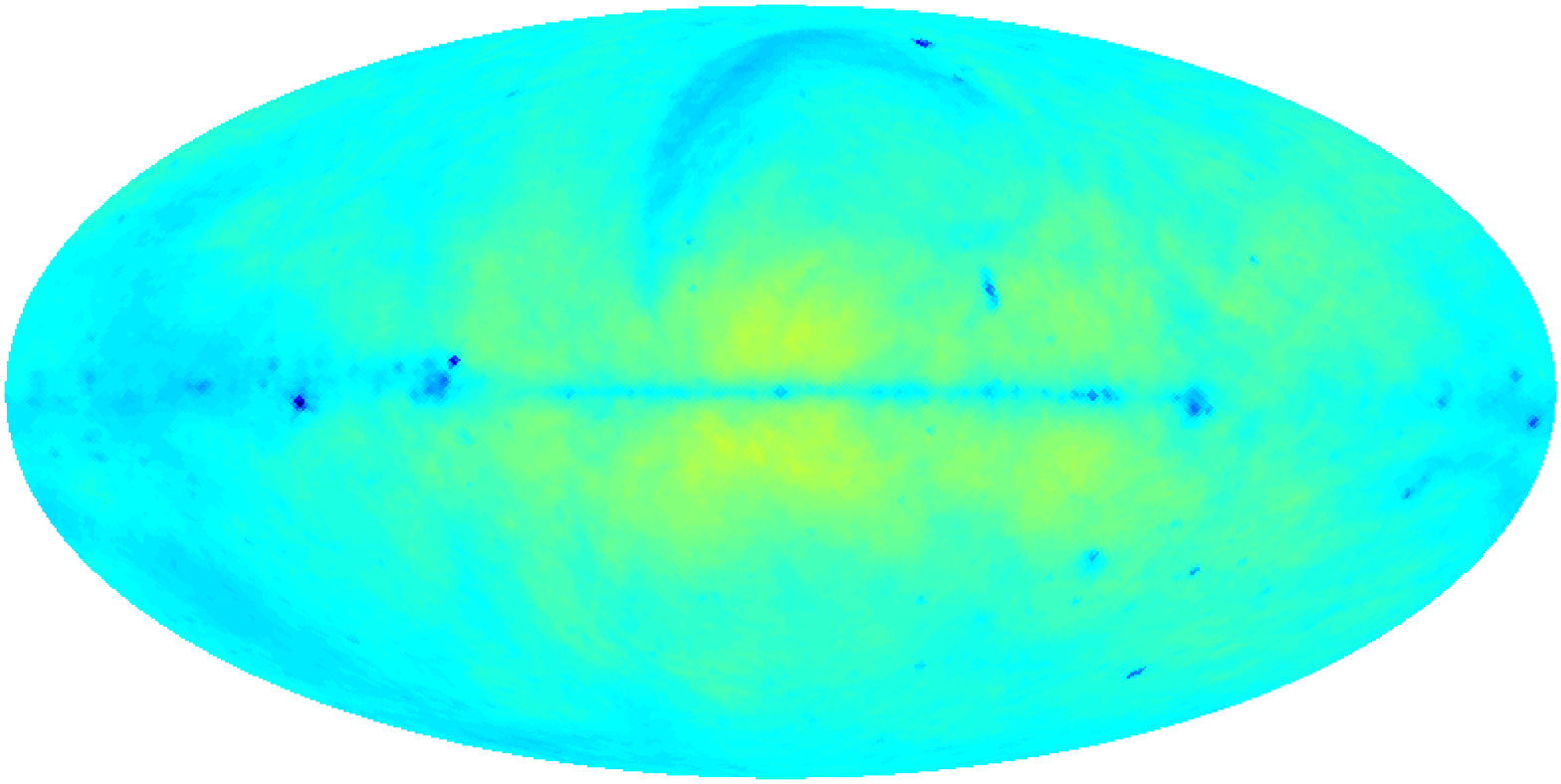}
\includegraphics[height=2.5 cm]{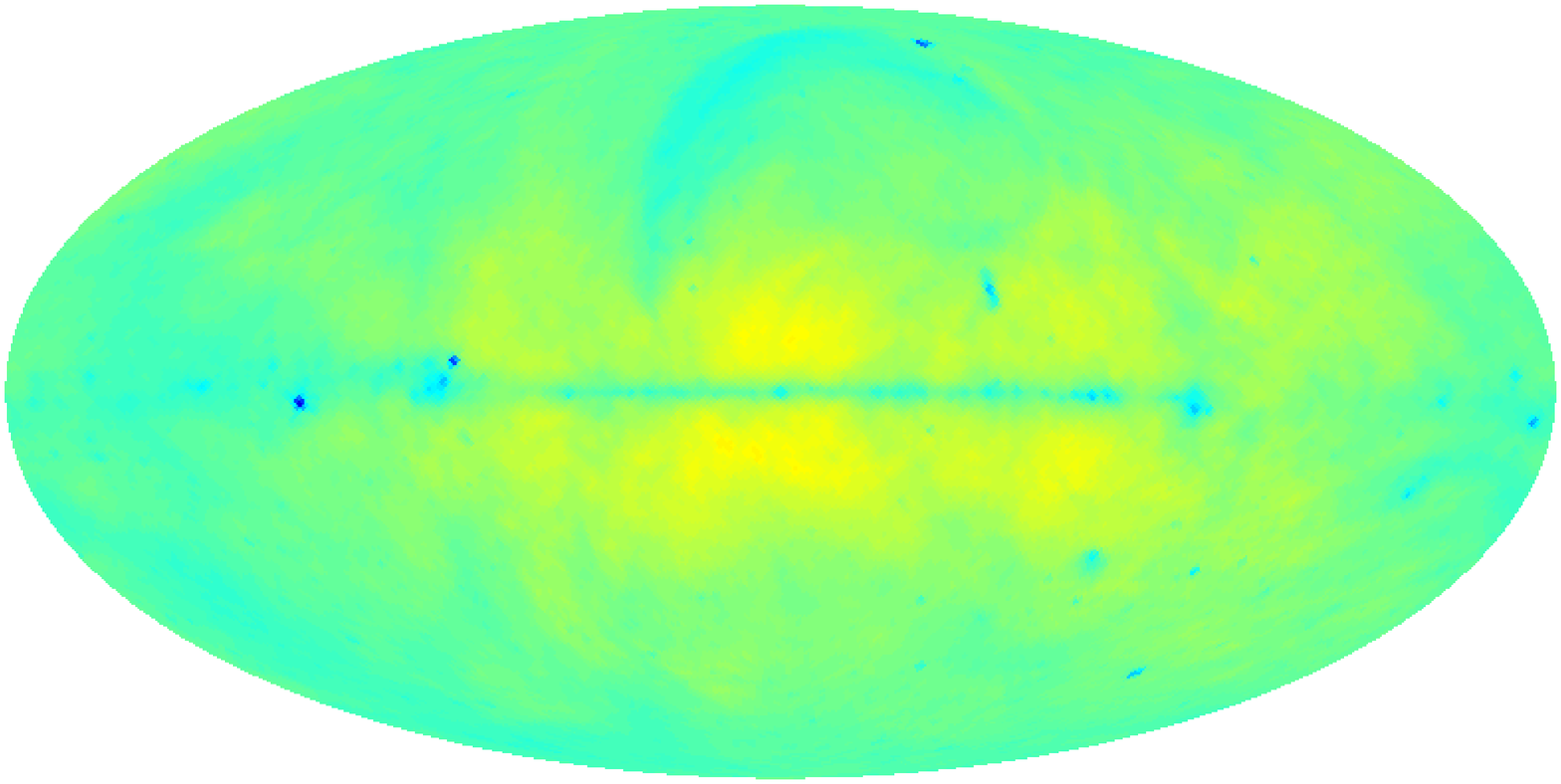}
\includegraphics[height=2.5 cm]{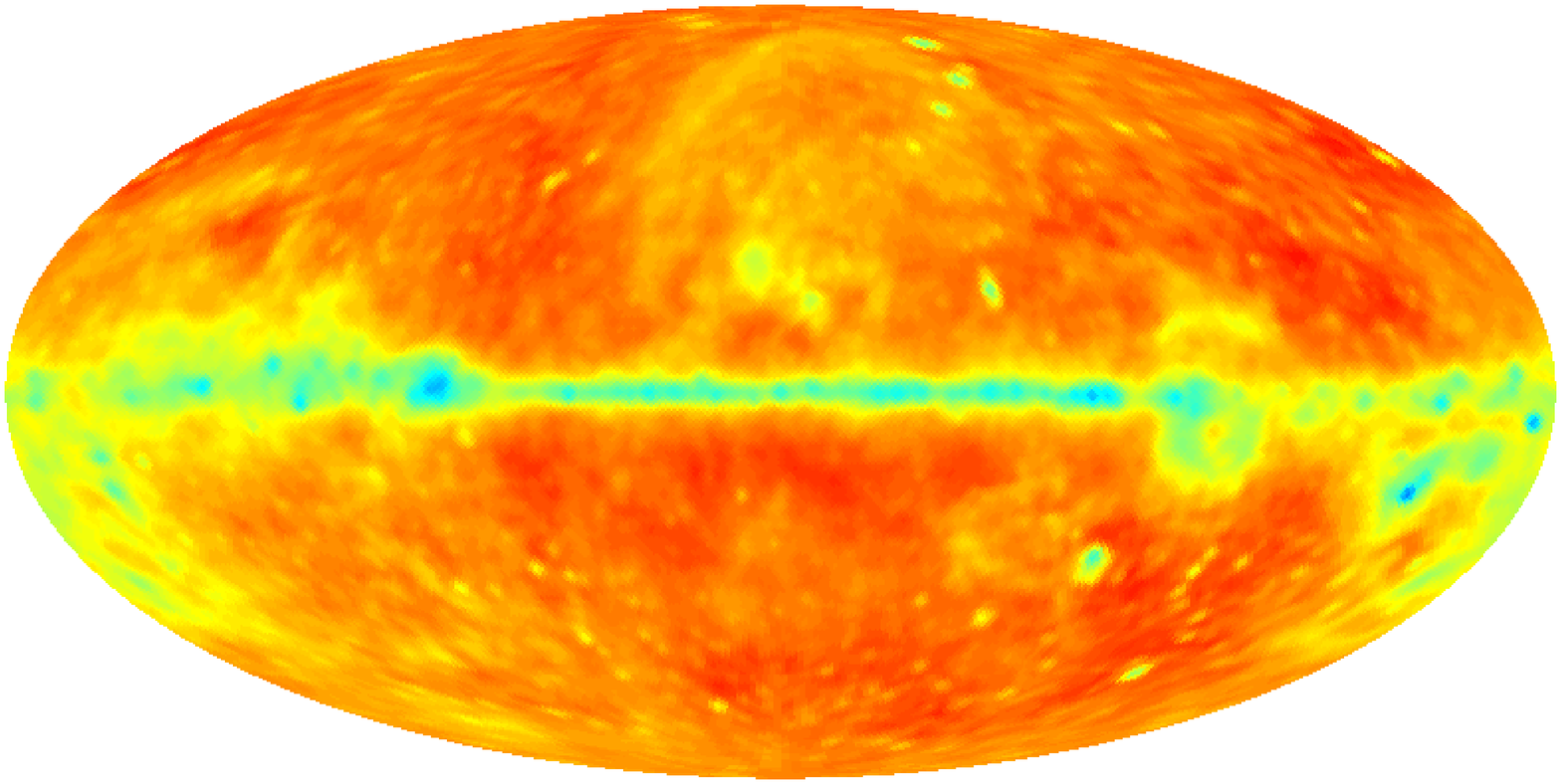}

\includegraphics[height=2.5 cm]{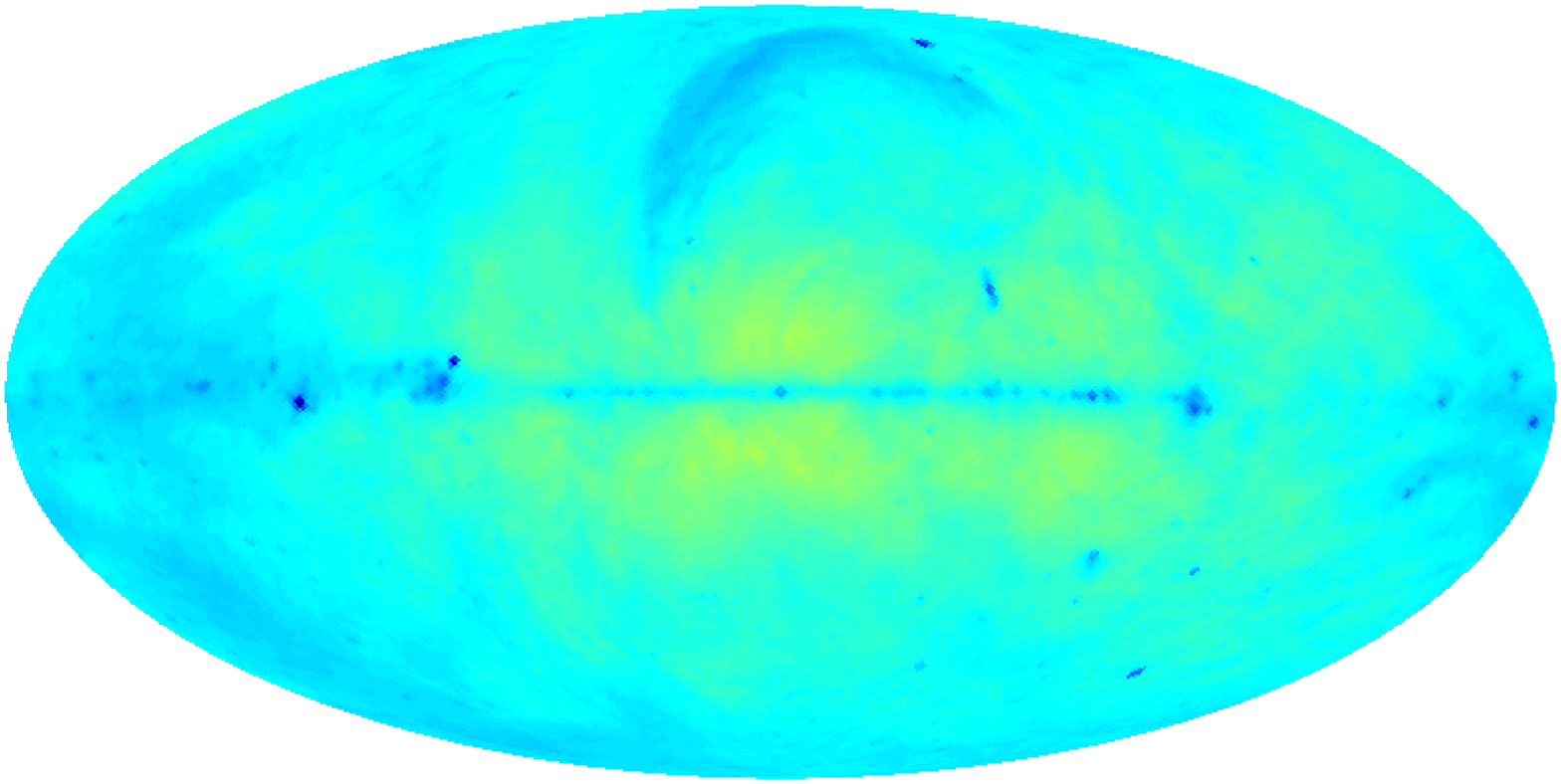}
\includegraphics[height=2.5 cm]{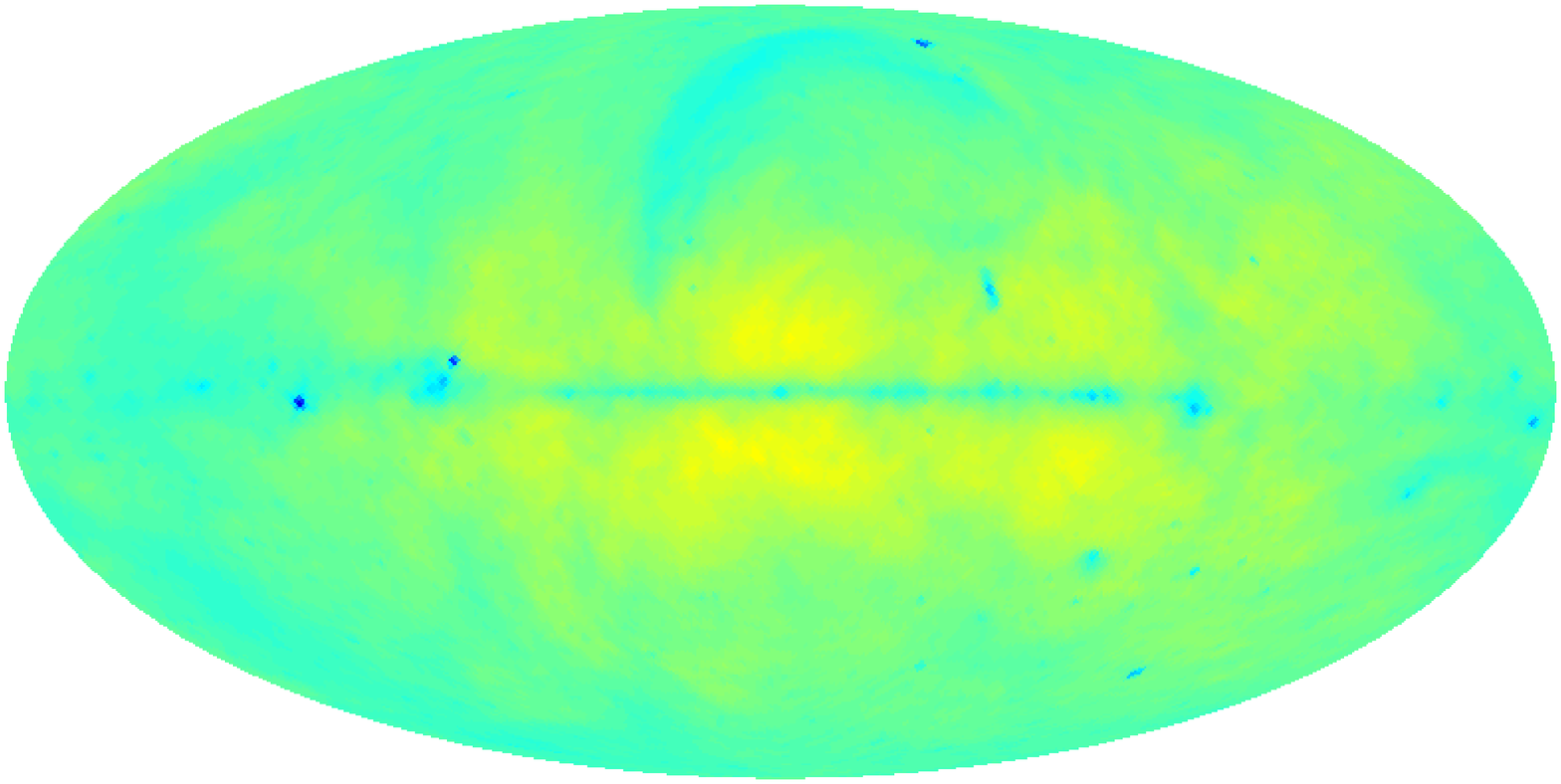}
\includegraphics[height=2.5 cm]{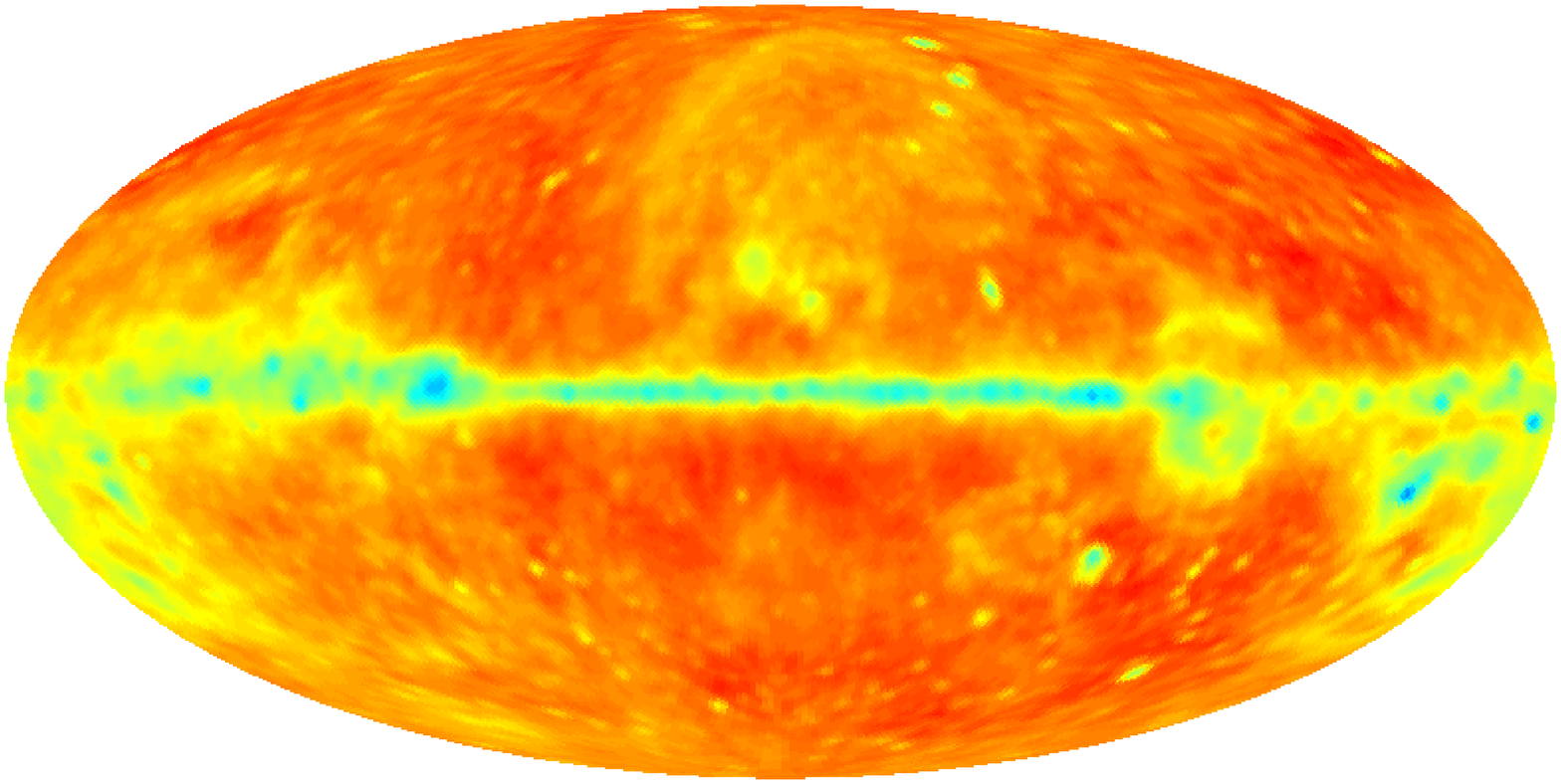}
\includegraphics[height=0.4 cm]{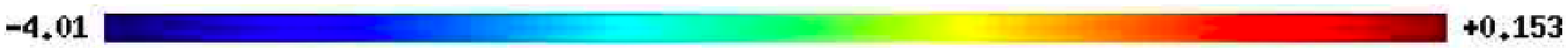}
\vspace{.5cm}
\caption{Excess maps, i.e. contours of predicted to observed radio flux, for decaying dark matter with $m_X=100\,$GeV, $\tau_X=10^{26}\,$s, and an injection spectrum $dN_e/dE = \delta(E-m_X)$. Results for the five different diffusion models (from top to bottom) of Tab.~\ref{tab:prop_model} and for three survey maps at 408 MHz, 1.42 GHz and 23 GHz (from left to right) are shown. Note the logarithmic color scaling for the excess, where
warmer color indicates larger excesses.}
\vspace{.5cm}
\label{fig:excess}
\end{figure}

Assuming a propagation model and dark matter profile, the radio emission produced by dark matter decay can be obtained for any given decay spectrum. 
Then an excess map can be calculated in comparison with observed radio maps, defined as the map of the ratio of predicted to observed radio flux in a given direction.
One can scan the whole excess map pixel by pixel until the largest excess is obtained. 
This pixel, therefore, corresponds to the optimal direction for observation.

Figure~\ref{fig:excess} shows several examples of these excess maps at the frequencies 408 MHz, 1.42 GHz, and 23 GHz, respectively. 
For the sake of illustration, we assumed dark matter particles with $m_X=100$ GeV with an NFW halo profile
and decaying into one monochromatic electron or positron. 
The most important feature in Fig.~\ref{fig:excess} is that the best directions for dark matter constraints do not point towards the Galactic center region. 
Although the dark matter signal close to the center is always larger than elsewhere, the observed background flux overcompensates it. Moreover, the optimal direction is not the anti-center where backgrounds are the smallest. The optimal directions tend to be not far from the center, and most of them concentrate in the southern hemisphere as many complex components such as giant molecular clouds and the north polar spur inhabit the northern hemisphere. The location of the warmest color which indicates the largest excess not only depends on which propagation model and halo profile is assumed, but also
depends on frequency. 

For constructing the response function, we have to perform different simulations with mono-energetic energy spectra at different injected energies $E_0$. For each of the resulting excess maps the optimal direction for observation is slightly different.
We don't want to provide more than one response function per observed frequency so we have to fix one particular direction.
In order to do so, we add up all the excess maps for different energies and search for the optimal direction.
The selected direction is then optimized for a perfectly flat spectrum whereas it may be a bad choice for
strongly peaked  or hard spectra. 
Fortunately, our calculations show that different selections for the optimal directions do not change the response function dramatically, at most by a factor of two in the worst scenarios.

\begin{figure}[tp]
\centering
\includegraphics[height=5.6 cm]{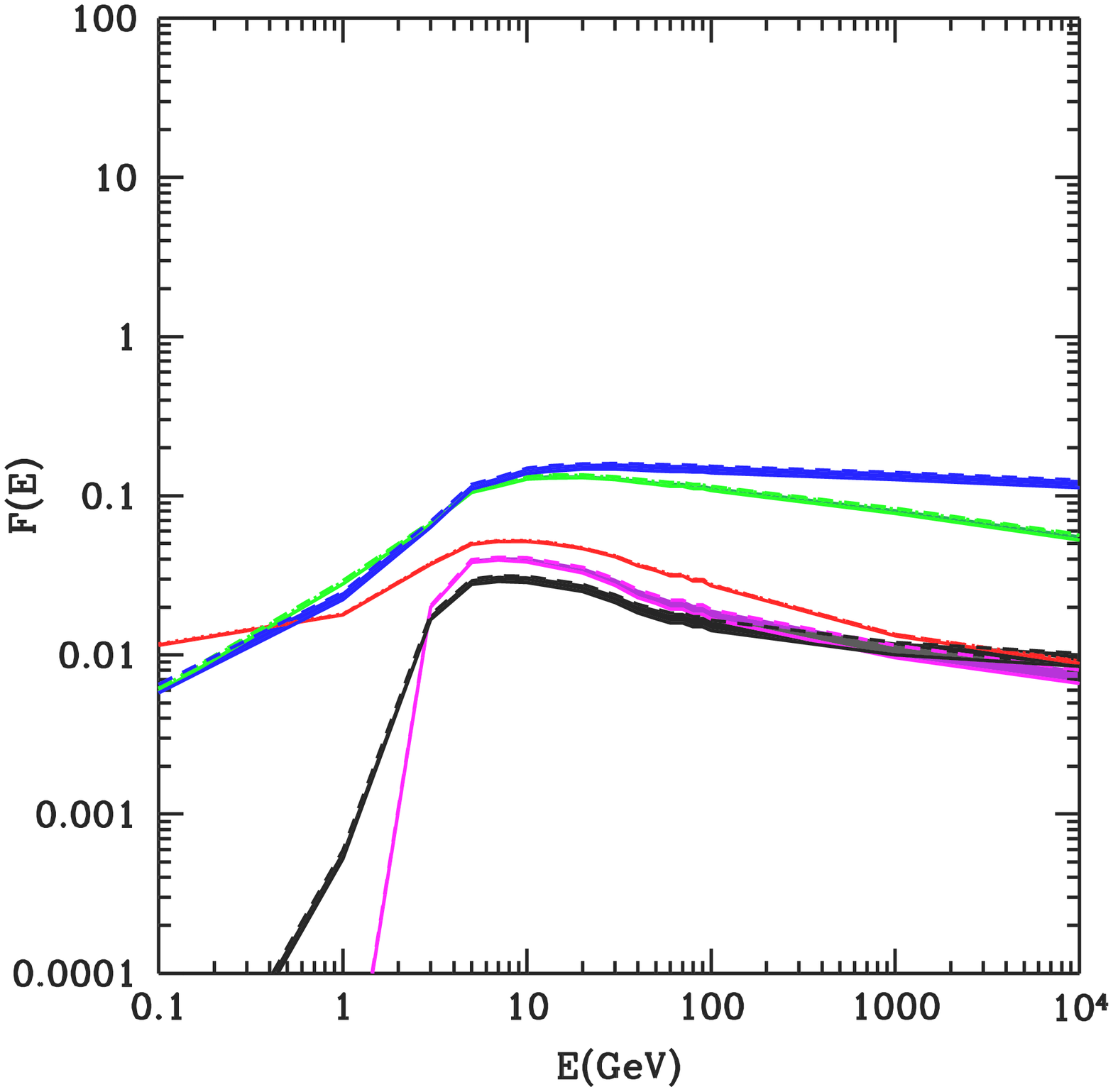}
\includegraphics[height=5.6 cm]{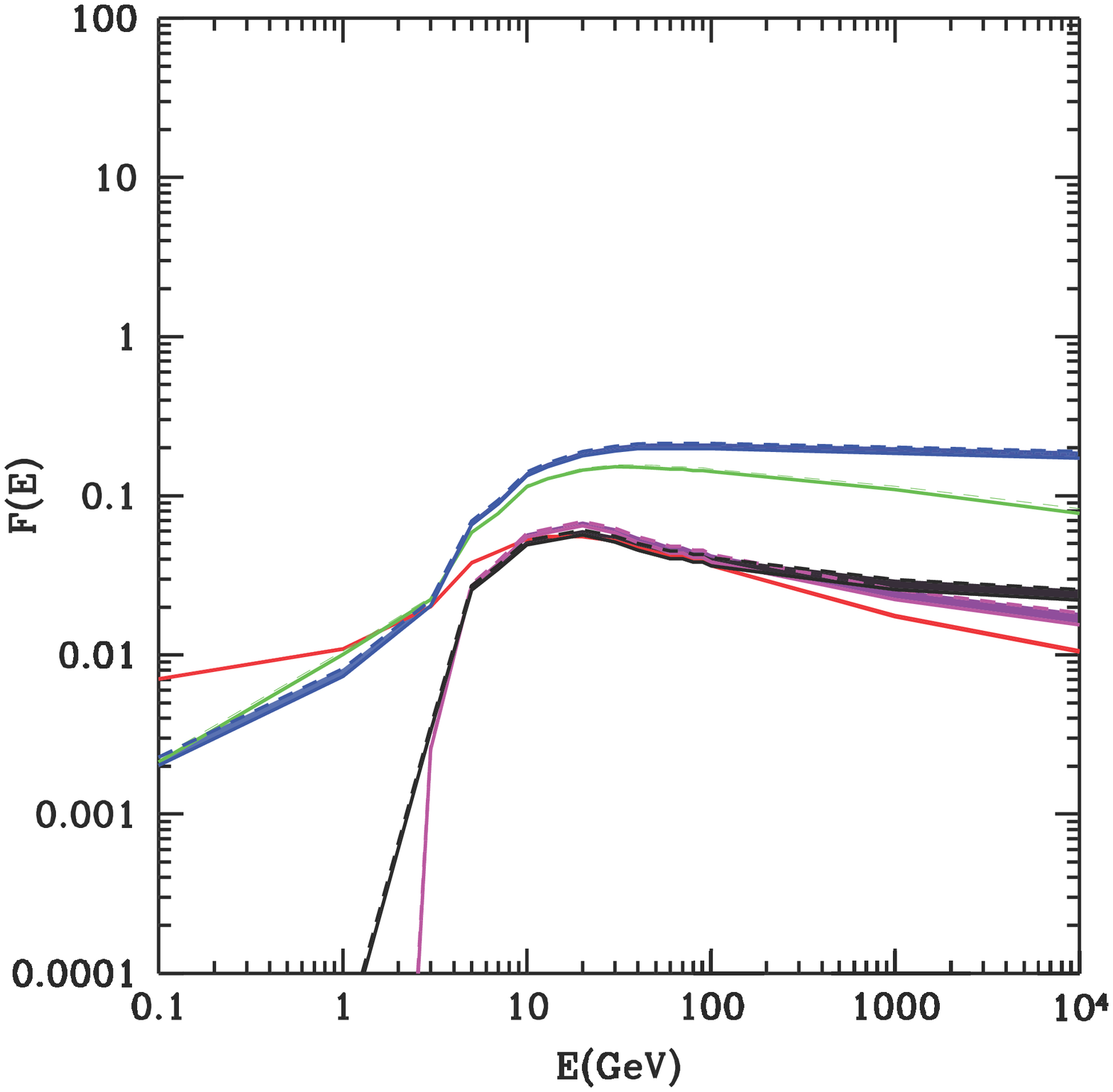}
\includegraphics[height=5.6 cm]{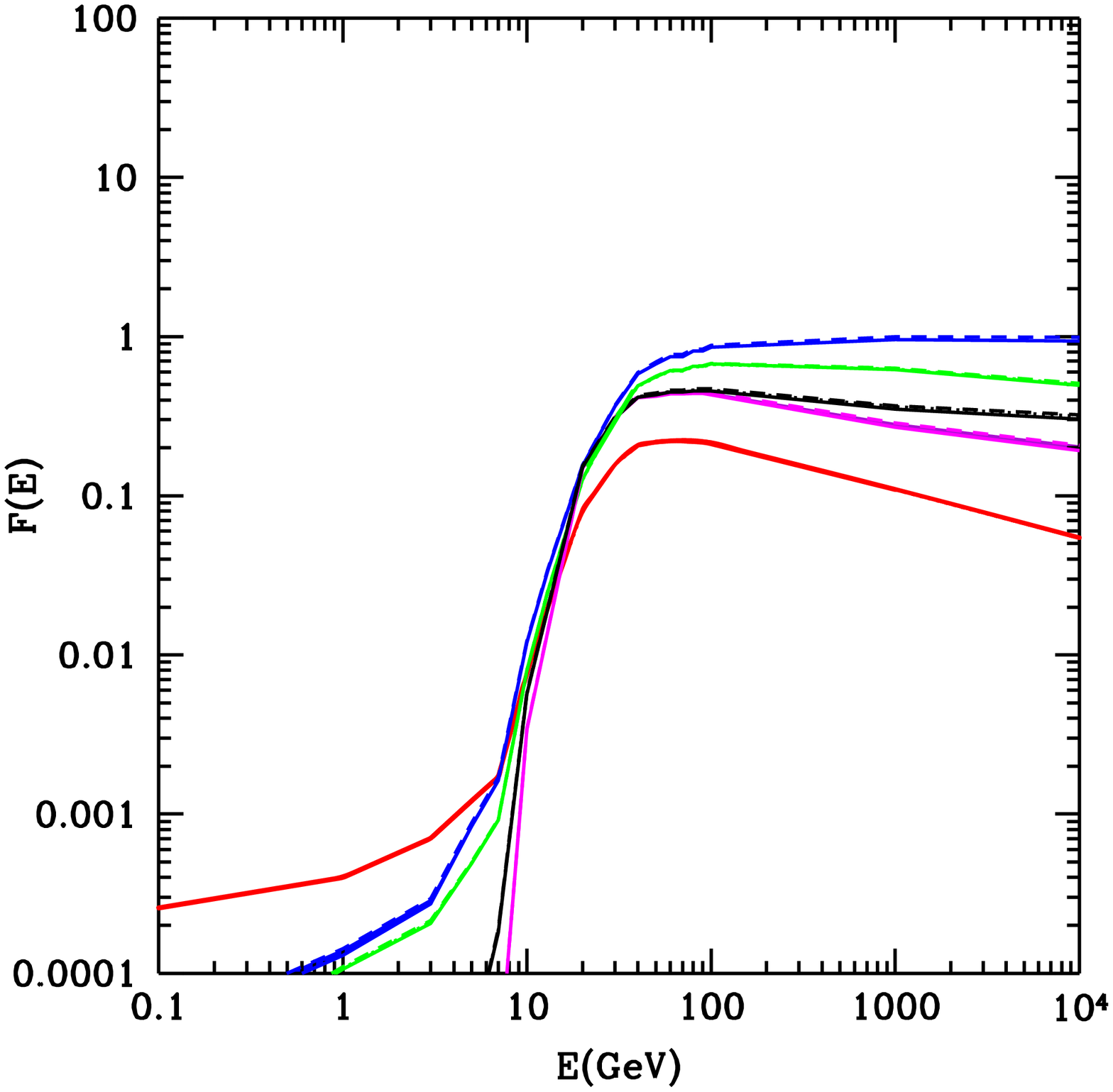}
\caption{The model dependence of the response function based on radio emission, $F^J_r$, is shown. 
The response function based on the observed radio sky at 408 MHz, 1.42 GHz and 23 GHz (from left to right), respectively, are given. 
The red, green, blue, magenta and black bands denote the MIN, MED, MAX, DC, and DR models of Tab.~\ref{tab:prop_model}, respectively. The width of the bands represents the variation within the Kra, Iso and NFW halo profiles of Tab.~\ref{tab:halo}.
\red{The optimal directions are $(\phi,\theta)=( 291^\circ,-13.9^\circ), (291^\circ,-13.9^\circ), (233^\circ,25^\circ)$ for the
three considered radio frequencies, respectively.
Analytical fits to these curves are presented in appendix~B.}}
\label{fig:response}
\end{figure}

In Figure~\ref{fig:response} we present our response functions for the three synchrotron frequencies, showing the dependence on the propagation models and halo profiles. 
The optimal directions are shown in the caption. 
As indicated before, the uncertainty of the response function is dominated by the propagation model
whereas the influence of the dark matter halo profile is small.
This is because the best direction points far from the Galactic center, see Fig.~\ref{fig:excess}, where the different halo models considered are similar. 
On general grounds, if the optimal direction is close to the Galactic center, the excess emission will be more sensitive to the halo profile since there the dark matter density is more uncertain in the absence of sufficiently high resolution numerical simulations.


\red{As shown in Fig.~\ref{fig:response}, for the radio excess maps, the MAX propagation model always gives the strongest constraints. The DC and DR models, which exhibit similar behavior, one clearly sees an exponential cut off at low injection energies, whereas
the response functions for the MIN, MED and MAX models are dropping more slowly with decreasing energy.
This is not surprising since, in the latter case, re-acceleration shifts lower energy electrons to higher energies.
The drop at low energies in these models is strongest at the highest frequencies at which re-acceleration
of the corresponding higher energy electrons is less efficient.
We note that in order to reproduce the observed B/C data, the re-acceleration zone in the MIN, MED and MAX models
should be limited to a slab of height $h_{\rm reac}\simeq0.1\,$kpc, comparable to the height of
the gaseous disk. If the re-acceleration region would extend to the full height $L$ of the diffusive
region, the response function would be flatter and its values would be higher by about a factor of 3
above a few tens of GeV.}

\begin{figure}[tp]
\centering
\includegraphics[height=5.2 cm]{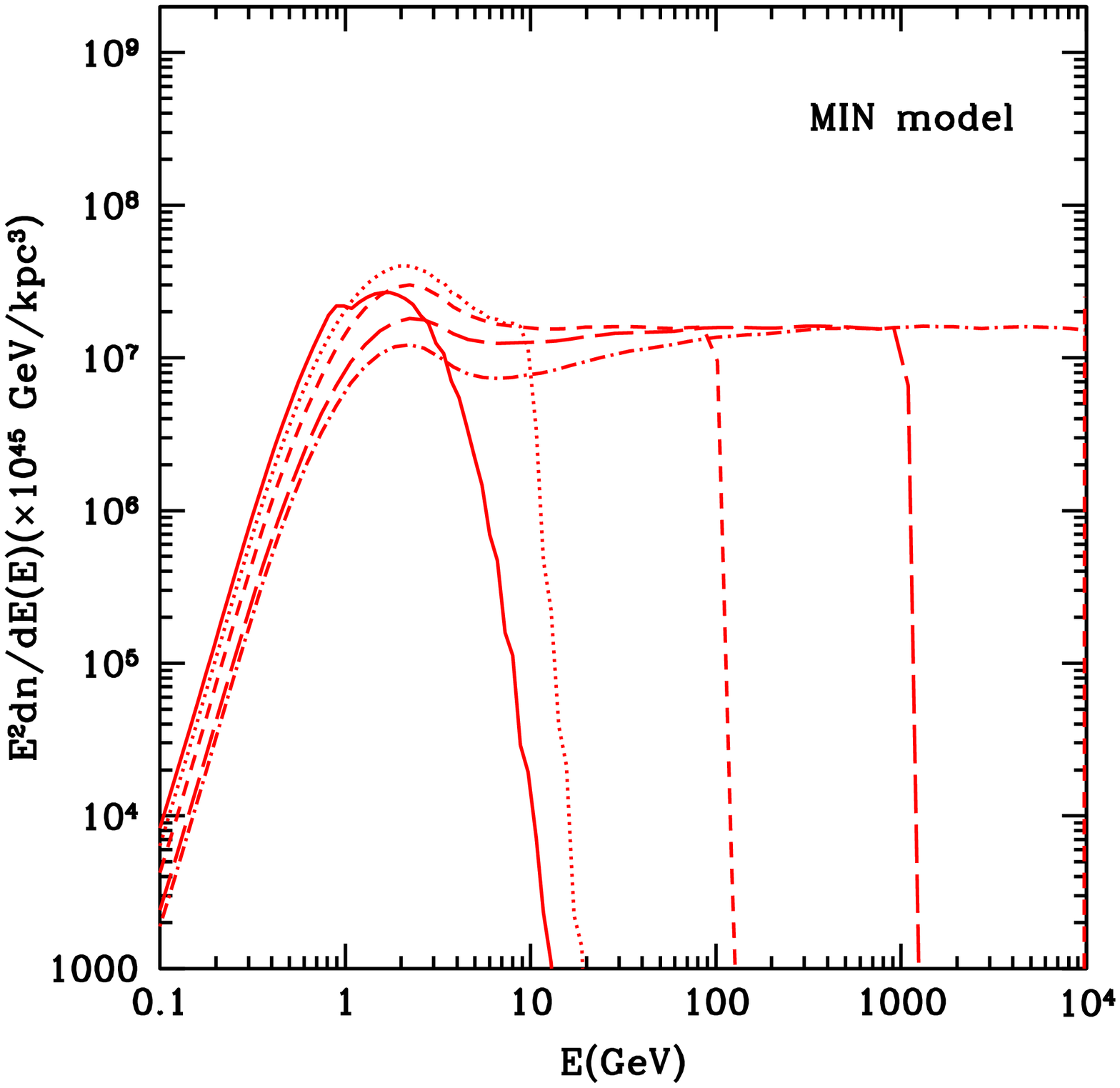}
\includegraphics[height=5.2 cm]{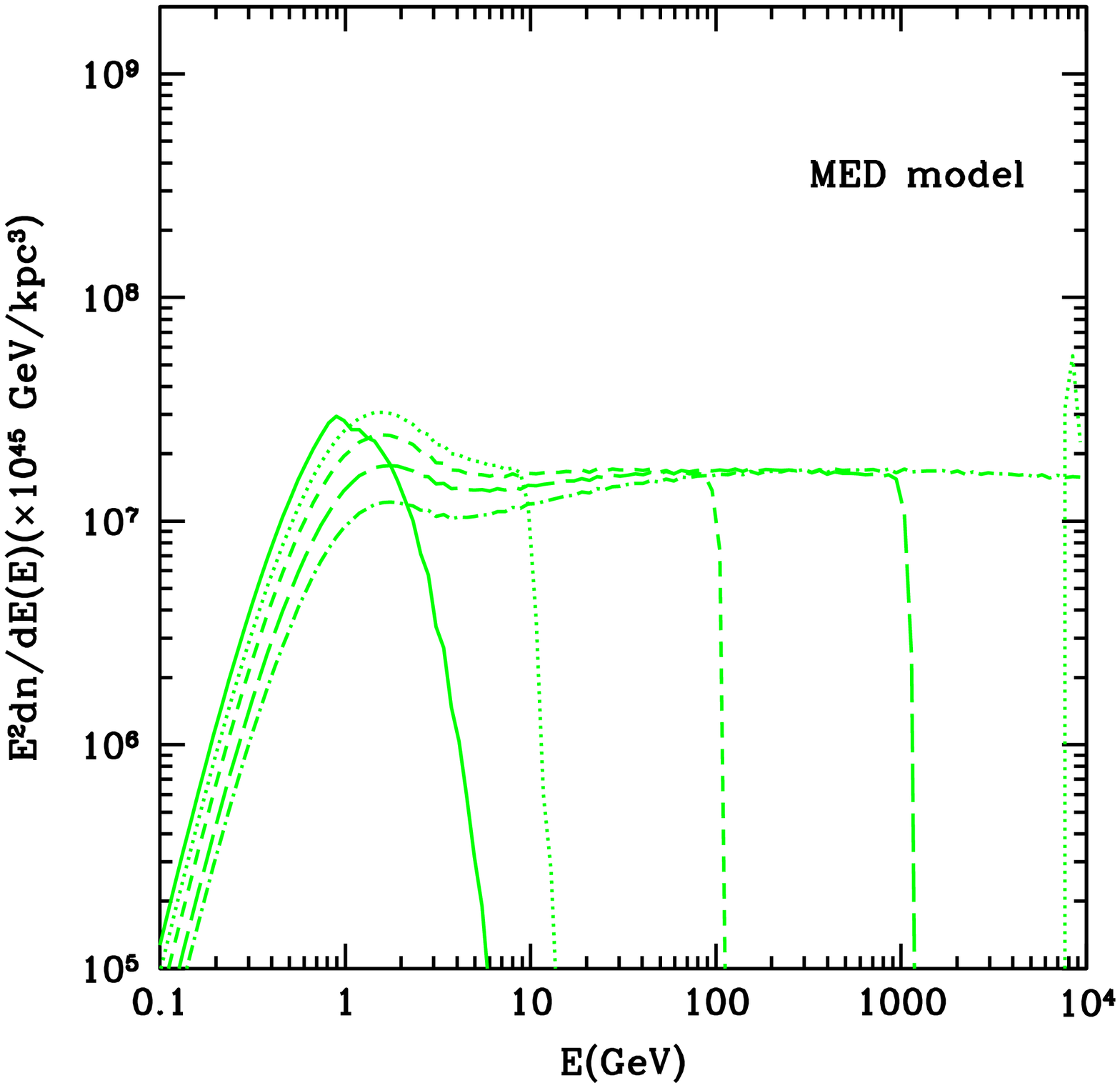}
\includegraphics[height=5.2 cm]{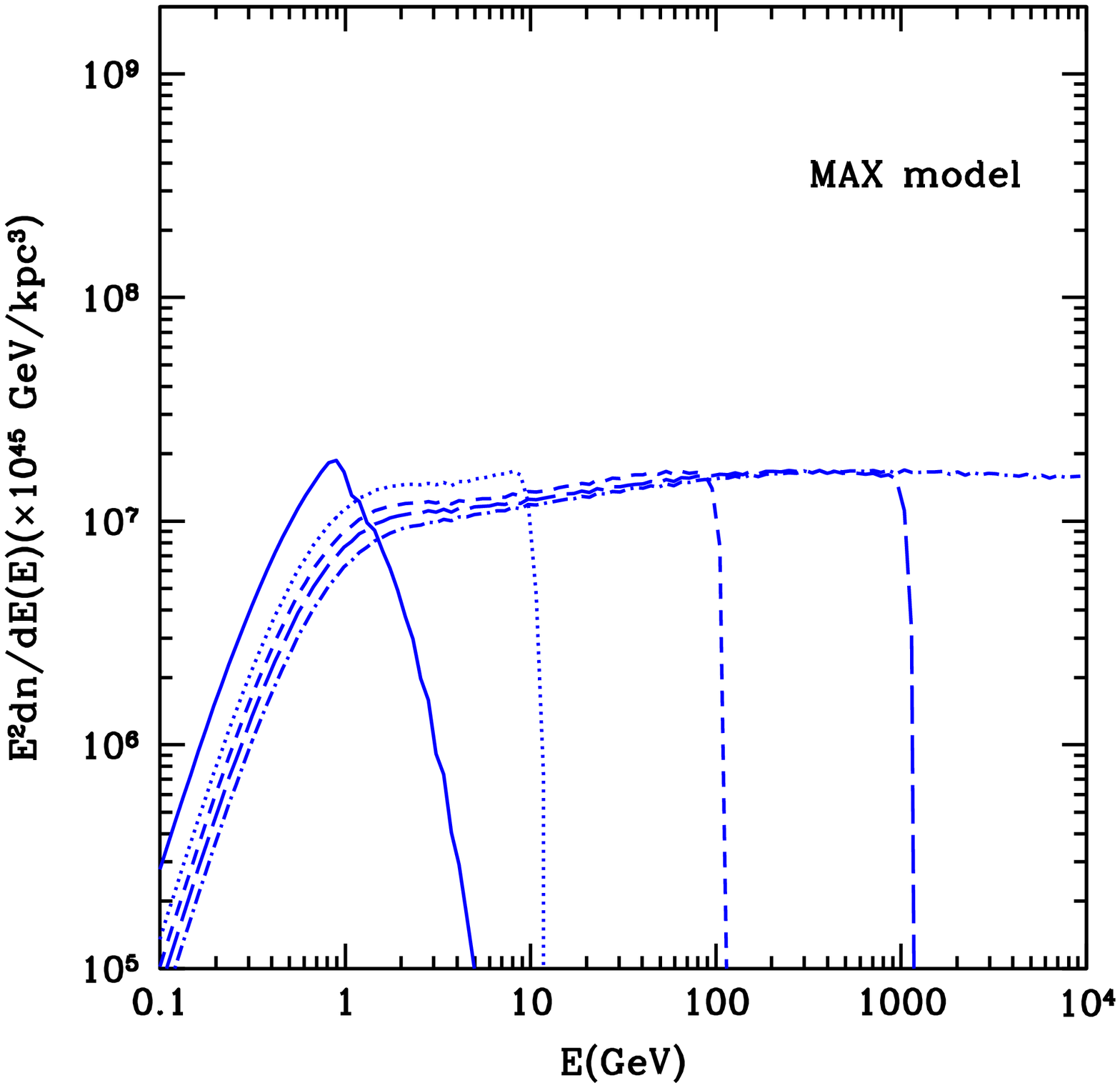}

\includegraphics[height=5.6 cm]{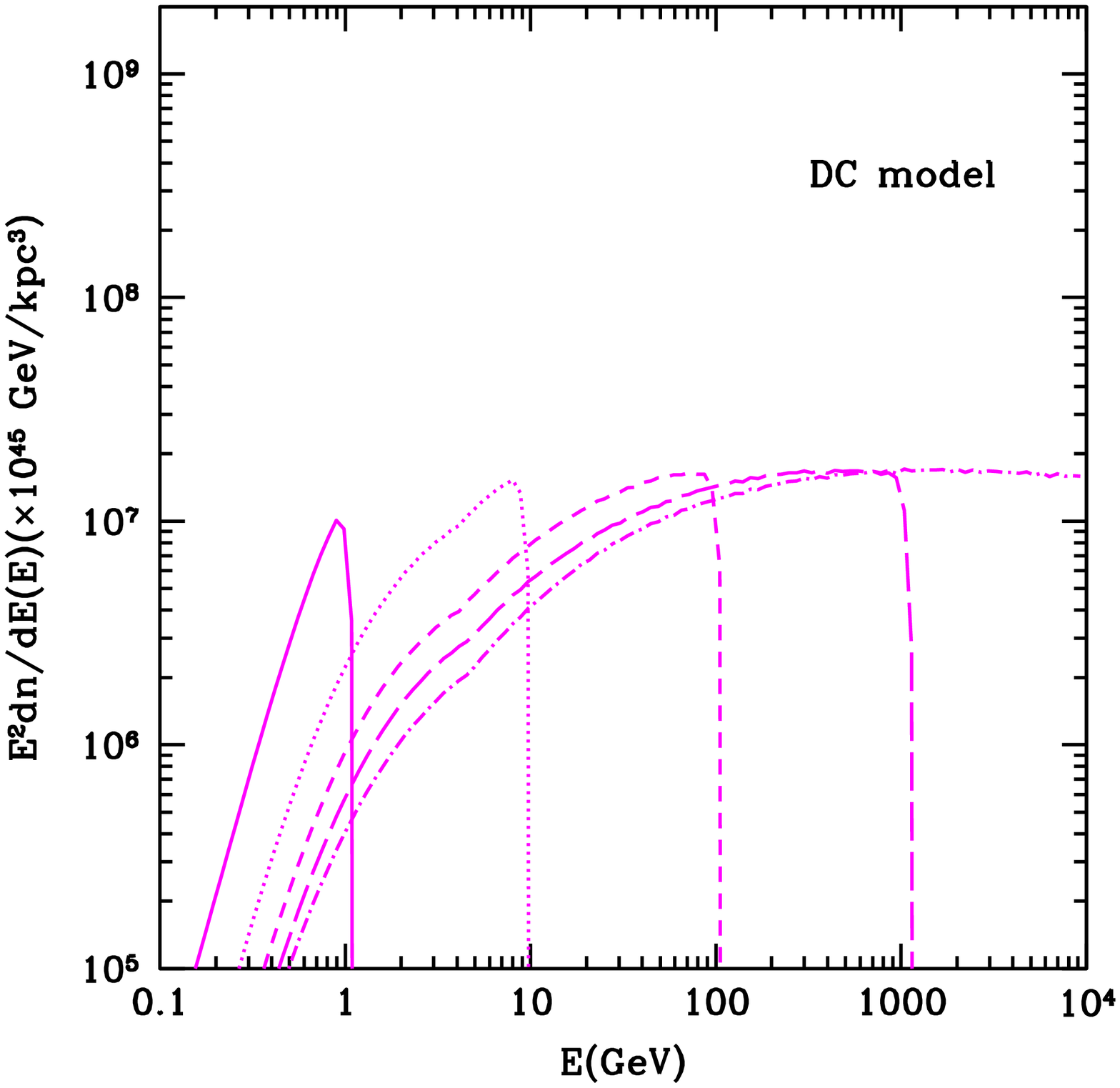}
\includegraphics[height=5.6 cm]{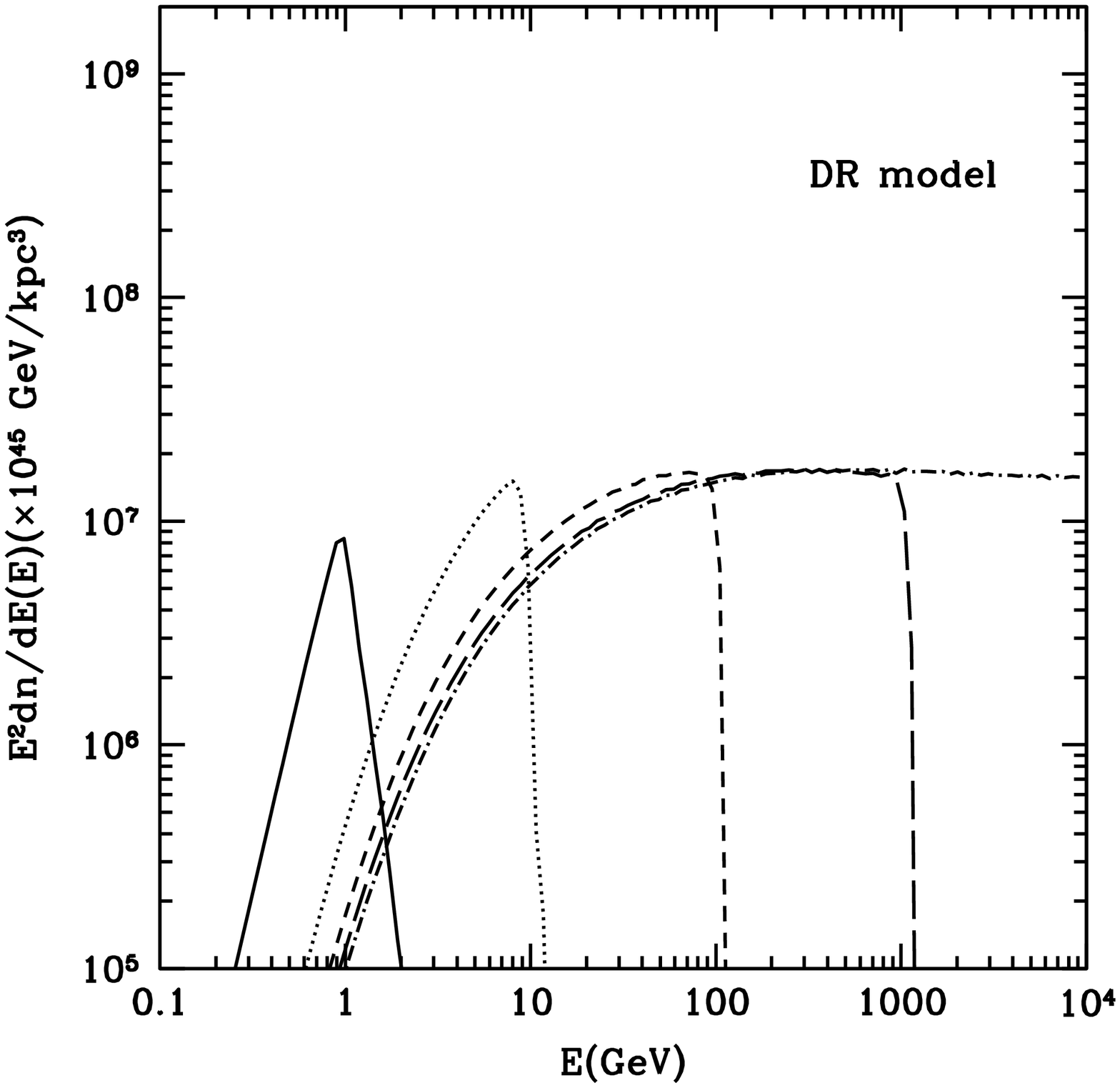}
\caption{The electron or positron spectra versus energy at $r=1\,$kpc, $\red{z=0.2}\,$kpc in the NFW halo model. The solid, dotted, short dash, long dash and dotted-short dash line represent an injection energy of 1 GeV, 10 GeV, 100 GeV, 1 TeV and 10 TeV, respectively. Color keys are as in Fig.~\ref{fig:response}.}
\label{fig:spectrum_E}
\end{figure}

To illustrate these points we show in Fig.~\ref{fig:spectrum_E} electron spectra in the galactic disc at
1 kpc from the center and 0.2 kpc above the disk, for different propagation models and injection energies. 
When re-acceleration is included, we get a noticeable bump in the  spectrum at a few GeV. Above these energies, the energy loss generated from inverse Compton scattering and synchrotron emission dominates over energy gain by re-acceleration, and below a few GeV, re-acceleration overcompensates the energy losses. Thus a visible bump appears  when the electrons accumulate in an energy region where re-acceleration and energy losses offset each other. It seems that the amplitude and the position of the bump is independent of the injection energy below a few GeV. The large amount of electrons and positrons accumulating in this bump region induce most of the radio signals around GHz frequencies.
\red{That is why the shape of the response function drops more gradually in the lower energy region
in the MIN, MED and MAX models compared to the exponential drop in the DC and DR models.}

In the DC model, due to the absence of re-acceleration, the energy spectra appear as a sawtooth shape as the number of propagated electrons above the injection energy have a sharp cutoff. 
Similarly, in the DR model, since the electrons and positrons can not gain enough energy from re-acceleration due to the larger diffusion parameter and the smaller Alfven speed, see Eq.~(\ref{eq:Dpp}), the propagated energy spectrum above the injection energy tends to zero rapidly.

Another interesting property of the response function is that it tends to fall at high energies.
This can be understood as follows: Higher energy electrons either loose energy more quickly, or, if their
energy loss length is still larger than the half height of the diffusion zone, can propagate further and can thus
escape from the diffusion zone more readily. The diffusion length can be estimated as 
$\sqrt{D_{xx}(E)t_{\rm loss}(E)}\approx$ a few kpc, and is comparable to the the thickness of the diffusion zone
in the MIN and MED models. As a consequence, in Fig.~\ref{fig:spectrum_E} for the MIN and MED models, the
propagated spectrum is indeed smaller by factors of a few for the highest injection energies. 
In the case of the MAX model, the half height of the diffusion zone is considerably larger than the typical diffusion scale so
that the injection energy has a weak impact on electron spectrum below 100 GeV.

We note that the choice of the diffusion models and the injection energy affect the shape of the electron spectrum significantly only below 100 GeV, where the influence of the diffusion mechanism on electron and positron propagation is still important. For energies above 100 GeV, energy loss dominates the electron spectrum. The spectrum
in this energy range thus should only depend on the ISRF and magnetic field and not significantly on the diffusion parameters, as is confirmed by Fig.~\ref{fig:spectrum_E}. The plots for the DC and DR models, for which 
re-acceleration is insignificant, confirm our qualitative analysis. For instance, for 10 TeV injection, a flat spectrum appears at energies above 100 GeV, below which the spectrum steadily drops due to electron diffusion.

The response function reaches its maximum around the critical energy of Eq.~(\ref{eq:E_crit}) corresponding to the energy at which electrons emit photons of the considered frequency $\nu$ (namely about 5 GeV for 408 MHz, 9 GeV for 1.42 GHz and 20 GeV for 23 GHz, for slowly varying magnetic field of $\sim5\mu$G strength in the regions of interest).

We also note that the response function for the DC and DR models tend to give stronger constraints than the MIN scenario for the 23 GHz map. Since the thickness of the diffusion zone in the MIN model is only 1 kpc,
there is no strong emission from directions far from the Galactic center, similarly to Fig.~\ref{fig:1.42G_models}. As a result, the optimal direction at 23 GHz points to high latitude.

%
%

\section{Response Function for Positron Flux} \label{sec:res-positron}

\begin{figure}[tp]
\centering
\includegraphics[height=9.5 cm]{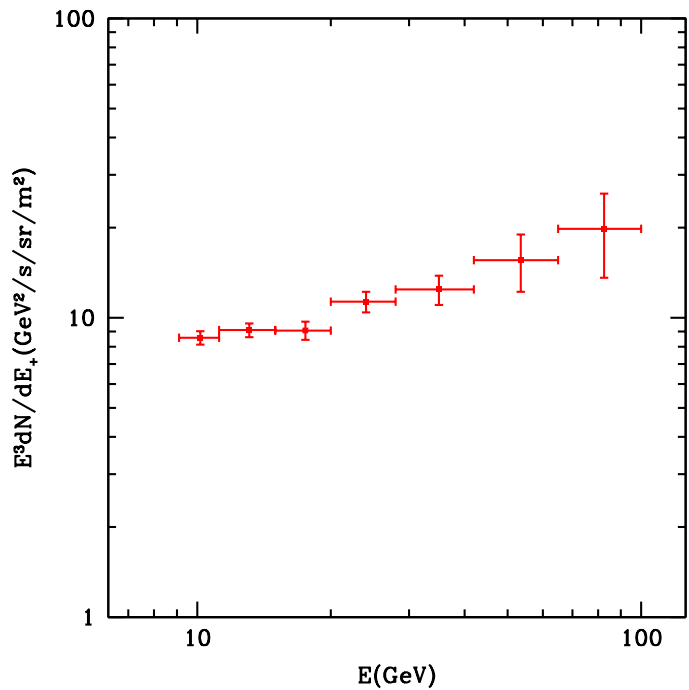}
\caption{The positron flux observed at Earth as obtained by multiplying the
$e^++e^-$ flux observed by FERMI~\cite{Abdo:2009zk} with the positron fraction measured
by PAMELA~\cite{Adriani:2008zq,Adriani:2008zr}, see text.}
\label{fig:e+flux}
\end{figure}

Recently, PAMELA reported a relatively large positron fraction in the electron/positron flux above 10 GeV~\cite{Adriani:2008zr}. Possible
explanations include as yet unknown nearby astrophysical sources or the decay or annihilation of dark matter.
However, decaying dark matter models can be constrained by requiring the predicted positron flux to be
smaller than the observed one. This can again be expressed in terms of a response function along the lines of Sec.~\ref{sec:response}.
In order to convert the positron fraction given by PAMELA data~\cite{Adriani:2008zr} into the positron flux,
we \red{multiply it with } the latest $e^+e^-$ flux observed by the Fermi Telescope \cite{Abdo:2009zk} and note that
the parametrizations for the Galactic electron flux in Ref.~\cite{Baltz:1998xv,Moskalenko:1997gh} is
larger than the new Fermi data by a factor of about 1.5 below about 30 GeV.
\red{Compared with the PAMELA data, the statistical errors of the Fermi data is sub-dominant
because of the finer energy binning and the smaller statistical error of the flux. Therefore, for the
statistical error of the positron flux we take into account only the statistical error of the PAMELA
positron fraction data. The Fermi data are well fit by a simple power law expression $J_e=172.37\,E^{-3.04 }\,\rm{s^{-1}m^{-2}sr^{-1}GeV^{-1}}$. The resulting ``observed'' positron flux is shown in Fig.~\ref{fig:e+flux}.
The strongest constraints come from the positron flux at the lowest energy where the statistical error
is negligible.}

\begin{figure}[tp]
\centering
\includegraphics[height=5.5 cm,trim=1 5cm 1 3cm,clip]{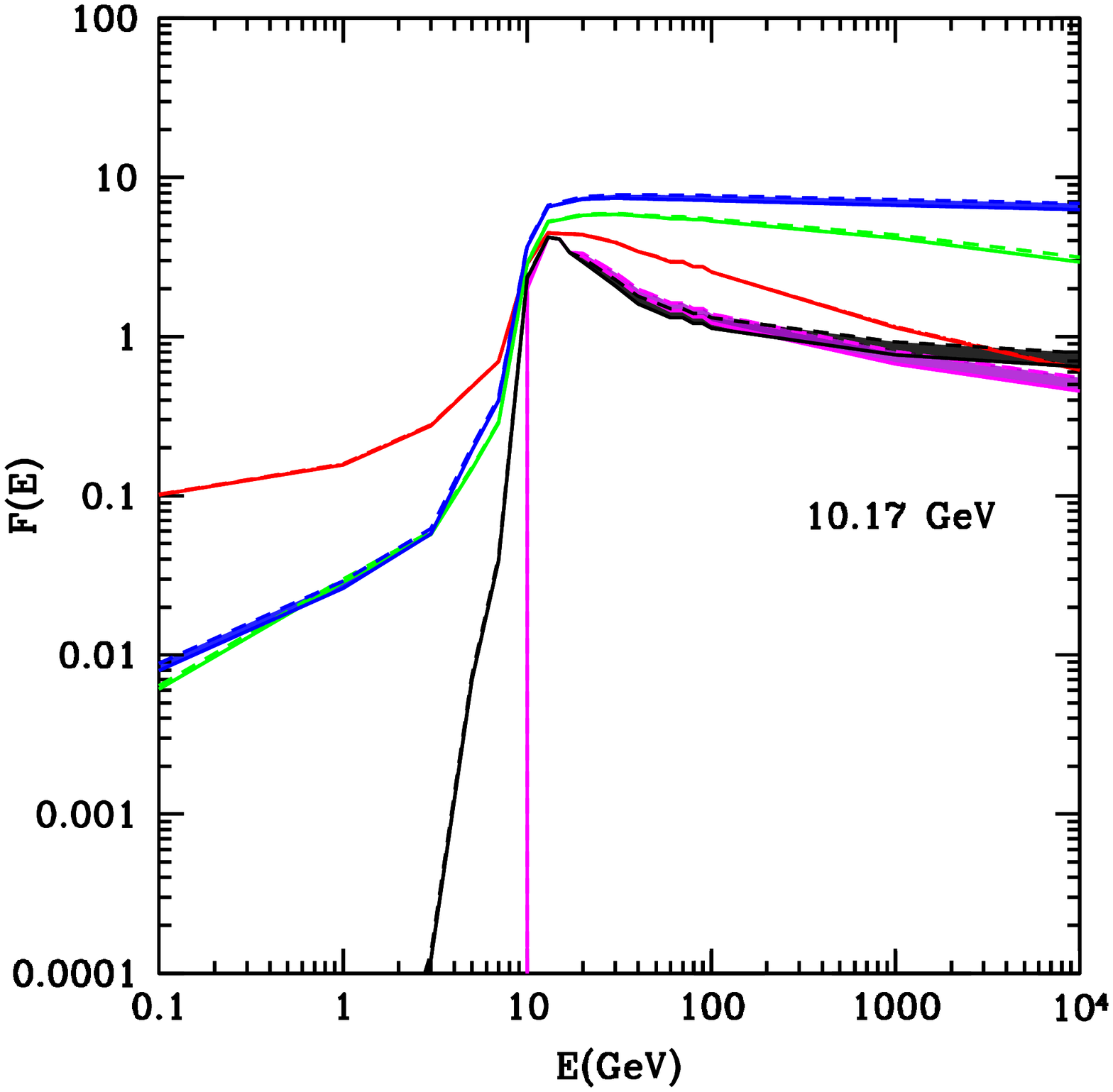}
\includegraphics[height=5.5 cm,trim=1 5cm 1 3cm,clip]{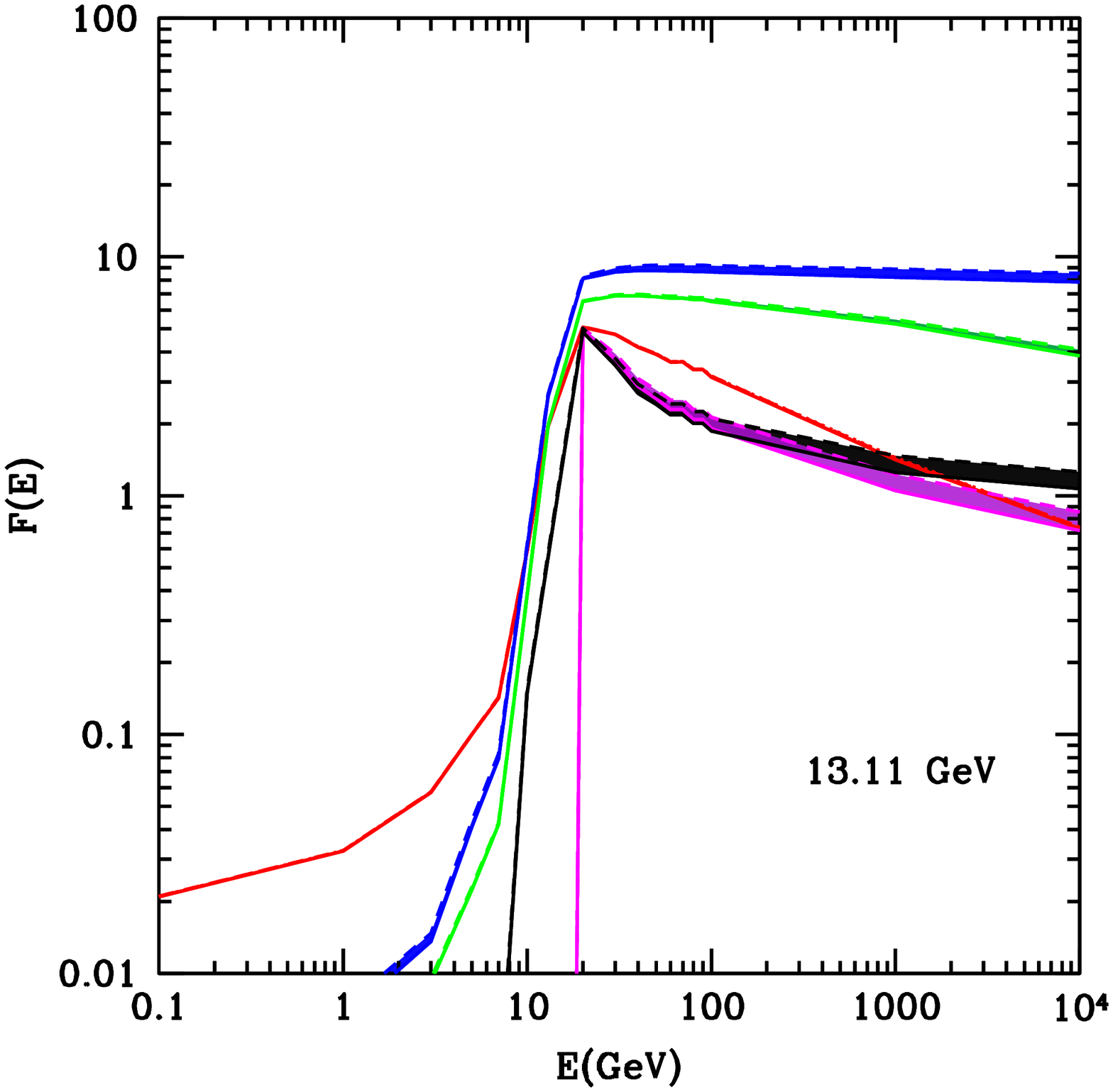}
\includegraphics[height=5.5 cm,trim=1 5cm 1 3cm,clip]{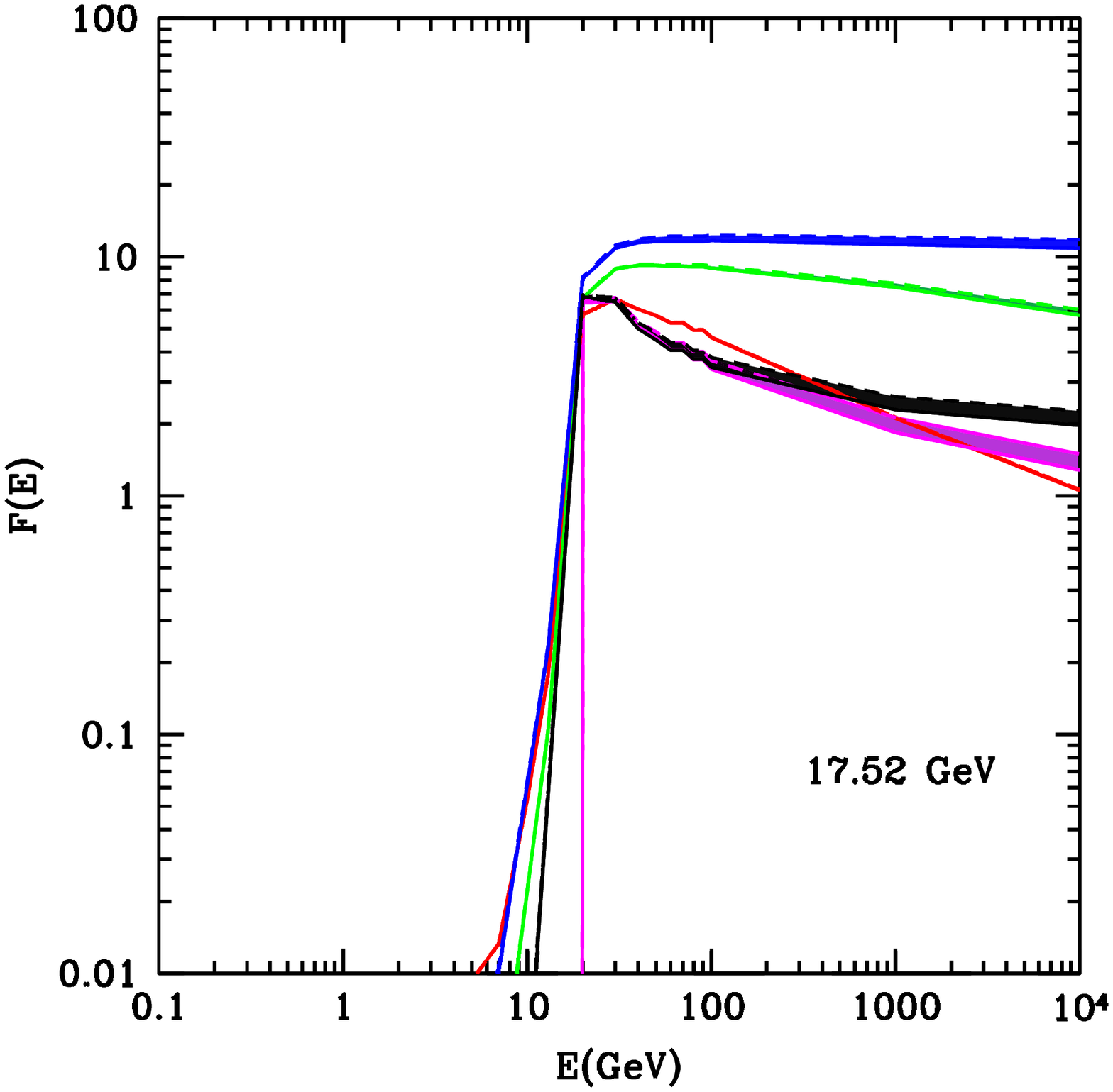}

\includegraphics[height=5.5 cm,trim=1 5cm 1 3cm,clip]{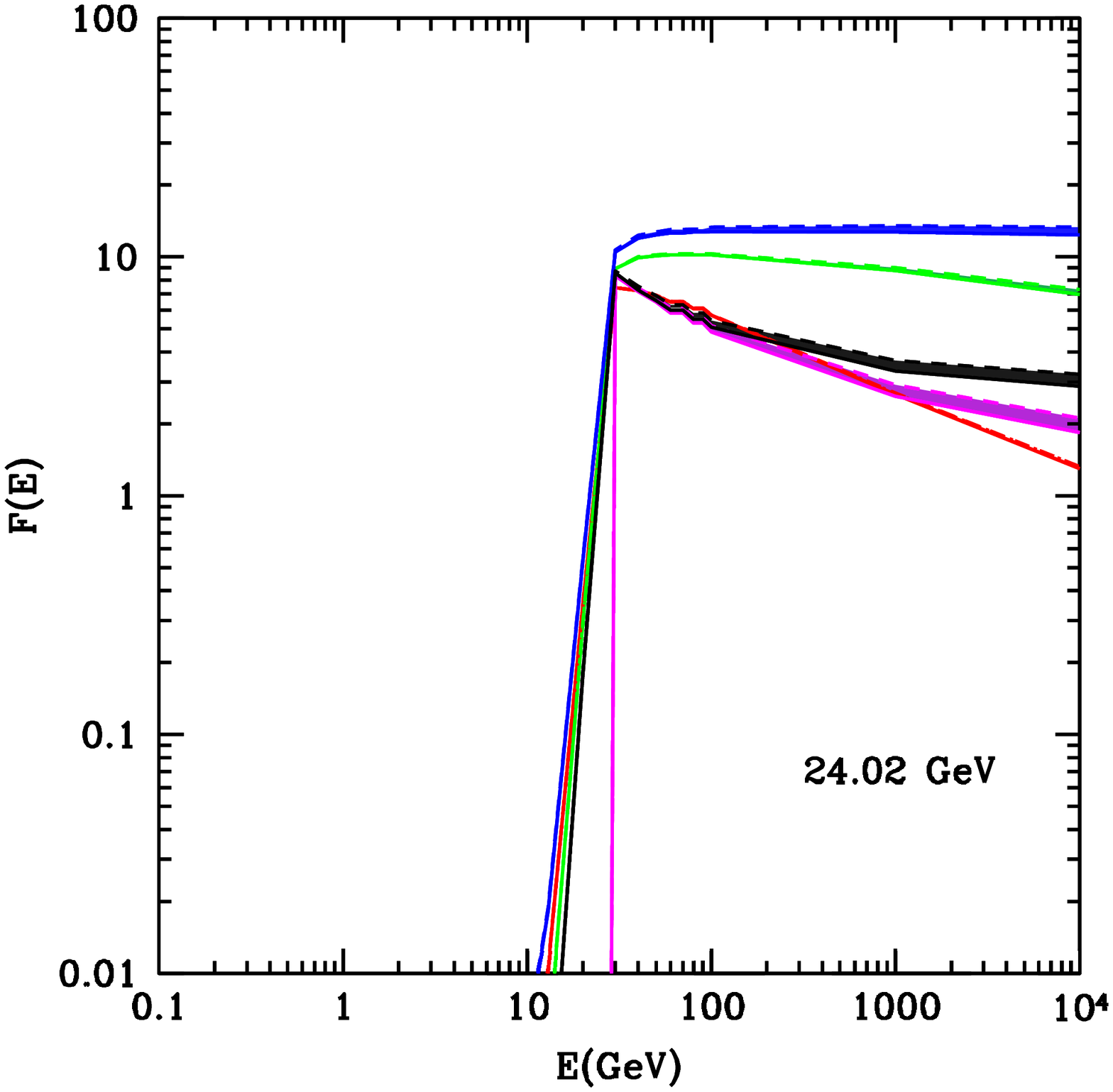}
\includegraphics[height=5.5 cm,trim=1 5cm 1 3cm,clip]{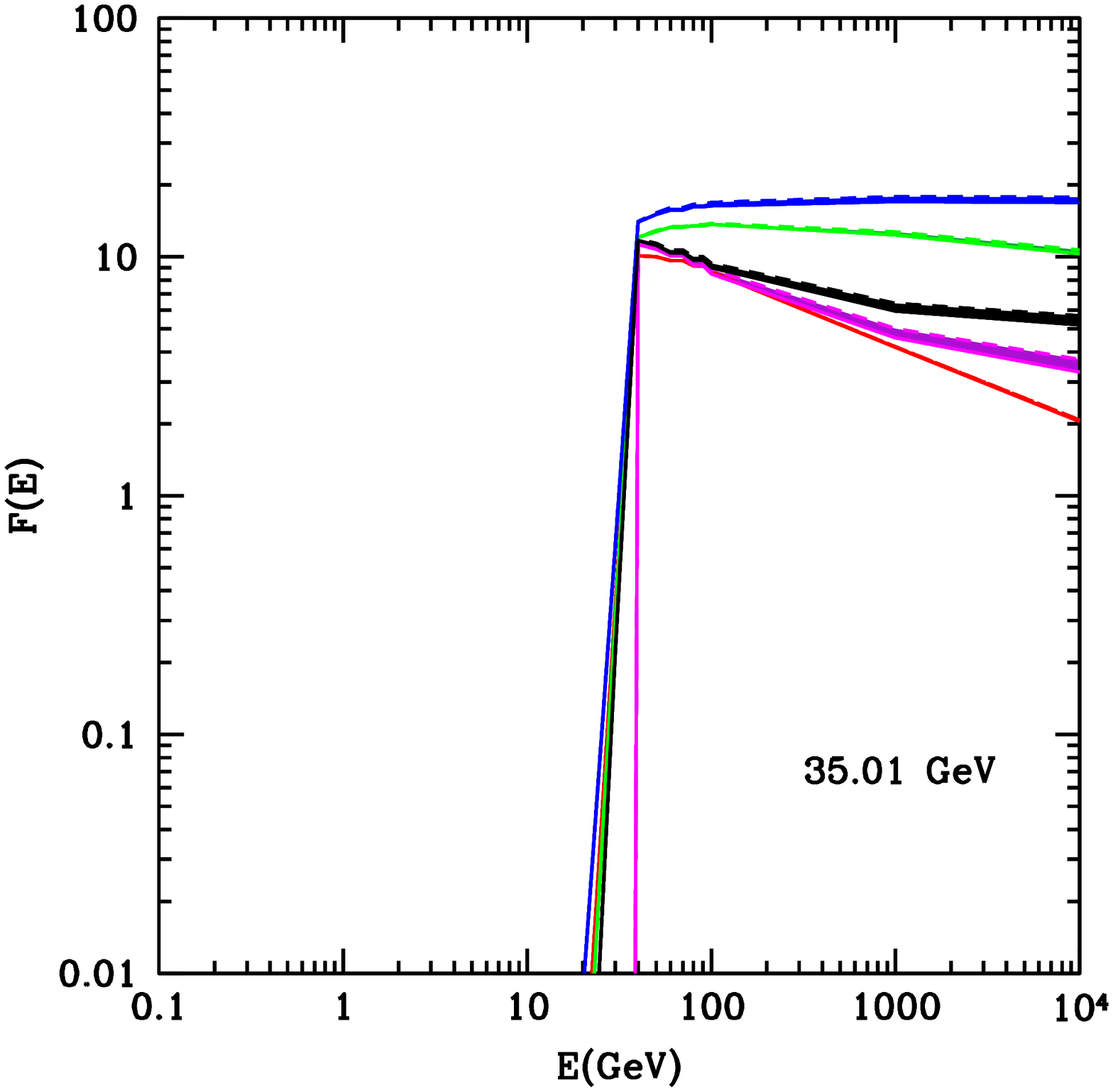}
\includegraphics[height=5.5 cm,trim=1 5cm 1 3cm,clip]{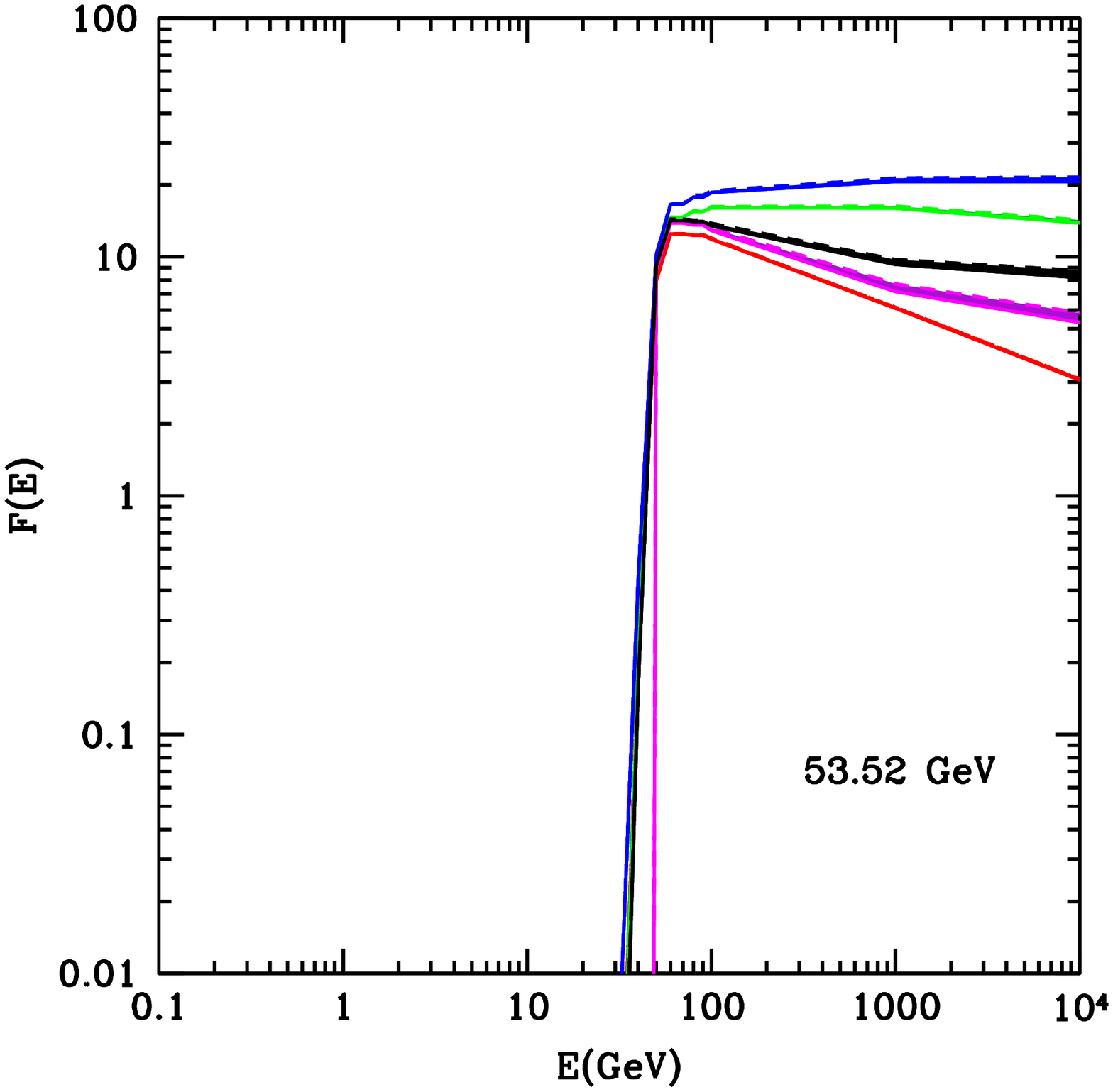}

\includegraphics[height=5.5 cm,trim=1 5cm 1 3cm,clip]{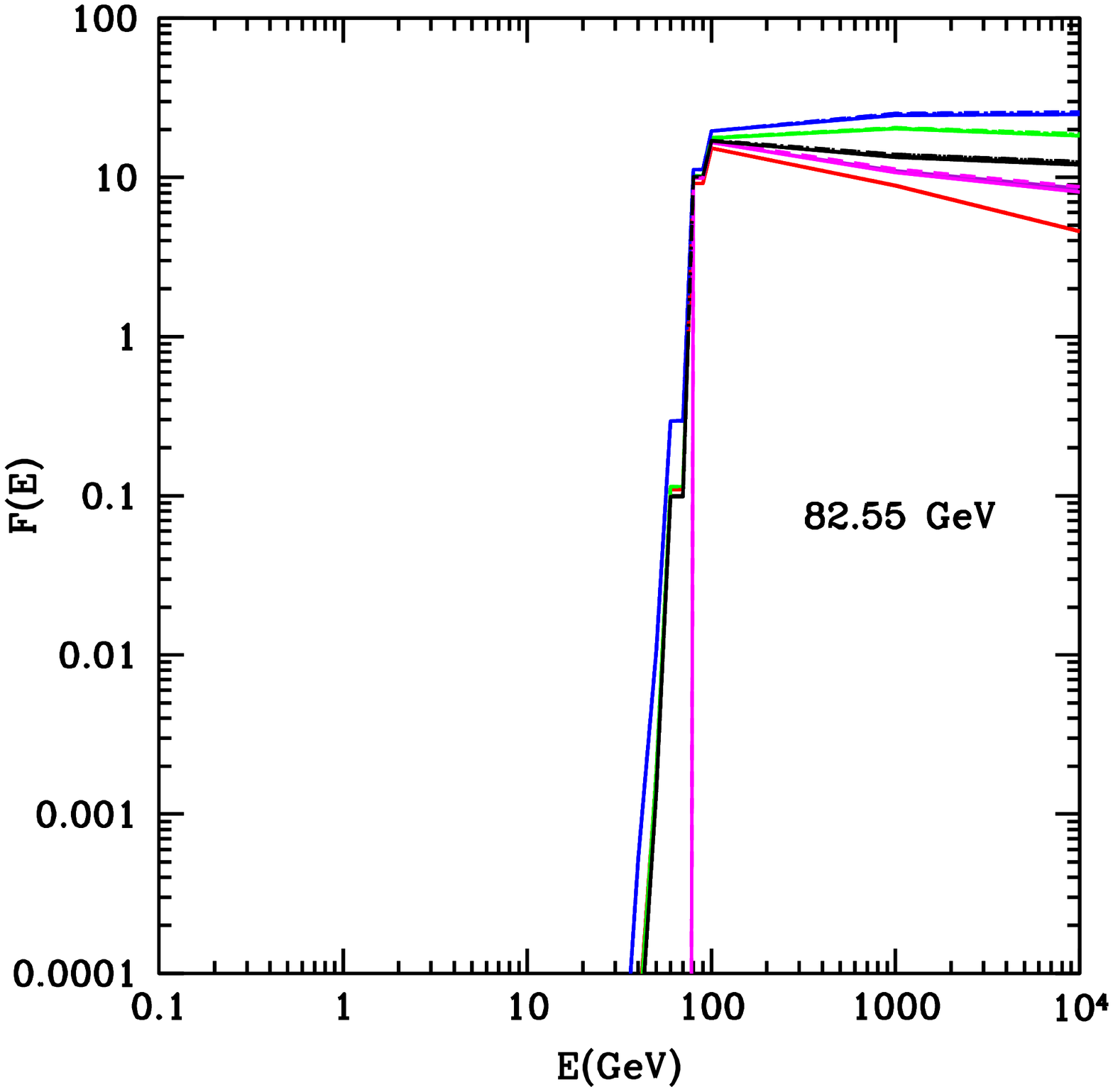}
\caption{The response function $F_p(E)$ resulting from the observed positron flux given by PAMELA~\cite{Adriani:2008zr} for various energies at which the positron flux was observed.
The model dependence is also shown. The color key is as in Fig.~\ref{fig:response}.
Analytical fits to these curves are presented in appendix~B.}
\label{fig:e+bound}
\end{figure}

The response functions based on the positron flux measured by PAMELA~\cite{Adriani:2008zr} are shown in Fig.~\ref{fig:e+bound}. We only consider the high energy region above 10 GeV where the solar wind has no
significant influence. 
Compared to the synchrotron response functions the positron response functions are generally larger
and, therefore, in general lead to stronger constraints. 

We shall note however that this would not be the case in CP-non-symmetric dark matter decay models in which the positron flux is suppressed with respect to the electron flux. 
In this situation one should note that synchrotron constraints based only on the electron density are stronger than those based on the locally observed electron flux itself, since the electron flux is about ten times larger than the positron flux.

Fig.~\ref{fig:e+bound} shows a prominent feature in the response function based on the PAMELA data in comparison
with the response function based on the radio emission: It depends mostly on the diffusion model but little on the dark matter halo profile. This is because
high energy positrons mostly come from nearby sources within $\sim 1\,$kpc where different halo profiles
yield very similar dark matter densities. 
Further, the constraints are weakly affected by the magnetic field
since energy losses are dominated by the background radiation fields.

In addition Fig.~\ref{fig:e+bound} shows that the response function cuts off below the energy at which the positron flux is observed in the DC and DR models where powerful re-acceleration is absent, as one would expect since electrons essentially can only loose energy in this situation. On the other hand, the response function tends to peak where the injection energy approaches the observed energy. Above that energy the response function gradually falls off due to the faster diffusion effects of higher energy positrons, similarly to the behavior of the radio based response function discussed before. The MAX scenario predicts the largest locally observed positron flux since the stronger re-acceleration in the MAX models shifts the predicted peak of the energy spectrum to larger energies, similarly to Fig.~\ref{fig:spectrum_E}. We should note that the amplitude of the bump in Fig.~\ref{fig:spectrum_E} in the MAX model is somewhat lower than in the MIN and MED scenarios due to the larger diffusion parameter. However, since this bump is shifted to larger energies in the MAX model, the MAX scenario still gives the strongest constraint.

%
%

\section{Constraints on Dark Matter Models}
\label{sec:constraint}

The response functions developed in the previous sections can provide interesting constraints on any dark matter decay model. 
In this section we discuss two simple examples that can be applicable to a number of realistic scenarios.

Let us first consider an extremely simple case, direct decay into and electron/positron pair $X\to e^+ e^-$
with branching ratio $b_e$. This will give conservative constraints for models that also allow
decays into other fermion pairs such as~\cite{Redondo:2008ec,Chen:2008yi}. 
Moreover, it is certainly the simplest scenario that could account for the PAMELA excess~\cite{Nardi:2008ix,Ibarra:2008jk}.
The spectrum of injected electrons is then $dN_\pm/dE_0=b_e\delta(E_0-m_X/2)$ and the constraints from
the radio and positron fluxes are therefore 

\begin{equation}\label{eq:constraint_models}
\fl \frac{\tau}{10^{26}\ {\rm s}} \geq 2b_e\, \left(\frac{100\, {\rm GeV}}{m_X}\right) F_r(m_X/2)\quad ; \quad
\frac{\tau}{10^{26}\ {\rm s}} \geq b_e\,\left(\frac{100\, {\rm GeV}}{m_X}\right) F_p(m_X/2) \  .
\end{equation}

\begin{figure}[tbp] 
\centering
{
\psfragscanon
\psfrag{a}[][l]{$m_X$ [GeV]}
\psfrag{b}[][l]{$\tau \times b_e^{-1}$ [$10^{26}$ sec.]}
\psfrag{c}[][l]{$X\to e^+e^-$}
\includegraphics[width=7cm]{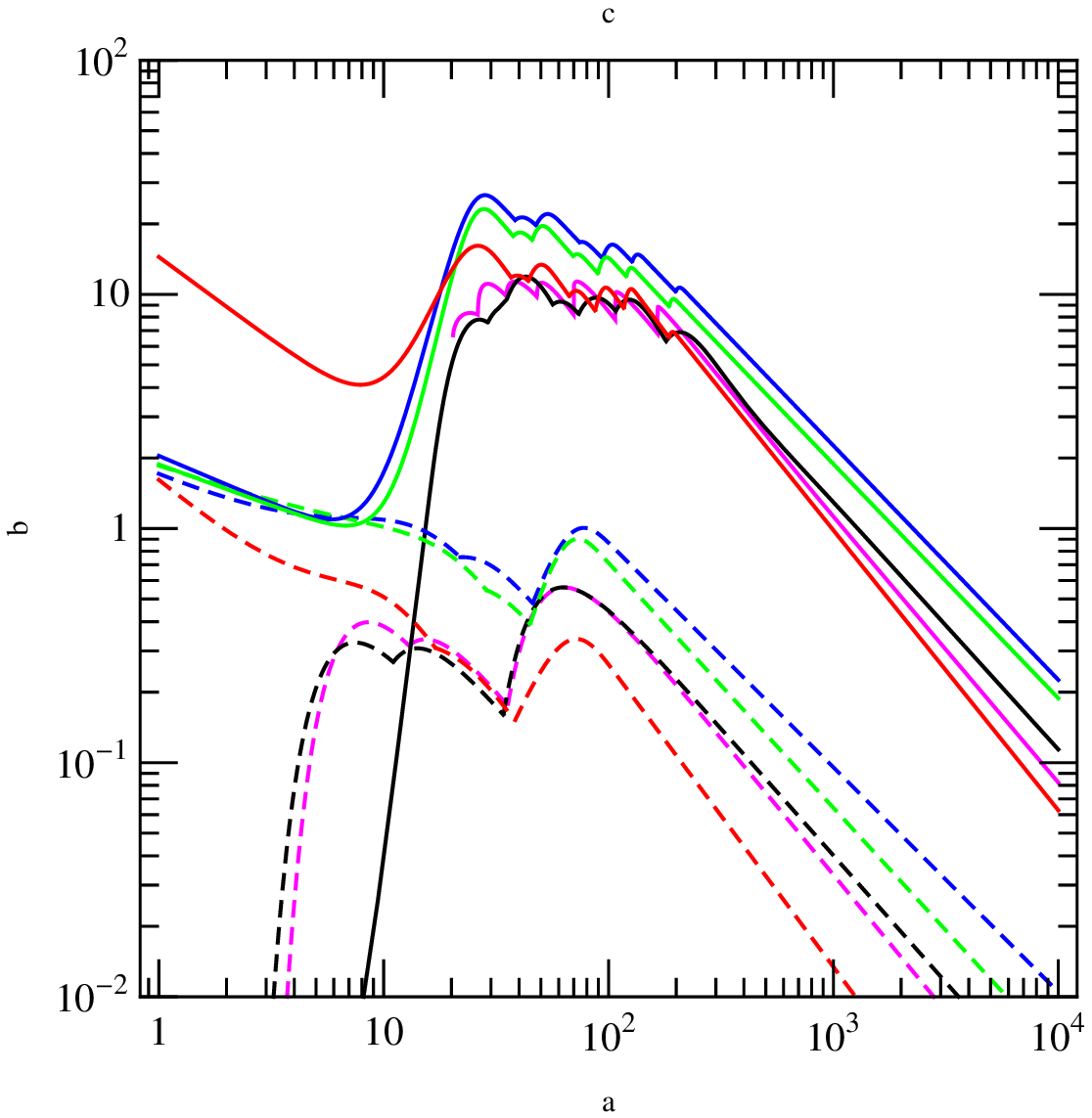}
}
{
\psfragscanon
\psfrag{a}[][l]{$m_X$ [GeV]}
\psfrag{b}[][l]{$\tau \times b_{ll{\rm inv}}^{-1}$ [$10^{26}$ sec.]}
\psfrag{c}[][l]{$X\to l^+l^-$+ inv.}
\includegraphics[width=7cm]{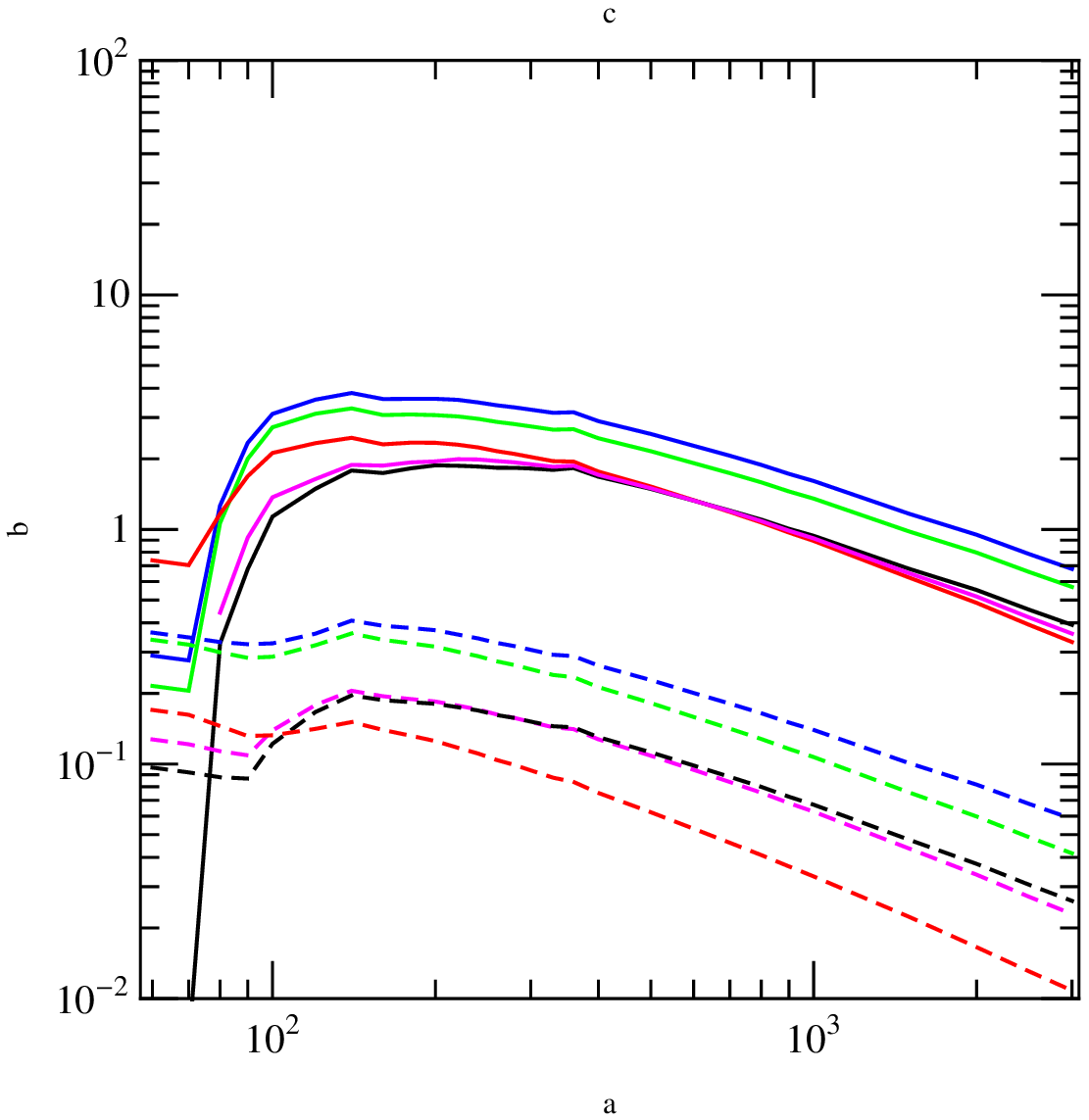}
}
\caption{Bounds on a decaying dark matter particle for the decay mode $X\to e^+ + e^-$ (left panel)
and $X\to l^+ l^- +$invisible (right panel), see the text for details. 
The bound is on the lifetime of the dark matter particle divided by the branching ratio of the relevant mode.
The color key for the propagation model is as in Fig.~\ref{fig:response}.
Constraints from radio emission are dashed and from the PAMELA positron flux
(with normalization given by the newest FERMI data) are solid. 
Each constraint line is based on either the synchrotron response function for the three
frequencies (408 MHz, 1.420 GHz and 23 GHz) or on positron response function for the
7 PAMELA energy bins.}
\label{fig:bounds}
\end{figure}

These bounds are shown on the left panel of Fig.~\ref{fig:bounds} for all the different
propagation models considered. 
We see that in general the positron data provide stronger constraints on this particular model
than radio data, unless the dark matter mass is smaller than the PAMELA energy window i.e. $m_X < 11$ GeV,
and at the same time there is no strong re-acceleration, as is the case in the DC and DR propagation
models.

As a more realistic case, we have considered the scenario developed in Ref.~\cite{Chen:2008yi}.
This framework involves an extra U(1) gauge symmetry under which the standard model fields have no charge, a \emph{hidden} U(1).
Despite being secluded, interactions of this hidden sector with the standard model particles are realized
through a tiny kinetic mixing~\cite{Holdom:1985ag} $\chi$ with the hypercharge U(1)$_Y$. 
In particular, the hidden gauge boson $A'$ couples to all hypercharged particles with an additional
suppression factor $\chi$ which can have extremely small values, cf.~\cite{Dienes:1996zr,Abel:2003ue,Abel:2006qt,Abel:2008ai}.
If this hidden gauge boson accounts for the dark matter and has a mass around $\sim 200$ GeV
it can decay into lepton, quark and $W$-boson pairs.
Furthermore, if the $A'$ lifetime is of order $10^{26}$ sec,
the $\gamma-$rays and positrons resulting from the decay can explain the EGRET and PAMELA excesses.
Convolving our response functions with the positron spectrum provided in~\cite{Chen:2008yi} we find
that, based on the continuum component alone, this model can be ruled out in the MIN, MED and MAX propagation scenarios due to the strong re-acceleration effects. 
The direct decay into $e^+e^-$, has a branching ratio ranging from 0.05 to 0.14 in the mass range $100$ GeV $\geq m_{A'}\geq 300$ GeV  and is directly constrained by the left panel of Fig.~\ref{fig:bounds}%
\footnote{This scenario is somehow similar to the gravitino with R-parity violation described in~\cite{Takayama:2000uz,Buchmuller:2007ui}, 
whereas the decay branching ratio of hidden gauge boson to the W boson is highly suppressed
compared to the gravitino case.}.

Let us now consider the three body dark matter decay $X\to l^+ + l^-+$invisible where $l^\pm$ are Standard Model
leptons and \emph{invisible} stands for a nearly massless fermion ($m_X\gg m_{\rm inv}$). 
This decay arises for instance in supersymmetric versions of the hidden U(1) extension of the Standard Model outlined above, 
which also include the \emph{hidden gaugino}, the supersymmetric partner of the hidden gauge boson~\cite{Ibarra:2008kn}.
As a concrete example consider the scenario recently described in~\cite{Ibarra:2009bm}: 
Dark matter is made of bino-like neutralinos $\chi_1^0$ which are the lightest supersymmetric particles of the Standard Model sector. 
The neutralino mass is taken to be around $m_{\chi^0_1}=300$ GeV which gives the correct relic dark matter
abundance while the hidden gaugino $\tilde X$ is supposed to be a sub-dominant component of dark matter.
If the hidden gaugino is lighter than the neutralino, the decay channel $\chi_1^0\to l^+ l^- \tilde X$ is available,
whereas the reverse process $\tilde X\to l^+ l^- \chi_1^0$ can be realized in the opposite case. 
These decays are mediated by heavy sleptons. Since squarks are usually heavier than sleptons,
decays into quark pairs are suppressed.

In order to study bounds on this decay we focus onto the democratic case, i.e. equal branching ratio for decaying into lepton pairs of the three different families.
We used PYTHIA~\cite{Sjostrand:2006za}  to simulate the final electron and positron decay spectra, and our response functions to give the bounds shown in the right panel of Fig.~\ref{fig:bounds}.
These decay channels can be also found in other contexts like in~\cite{Shirai:2009kh} where a massive B-L gauge boson mediates the kinetic mixing, in grand unified theories~\cite{Arvanitaki:2008hq,Arvanitaki:2009yb},
or in general SUSY models with slightly broken R-parity with or without considering supergravity~\cite{Ishiwata:2009vx,Shirai:2009fq,Chen:2009ew}.

%
%

\section{Conclusions}
\label{sec:conclusion}
In our study we have computed the prediction for two important signatures from decaying dark matter, namely
the synchrotron emission in our Galaxy and the positron flux. In view of the recent experimental observations such as the radio full sky surveys and the new PAMELA data, we have introduced useful response functions that can be applied to constrain any interesting decay models. 
Robust constraints can be obtained in terms of convolving the response function with the specific decay spectrum into electrons and positrons.
Our results show that the resulting constraints depend mostly on the set of  propagation parameters rather than the halo profiles and 
reveal re-acceleration as \red{an important } ingredient. 

We have finally applied our method to provide model independent constraints on two widely discussed decay modes and studied  
the implications for some more concrete realizations. In some cases, specifically when choosing the propagation models with largest re-acceleration parameters our methods rule out recently proposed decaying dark matter models.

We shall emphasize here that the methods showed in this paper can be used for other observables like gamma radiation, high energy neutrinos, etc... and can be also used for the case of annihilating dark matter.

Dark matter decays or annihilations can also affect the extragalactic radio background. Compared with other astrophysical sources, dark matter could lead to discriminable anisotropic features in our radio sky~\cite{Zhang:2008rs}. Using future radio observation such as by SKA~\cite{Jackson:2004vw}, possible signals
of dark matter annihilation or decay could be discovered.
In addition, new constraints could result from the $\gamma-$ray flux observations by the Fermi Telescope~\cite{fermi}. Moreover, decay would also affect the expansion history of the Universe because of the change of the equation of state\cite{Gong:2008gi,DeLopeAmigo:2009dc}, and potentially leave an imprint on the Universe. Considering the heating and ionization effects on baryonic gas during the dark age~\cite{Chen:2003gz,Zhang:2006fr,Zhang:2007zzh}, future 21 cm observation~\cite{lofar,Furlanetto:2006wp,Valdes:2007cu} could discover visible evidence for
dark matter decay or annihilations.   

\section*{Acknowledgements}
The authors want to warmly thank Christoph Weniger for fruitful conversations and his help 
with PYTHIA. 
This work was supported by the Deutsche Forschungsgemeinschaft
 (SFB 676 ``Particles, Strings and the Early Universe: The
 Structure of Matter and Space-Time and GRK 602 ``Future
 Developments in Particle Physics'') and the European Union under the ILIAS project (contract No.\ RII3-CT-2004-506222).

\appendix

%
%

\section{\J{Numerical Solution of the Propagation Equation}}

\J{In order to solve the diffusion equation for monochromatic injection of electrons, required for our
response functions, we have developed our own numerical code. 
The main features are the same as in the GALPROP code~\cite{Strong:1998pw}. 
We discretize the parameter space $({\bf r},p,t)$ using cylindrical coordinates for the position in the galaxy 
${\bf r}=(r,z)$. The diffusion zone is confined to be a flat cylinder with radius $r_{\rm max}$ and height $2L$.
The number of bins used for the simulations in this paper is 60 in $r$ and $z$ and 
80 in $p$ (this last one in logarithmic scale).

Neumann boundary conditions are imposed at the origin ($r=0$, $p=0$) since there is no net flux across these interfaces. 
In this point our code differs from GALPROP, since there no boundary conditions in $p$ are used. 
Our code should then provide more accurate solutions at very low energies. 

We impose Dirichlet boundary conditions at the external surface of the diffusion zone by setting the 
electron/positron density to zero there.
However, electrons are also produced outside the diffusion zone and some of them propagate into this region again. 
Because of this, the number of electrons in the stationary solution (within the diffusion zone) would be in reality a bit
higher than our numerical results.
In order to quantify this effect, a more consistent boundary condition is investigated Appendix A.3.  

The stationary solution is looked for by using the Crank-Nicholson implicit updating scheme.
The time intervals start with $10^8$ year and are decreased to refine the solution up to a minimum of $10^2$ years.
We have cross-checked our results using GALPROP in a number of relevant examples and found the agreement satisfactory}.

In the rest of this appendix we provide details on the parameters entering the diffusion-convection-loss Eq.~(\ref{eq:transport}) 
that we use to compute the response functions.

\subsection{Diffusion, Convection, and Re-acceleration}
The diffusion term reflects the spatial propagation of cosmic rays through the tangled Galactic magnetic fields. The diffusion coefficient $D_{xx}({\bf r},p)$ is assumed to be constant within the slab considered and is
described by using a rigidity dependent function,
\begin{equation}\label{eq:Dxx}
D_{xx} = \beta D_0\left(\frac{R}{{\rm GV}}\right)^\delta
\end{equation}
where $\beta = v/c$ is the velocity and $R$ is the rigidity of the particle defined by $R = pc/Ze$ in terms of
momentum $p$ and electric charge $Ze$. The normalization $D_0$ and the spectral index $\delta$ can be determined
from Boron-to-Carbon ratio data~\cite{Maurin:2001sj}.

For case of re-acceleration the momentum diffusion coefficient $D_{pp}$ is related to the spatial diffusion coefficient $D_{xx}$ using the formula given in Ref.~\cite{seo:1994},
\begin{equation}\label{eq:Dpp}
D_{pp} = \frac{4p^2v^2_A}{3\delta(4-\delta^2)(4-\delta)wD_{xx}}\,,
\end{equation}
where $v_A$ is the Alfven speed, and $w$ is the ratio of magnetohydrodynamic wave energy density to the magnetic field energy density, which characterizes the level of turbulence. We take $w=1$ (since it can be subsumed in $v_A$).
\red{The re-acceleration term Eq.~(\ref{eq:Dpp}) is restricted
to a slab of scale height $h_{\rm reac}$ which is in general associated with the gaseous disk
and, therefore, smaller than the scale height of the diffusive region~\cite{Maurin:2002hw}, see 
Tab.~\ref{tab:prop_model} below.}

The convection velocity ${\bf V_c}$ is assumed to be cylindrically symmetric and to point in
the $z$-direction perpendicular to the Galactic plane. The divergence of this velocity gives rise to an energy loss term connected with the adiabatic expansion of cosmic rays. The energy loss term $\dot p$ is due to
interactions of the cosmic rays with the interstellar medium (ISM), interstellar radiation field (ISRF)
and synchrotron radiation in the Galactic magnetic field. 
The ionization, Coulomb interactions, bremsstrahlung, and inverse Compton losses are also taken into account~\cite{Strong:1998pw} and play an important role in case of re-acceleration.

The largest uncertainties in the predicted fluxes come from poorly known propagation parameters,
in particular the possibility of re-acceleration of produced electrons and positrons.
The corresponding uncertainty can reach one order of magnitude. We list in Tab.~\ref{tab:prop_model}
five different combinations of propagation parameters for the models MIN, MED, MAX, DC and DR proposed in Ref.~\cite{Strong:1998pw,Donato:2001ms,Donato:2003xg, Maurin:2002hw}, which are compatible with the observed B/C ratio.
In the present paper, we will not study certain more extreme propagation models such as the ones
discussed in Refs.~\cite{Gebauer:2007mg,Gebauer:2008vq,deBoer:2009rg} which consider relatively large
convection terms and anisotropic diffusion with coefficients that are different for
the radial and the cylindrical directions. We leave the study of such models to future work.


\begin{table}[tb]
\begin{center}
\begin{tabular}{|c||c|c|c|c|c|c|c|c|}
\hline
Model  & $\delta$\footnotemark[1] & $D_0$      & $R$  & $L$ & $V_{c}$ &$dV_c/dz$   & $V_{a}$  & $\red{h_{\rm reac}}$\\
       &                          & [kpc$^2$/Myr]&[kpc] & [kpc] & [km/s] &km/s/kpc     & [km/s] &[kpc]\\
\hline
MIN  & 0.85/0.85 &  0.0016 &20 &1  & 13.5 &0 &  22.4 & 0.1\\
MED  & 0.70/0.70 &  0.0112 &20 &4  & 12   &0 &  52.9 & 0.1\\
MAX  & 0.46/0.46 &  0.0765 &20 &15 &  5   &0 & 117.6 & 0.1\\
DC   & 0/0.55    &  0.0829 &30 &4  &  0   &6 &  0 & 4\\
DR   & 0.34/0.34 &  0.1823 &30 &4  &  0   &0 & 32 &4 \\
\hline
\end{tabular}

\footnotetext[1]{Below/above the break rigidity $\rho_0=4$ GV.}
\
\caption{Typical combinations of diffusion parameters that are consistent with the B/C
analysis. The first three propagation models correspond respectively to minimal, medium and maximal primary antiproton fluxes, abbreviated by MIN, MED, and MAX, respectively. In the DC model, the secondary
$e^{\pm}$, $p$ and $\bar{p}$ fluxes fit the data well, and the DR model can easily reproduce the
energy dependence of the B/C data.}
\vspace{-.5cm}
\label{tab:prop_model}
\end{center}
\end{table}

\subsection{Dark matter halo profile}

In the galactic center the radiation density is $u_\gamma\simeq10\,{\rm eV}\,{\rm cm}^{-3}$, thus the
diffusion length with diffusion coefficient Eq.~(\ref{eq:diff_coeff}) during an energy loss time
Eq.~(\ref{eq:loss}) is
\begin{equation}\label{eq:d_diff}
  d_{\rm diff}\simeq0.45\,\left(\J{R}/{\rm GeV}\right)^{-0.2}\,{\rm kpc}\,,
\end{equation}
which at a distance of $\simeq8.5\,$kpc from the Galactic center corresponds to
an apparent angle
\begin{equation}\label{eq:alpha_diff}
  \alpha_{\rm diff}\simeq3.1^\circ\,\left(\J{R}/{\rm GeV}\right)^{-0.2}\,.
\end{equation}
This is smaller, but not much smaller than the off-set from the Galactic center
of the typical sky directions we use to establish the constraint Eq.~(\ref{eq:constraint})
from radio emission. We, therefore, expect a moderate dependence of the radio emission based
constraints on any possible spikes
concentrated at the Galactic center. On the other hand, dark matter profiles
normalized to the dark matter density at the solar system differ significantly
at $\simeq1\,$kpc from the Galactic center. Some dependence on the dark matter
profile can thus not be avoided.

In this work, we will adopt three spherically symmetric density profiles for the
dark matter distribution. Recent N-body simulations suggest that a radially symmetric dark matter halo
profile can be parametrized by~\cite{Bertone:2004pz}
\begin{equation}\label{eq:dm_profile}
\rho_X(r) = \frac{\J{\rho_0}}{(r/\J{r_0})^{\gamma}[1+(r/\J{r_0})^{\alpha}]^{(\beta-\gamma)/\alpha}}\,.
\end{equation}
In Tab.~\ref{tab:halo}, we give the values of the parameters ($\alpha,\beta,\gamma$)
for some of most widely used profiles such as the Kra~\cite{Kra}, NFW~\cite{NFW} and
Iso~\cite{Bergstrom:1997fj} profiles. The constant
$\rho_0$ can be normalized to the dark matter density of $0.3~\rm{GeV/cm^3}$ at the
local solar system ($r=8.5$ kpc).

\begin{table}[t]
\begin{center}
\begin{tabular}{|c||c|c|c|c|}
\hline
model  & $\alpha$ & $\beta$  & $\gamma$ & $\J{r_0}(\rm kpc)$ \\
\hline
Kra  &     2      &  3      &0.4       &10   \\
Iso &     2    &  2      & 0       &3.5   \\
NFW  &     1      &  3      & 1        &20   \\
\hline
\end{tabular}
\caption{Some of widely used dark matter halo density profiles used in the present study.} 
\vspace{-.7cm}
\label{tab:halo}
\end{center}
\end{table}

\subsection{Boundary conditions}

In solving Eq.~(\ref{eq:transport}), traditionally, ones imposed the Dirichlet boundary condition
$n_e(\J{r},z=\pm L)=0$, $n_e(\J{r}=\J{r_{\rm max}},z)=0$, at which the particles can freely escape. However, electrons and positrons are also produced by decays outside the diffusion zone in the Galactic halo. 
In this environment, they can propagate along straight lines since inverse
ultra-relativistic inverse Compton scattering on low energy photons is boosted in the extreme forward direction. Furthermore, since $u_\gamma\simeq1\,{\rm eV}\,{\rm cm}^{-3}$ and becomes eventually dominated by the CMB far above the Galactic plane, according to Eq.~(\ref{eq:loss})
the energy loss length is $\ga100\,$kpc up to TeV energies. Thus, energy loss can be
neglected on halo scales. 
At cylindrical distance $\J{r}$ from the Galactic center, the total flux from the halo into the diffusion zone at
its boundary at $z=\pm L$ is given by
\begin{eqnarray}\label{eq:halo_flux}
  \fl j_e(E,\rho)_{\rm halo}&\simeq&\frac{1}{4\pi}\frac{1}{m_X\tau_X}\frac{dN_e}{dE}(E)\\
  &&\hskip-3cm\times\int_0^{\pi/2}d\theta\cos\theta\sin\theta\int_0^{2\pi}d\phi\int_0^\infty \J{ds}\, 
  \J{\rho_X}\left(\sqrt{\J{r}^2+L^2+l^2+2\J{s}(\J{r}\sin\theta\sin\phi+L\cos\theta)}\right)
  \nonumber\,,
\end{eqnarray}
where the dark matter profile $\rho_X(r)$ is assumed to be spherically symmetric.

In Eq.~(\ref{eq:halo_flux}), the integration is performed over the hemisphere above or below
the diffusion zone where the flux from a given direction is a line of sight integral over $\J{s}$.

\begin{figure}
\centering
\includegraphics[width=0.6\textwidth]{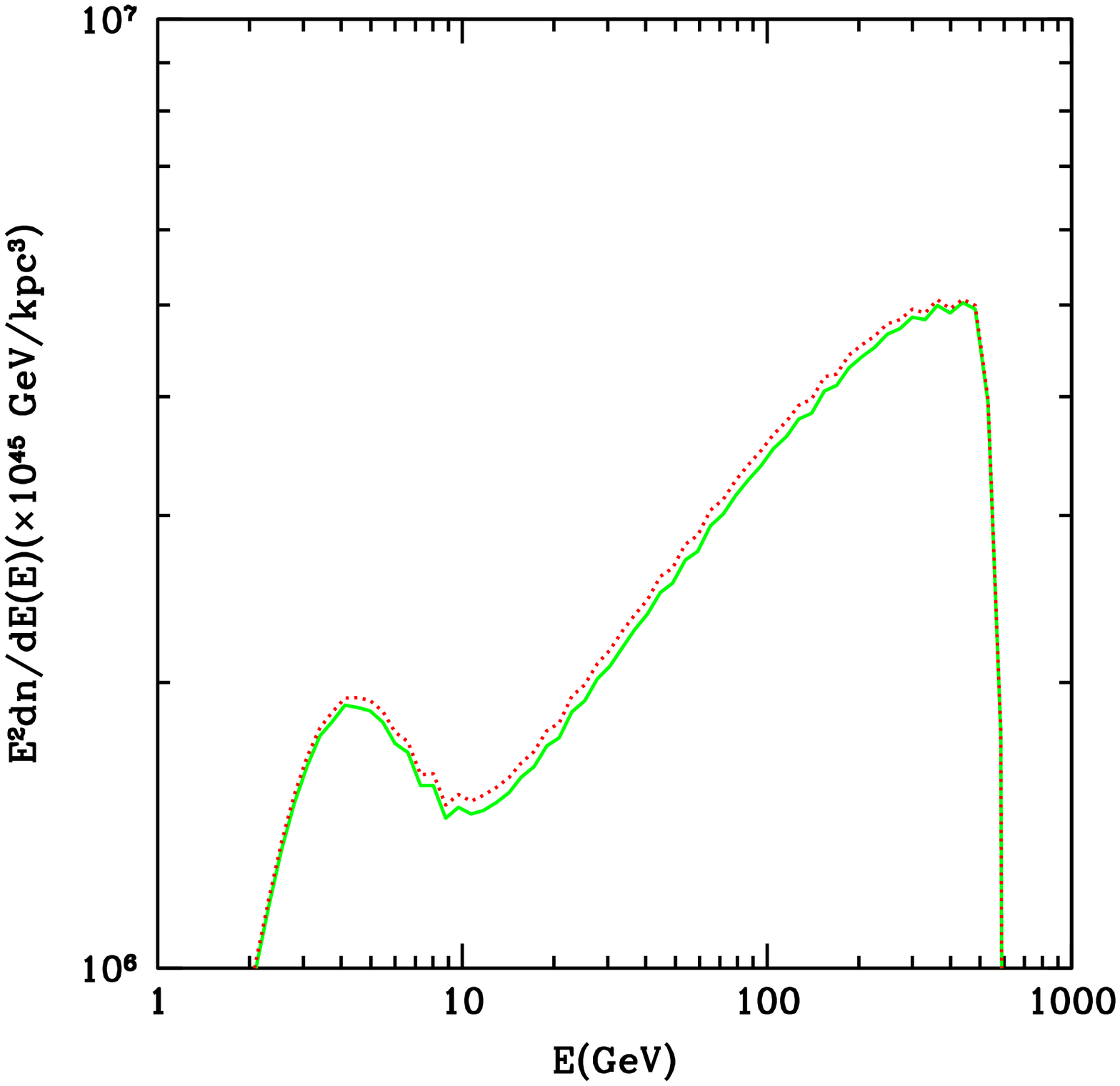} 
\caption{The dependence of electron density at Earth on boundary conditions in the MIN propagation model,
for injection of one electron of 500 GeV energy. The red line refers to the Robin boundary condition
Eq.~(\ref{eq:boundary}), and the green one to the Dirichlet condition.}
\label{fig:boundary}
\end{figure}

Continuity of the flux at the diffusive halo boundary then requires
\begin{equation}\label{eq:boundary}
  \left|D_{zz}(E,\J{r},\pm L)\,\partial_z n_e(E,\J{r},\pm L)\right|
  =\frac{c_0}{4}\,n_e(E,\J{r},\pm L)-j_e(E,\J{r})_{\rm halo}\,,
\end{equation}
where $n_e(E,\J{r},z)$ is the local electron plus positron density per unit energy whose distribution is
simulated in the propagation code. Eq.~(\ref{eq:boundary}) then
determines the boundary condition for the numerical simulation of the
electron-positron distribution. Although we use this third type boundary condition at $z=\pm L$ instead of the Dirichlet boundary condition, the Dirichlet boundary condition turns out to be a very good approximation since most decays occur within 1 kpc from the Galactic center. For example, in the MIN model, considering the
halo contribution, the radio signals would increase by only $10\%$ in Fig.~\ref{fig:boundary}, and
in other propagation models with a larger diffusion zone the enhancement is negligible,
as one would expect.

\subsection{Magnetic Field and Photon Energy Density}

Various techniques have been applied to the 
determination of the Galactic magnetic field. Detailed analysis of the rotation measures and dispersion
of pulsar emission has been carried out~\cite{B1994,Han:2006ci}. The work presented in Ref.~\cite{B1996} which is based on the large-scale data set on starlight polarization~\cite{B1970} with nearly 7000 stars show that the local field is parallel to the Galactic plane and follows the local spiral arms. A smooth Galactic magnetic field is also consistent with the conclusions
of Ref.~\cite{B1996} and can be parametrized as
\begin{equation}\label{eq:B1}
B(r,z) = 6\, e^{(-r/20 {\rm kpc})}e^{-|z|/5 {\rm kpc}}\,\mu{\rm G}.
\end{equation}
Random fluctuations are not included in the model. In order to quantify the influence of uncertainties
of the Galactic magnetic field on dark matter constraints, a second magnetic field model is also considered,
which is parametrized by
\begin{equation}\label{eq:B2}
B(r,z) = 5\,e^{(-(\J{r}-8.5{\rm kpc})/10 {\rm kpc})}e^{-|z|/2 {\rm kpc}}\, \mu{\rm G},
\end{equation}
The value of these parameters are adjusted to match the 408 MHz synchrotron distribution~\cite{Strong:1998fr}. 

The magnetic field profile close to the Galactic Centre is quite uncertain and could be considerably higher than a few $\mu$G~\cite{melia}. However, this would further reduce the diffusion angle Eq.~(\ref{eq:alpha_diff})
and is thus unlikely to have a significant influence on the radio constraints since the discussion in Sect.~\ref{sec:con-r} will show that the maximum excess of predicted
over observed signal does not occur within $\sim5^\circ$ of the Galactic center.

The ISRF distribution can be derived based on the IRAS and COBE infra-red data as well as by using
information on the stellar luminosity function. In the present work we use the model distributed with the GALPROP code~\cite{Strong:1998fr}. In this model, the ISRF energy density is about 10 ${\rm eV/cm^3}$
near the center and 5 ${\rm eV/cm^3}$ at the solar position.

\begin{figure}
\includegraphics[width=0.5\textwidth]{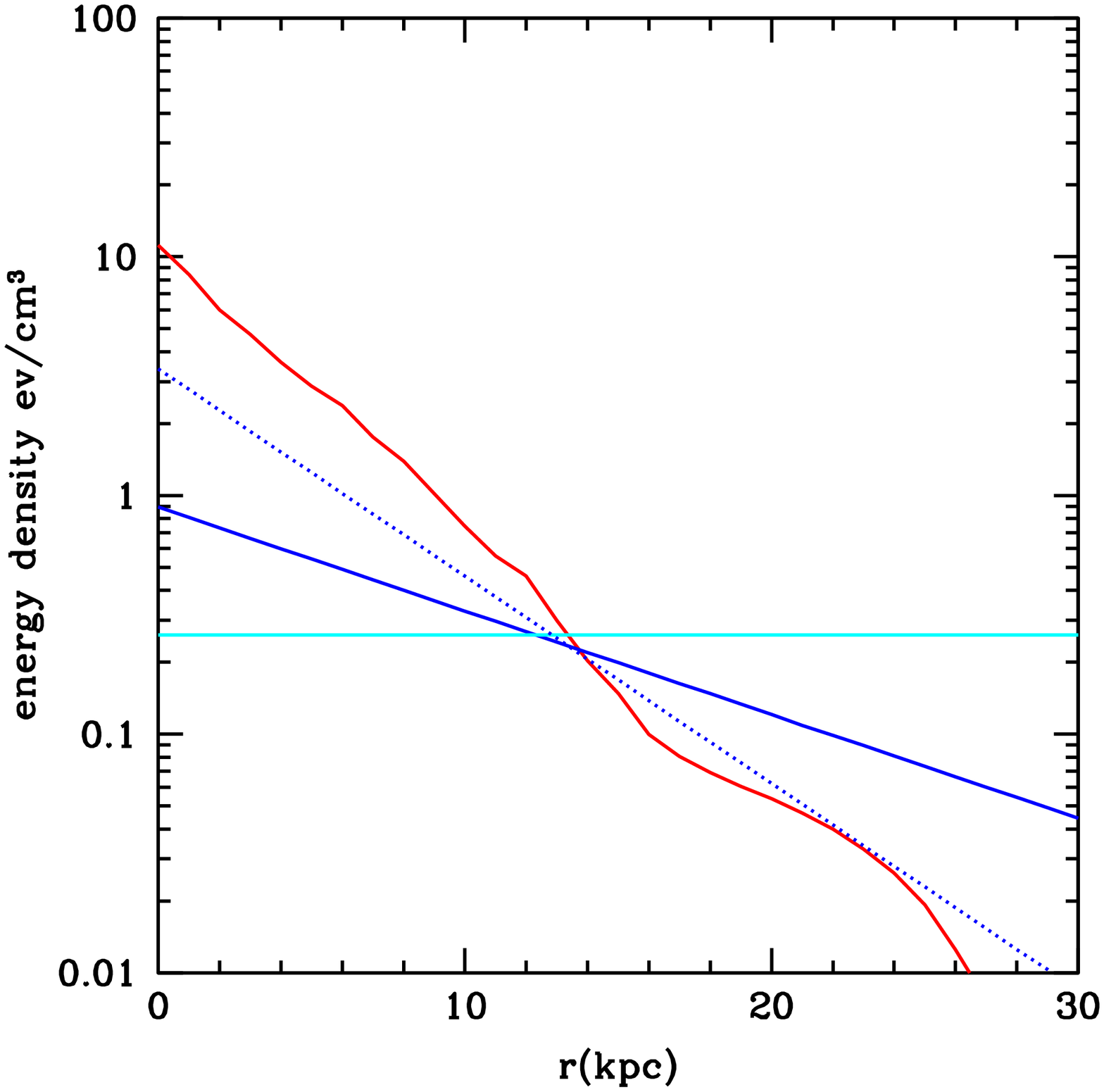}
\includegraphics[width=0.5\textwidth]{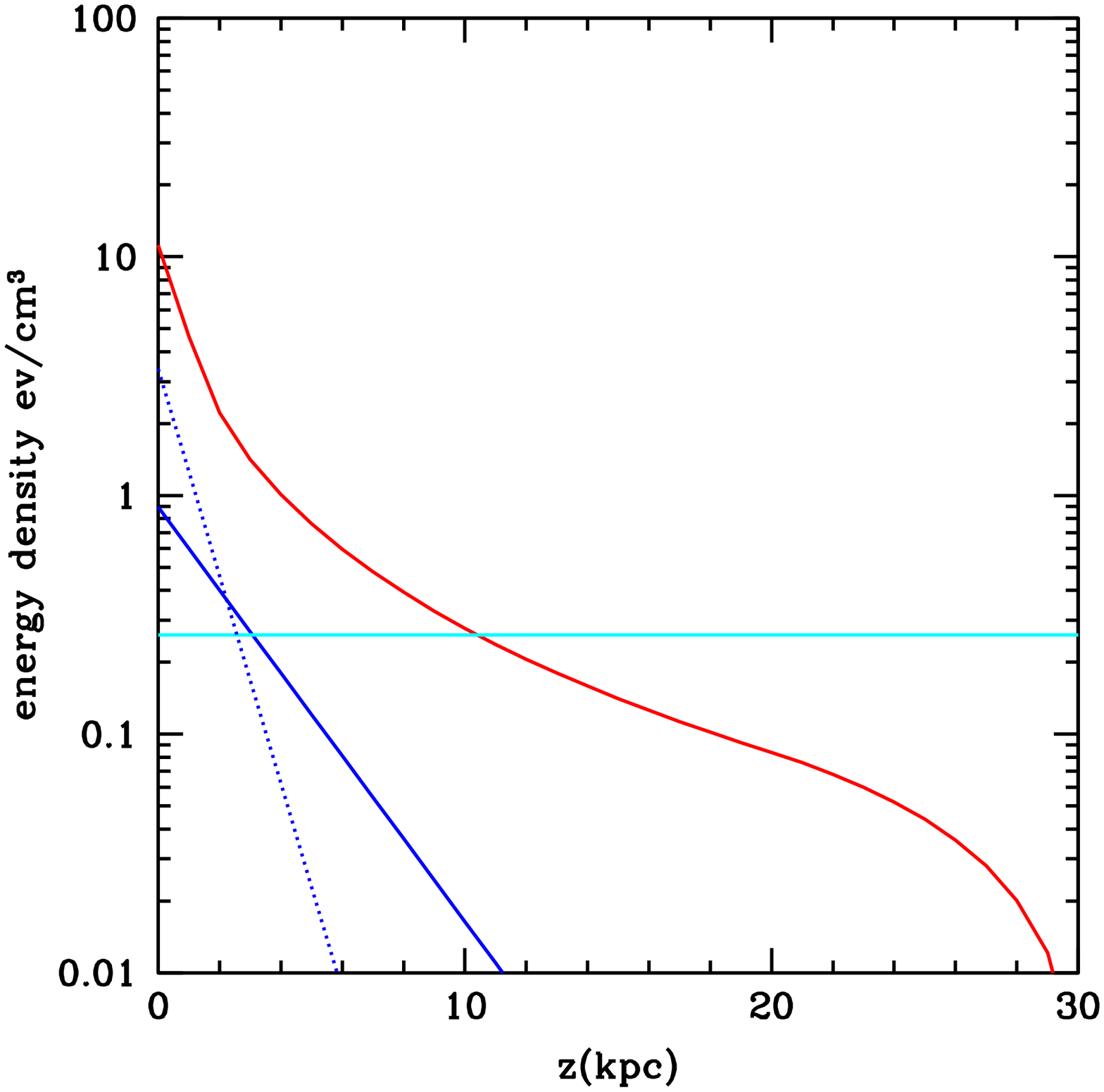}
\caption{The IRAS energy density as function of $r$ at $z=0$ and as function of $z$ at $\rho=0$. The contributions from stellar radiation, magnetic field from Eq.~(\ref{eq:B2}) and from Eq.~(\ref{eq:B1}) as well as from the
CMB are shown from top to bottom on the left side.}  
\label{fig:rz}
\end{figure}

In Fig.~\ref{fig:rz} we show the spatial distribution of photon energy density including CMB and magnetic field. For the electrons and positrons with energy above 1 GeV, inverse Compton scattering dominates electron energy loss since the stellar energy density $u_{\rm rad}$ is always much larger than the magnetic field energy density $u_B$. Since the synchrotron and inverse Compton processes have similar energy loss rates, the cooling time of electrons and positrons therefore are almost independent of the magnetic field. In other words, at very high energies where the diffusion length becomes large compared to the energy loss length, the number density of electrons and positrons is approximately determined by the strength of the interstellar radiation field only.

%
%

\section{Response Function Fits}
\subsection{Response Function for Radio Signals}

The response functions for different halo models are very similar and in first approximation proportional
to each other. For that reason, we will consider mostly the NFW halo in the following. The response
functions for other halos \J{are the same within roughly 10$\%$}.

For the NFW profile the synchrotron response functions $F_r(E_0)$ as a function of electron injection
energy $E_0$ in GeV can be fitted with the following expressions
\begin{eqnarray}\label{eq:fits_s}
F_r(E_0) &=&  N e^{-(E_*/E_0)^3}E_0^{-\delta}		{\rm \quad (DC,DR)} \\
F_r(E_0) &=&   N_1 E_0^{\delta_1}e^{-(E_*/E_0)^x}+N_2 E_0^{-\delta_2}(1-e^{-(E_0/E_*)^x})	{\rm \quad (MIN,MED,MAX)}
\nonumber\,,
\end{eqnarray}
where the fitting parameters are given in Tab.~\ref{tab:param1}.

\begin{table}[t!]
\begin{center}
\begin{tabular}{c|ccc}
DC		&	$\nu_1$		&	$\nu_2$ 		&	$\nu_3$		\\ \hline
$N(\times 10^2)$		&	$7.8$		&	$12$ 		&	$94$		\\ 
$E_*$	&	$3.2$		&	$5.7$ 		&	$23$		\\ 
$\delta$	&	$0.28$		&	$0.21$ 		&	$0.17$		\\ 
\end{tabular}
\hspace{0.5cm}
\begin{tabular}{c|ccc}
DR		&	$\nu_1$		&	$\nu_2$ 		&	$\nu_3$		\\ \hline
$N(\times 10^2)$		&	$4.7$		&	$8.4$ 		&	$68$		\\ 
$E_*$	&	$2.75$		&	$5.11$ 		&	$22.4$		\\ 
$\delta$	&	$0.20$		&	$0.14$ 		&	$0.08$		\\ 
\end{tabular}
\end{center}
\vspace{0cm}
\centering
{\begin{tabular}{c|ccc}
MAX	& $\nu_1$ & $\nu_2$ & $\nu_3$ \\ \hline
$N_1(\times 10^3)$	& $25$ & $8.2$ & $0.14$ \\
$\delta_1$	& $0 .58$ & $0 .56$ & $0 .51$ \\
$N_2(\times 10^2)$	& $19$ & $24$ & $74$ \\
$\delta_2$	& $0 .04$ & $0 .02$ & $0$ \\
$E_0$	& $6.59$ & $10.3$ & $31.2$ \\
$x$	& $2 .27$ & $2 .43$ & $3 .94$ \\
\end{tabular}
\begin{tabular}{c|ccc}
MED	& $\nu_1$ & $\nu_2$ & $\nu_3$ \\ \hline
	& $29$ & $11$ & $0.11$ \\
	& $0.68$ & $0.68$ & $0.67$ \\
	& $23$ & $25$ & $86$ \\
	& $0.15$ & $0.12$ & $0.05$ \\
	& $10.5$ & $12.4$ & $30$ \\
	& $2.4$ & $2.5$ & $4.5$ \\
\end{tabular}
\begin{tabular}{c|ccc}
MIN	& $\nu_1$ & $\nu_2$ & $\nu_3$ \\ \hline
 	& $18$ & $11$ & $0.4$ \\
  	& $0.2$ & $0.2$ & $0.2$ \\
 	& $10$ & $14$ & $83$ \\
	& $0.28$ & $0.29$ & $0.29$ \\
	& $4.5$ & $8.5$ & $31$ \\
	& $2.6$ & $2.3$ & $3.7$ \\
\end{tabular}}
\vspace{0.5cm}
\caption{Values of the parameters in the fits Eq.~(\ref{eq:fits_s}) to the synchrotron response function
for the different propagation models, assuming an NFW dark matter profile. The three frequencies 
$\nu_{1,2,3}$ are $408$ MHz, $1.42$GHz and $23$GHz, respectively.}
\label{tab:param1}
\end{table}

\subsection{Response Function for Positron Flux}

\begin{table}[h!]
\caption{Values of the parameters in the fits Eq.~(\ref{eq:fits_p}) to the positron response function
for the different propagation models and observation energies, assuming an NFW dark matter profile.}

\centering
\begin{tabular}{c|ccccccc}
DC	& $10.17$ & $13.11$ & $17.52$ & $24.02$ & $35.01$ & $53.52$ & $82.55$ \\  \hline
$N_1$	& $7.5$ & $9.1$ & $9.9$ & $8.9$ & $9.3$ & $7.7$ & $4.1$ \\
$\delta_1$	& $0.71$ & $0.38$ & $0.45$ & $0.28$ & $0.35$ & $0.44$ & $0.27$ \\
$N_2$	& $2.1$ & $3.2$ & $5.6$ & $7.8$ & $14.$ & $21.$ & $26.$ \\
$\delta_2$	& $0.19$ & $0.19$ & $0.19$ & $0.18$ & $0.18$ & $0.16$ & $0.14$ 
\end{tabular}\vspace{.5cm}

\begin{tabular}{c|ccccccc}
DR	& $10.17$ & $13.11$ & $17.52$ & $24.02$ & $35.01$ & $53.52$ & $82.55$ \\ \hline
$l_1$	& $15.$ & $16.$ & $19.$ & $22.$ & $26.$ & $32.$ & $39.$ \\
$w$	& $1.8$ & $1.5$ & $1.3$ & $1.1$ & $0.85$ & $0.73$ & $0.63$ \\
$N_2$	& $1.6$ & $2.4$ & $4.3$ & $6.2$ & $11.$ & $16.$ & $19.$ \\
$\delta_2$	& $0.13$ & $0.12$ & $0.12$ & $0.12$ & $0.11$ & $0.09$ & $0.059$ \\
$N_3$	& $120.$ & $130.$ & $5500.$ & $4100.$ & $27000.$ & $6.6\times 10^6$ & $670000.$ \\
$\delta_3$	& $0.8$ & $0.65$ & $1.6$ & $1.4$ & $1.6$ & $2.7$ & $1.9$ \
\end{tabular}\vspace{.5cm}

\centering
\begin{tabular}{c|ccccccc}
MIN	& $10.17$ & $13.11$ & $17.52$ & $24.02$ & $35.01$ & $53.52$ & $82.55$ \\ \hline
$N_1(\times 10^3)$	& $31$ & $7.5$ & $1.2$ & $0.13$ & $10^{-3}$ & $10^{-5}$ & $10^{-10}$ \\
$\delta_1$	& $0.59$ & $0.57$ & $0.56$ & $0.54$ & $0.53$ & $0.52$ & $0.52$ \\
$N_2$	& $8.6$ & $9.9$ & $13.$ & $13.$ & $18.$ & $20.$ & $23.$ \\
$\delta_2$	& $0.024$ & $0.016$ & $0.0083$ & $0$ & $0$ & $0$ & $0$ \\
$E_0$	& $12.$ & $17.$ & $23.$ & $32.$ & $45.$ & $60.$ & $93.$ \\
$x$	& $5.2$ & $5.2$ & $6.3$ & $6.7$ & $8.4$ & $11.$ & $12.$ \\
\end{tabular}\vspace{.5cm}

\centering
\begin{tabular}{c|ccccccc}
MED	& $10.17$ & $13.11$ & $17.52$ & $24.02$ & $35.01$ & $53.52$ & $82.55$ \\ \hline
$N_ 1(\times 10^3)$	& $29$ & $4.8$ & $0.36$ & $0.022$ & $10^{-4}$ & $10^{-7}$ & $10^{-10}$ \\
$\delta_1$	& $0 .65$ & $0 .64$ & $0 .64$ & $0 .63$ & $0 .62$ & $0.61$ & $0 .61$ \\
$N_ 2$	& $9 .3$ & $11.$ & $14.$ & $14.$ & $18.$ & $18.$ & $19.$ \\
$\delta_2$	& $0 .11$ & $0 .1$ & $0 .087$ & $0 .071$ & $0 .051$ & $0.022$ & $0$ \\
$E_ 0$	& $12.$ & $17.$ & $22.$ & $31.$ & $43.$ & $57.$ & $89.$ \\
$x$	& $6.$ & $6.$ & $7 .9$ & $8 .3$ & $11.$ & $14.$ & $16.$ \\
\end{tabular}\vspace{.5cm}

\centering
\begin{tabular}{c|ccccccc}
MAX	& $10.17$ & $13.11$ & $17.52$ & $24.02$ & $35.01$ & $53.52$ & $82.55$ \\ \hline
$N_1(\times 10^3)$	& $180$ & $37$ & $3.5$ & $0.22$ & $10^{-3}$ & $10^{-6}$ & $10^{-9}$ \\
$\delta_1$	& $0.29$ & $0.29$ & $0.29$ & $0.29$ & $0.29$ & $0.29$ & $0.29$ \\
$N_2$	& $11.$ & $14.$ & $21.$ & $25.$ & $36.$ & $45.$ & $33.$ \\
$\delta_2$	& $0.32$ & $0.33$ & $0.33$ & $0.32$ & $0.31$ & $0.29$ & $0.2$ \\
$E_0$	& $11.$ & $17.$ & $22.$ & $31.$ & $43.$ & $57.$ & $88.$ \\
$x$	& $4.5$ & $4.7$ & $6.7$ & $7.4$ & $10.$ & $14.$ & $16.$ \\
\end{tabular}\label{tab:param2}
\end{table}\vspace{.5cm}

The response functions for the positron flux based on the PAMELA data at a given
energy $E$ as a function of injection energy $E_0$ in GeV can be fitted with 

\begin{eqnarray}\label{eq:fits_p}
\fl F^{\rm DC}_p(E_0) &=&  N_1(E_0/E-1)^{\delta_1} e^{-(E_0/E)} + N_2 E_0^{-\delta_2}\nonumber		\\
\fl F^{\rm DR}_p(E_0) &=&  \frac{e^{-4(E_0/E)}}{1+E_0^{-1.3}e^{l_1- w E_0}}  + \left(N_3 E_0^{-\delta_3} e^{-E_0/E} +N_2 E_0^{-\delta_2}\right)e^{-2(E/E_0)^3}		\\
\fl F^{\rm MM}_p(E_0) &=&   N_1 E_0^{\delta_1}e^{-(E_0/E_*)^x}+N_2 E_0^{-\delta_2}(1-e^{-(E_0/E_*)^x})\nonumber\,,
\end{eqnarray}

where the values for the fitting parameters can be found in Tab.~\ref{tab:param2} for the seven PAMELA energy bins $E$.

The DC model does not include re-acceleration and can be modeled by a power law at high energies ($E_0\gg E$).
Our calculations show an enhancement near $E \sim E_0 $ which is fitted by the parameters with subscript $1$.

The DR model includes re-acceleration but the effect is not very strong. In the region $E_0<E$ the response function is an exponentially suppressed power law  while near the threshold and at high energies the behavior is similar to the DC model.

The MIN, MED and MAX model have strong re-acceleration and can be fitted at both high and low energies with power laws with small indices. The matching at energy $\sim E_0$ is included by the exponential factors with an parameter $x$.

\end{document}